\documentclass[10pt,apj]{emulateapj}
\usepackage[dvips]{color}

\usepackage{apjfonts}

\newcommand\kms{km s$^{-1}$}
\newcommand\msun{\ifmmode{M_{\odot}}\else $M_{\odot}$\fi}
\newcommand\rsun{\ifmmode{R_{\odot}}\else $R_{\odot}$\fi}
\newcommand\rgc{$R_{\rm GC}$}
\newcommand{\teff}{$T_{\rm eff}$} 
\newcommand{\fei}{Fe\,{\sc i}} 
\newcommand{\feii}{Fe\,{\sc ii}} 

\usepackage[dvips]{color}

\begin{document}

\title{ELEMENTAL ABUNDANCE RATIOS IN STARS OF THE OUTER GALACTIC DISK.\ IV.\ 
A NEW SAMPLE OF OPEN CLUSTERS\altaffilmark{1}}

\author{DAVID YONG}
\affil{Research School of Astronomy and Astrophysics, 
Australian National University, Weston, ACT 2611, Australia; 
yong@mso.anu.edu.au} 

\author{BRUCE W.\ CARNEY}
\affil{Department of Physics \& Astronomy, University of North
Carolina, Chapel Hill, NC 27599-3255, USA; email: bruce@physics.unc.edu}

\author{EILEEN D.\ FRIEL}
\affil{Department of Astronomy, Indiana University, Bloomington, IN 47405, USA; email: efriel@indiana.edu} 

\altaffiltext{1}{The data presented herein were 
obtained at the W.M. Keck Observatory, which is operated as a scientific 
partnership among the California Institute of Technology, the University 
of California and the National Aeronautics and Space Administration. The 
Observatory was made possible by the generous financial support of the 
W.\ M.\ Keck Foundation.}

\begin{abstract}

We present radial velocities and chemical abundances for nine stars 
in the old, distant open clusters Be 18, Be 21, Be 22, Be 32, 
and PWM 4. For Be 18 and PWM 4, these are the first chemical abundance 
measurements. 
Combining our data with literature results 
produces a compilation of some 
68 chemical abundance measurements 
in 49 unique clusters. 
For this combined 
sample, we study the chemical 
abundances of open clusters as a function of 
distance, age, and metallicity. 
We confirm that the metallicity gradient in the 
outer disk is flatter than the gradient in the vicinity 
of the solar neighborhood. We also confirm that the 
open clusters in the outer disk are metal-poor with enhancements 
in the ratios [$\alpha$/Fe] and perhaps [Eu/Fe]. 
All elements show negligible or small trends between [X/Fe] and distance 
($<$ 0.02 dex/kpc), 
but for some elements, there is a hint that the 
local (\rgc\ $<$ 13 kpc) and distant (\rgc\ $>$ 13 kpc) 
samples may have 
different trends with distance. 
There is no evidence for significant 
abundance trends versus age ($<$ 0.04 dex Gyr$^{-1}$).  
We measure the linear relation 
between [X/Fe] and metallicity, [Fe/H], and find that the scatter about 
the mean trend is 
comparable to the measurement 
uncertainties. 
Comparison with 
solar neighborhood field giants shows that the open 
clusters share similar abundance ratios [X/Fe] at a given 
metallicity. 
While the flattening of the metallicity gradient and enhanced 
[$\alpha$/Fe] ratios in the outer disk suggest a different chemical 
enrichment history to the solar neighborhood, 
we echo the sentiments expressed by Friel et al.\ that 
definitive conclusions await 
homogeneous analyses of larger samples of stars in 
larger numbers of clusters.  
Arguably, our understanding of the evolution of the outer disk from open 
clusters is currently limited by systematic abundance differences between various 
studies. 

\end{abstract}


\section{INTRODUCTION}

Our Galaxy's open clusters are valuable 
tools to study the disk \citep{friel95}. 
Accurate homogeneous distances and ages can be measured for samples of 
open clusters that span a wide range in parameters \citep{salaris04}. 
Additionally, metallicity estimates of open clusters 
can be readily obtained from 
a variety of methods (albeit with differing degrees of accuracy) 
thereby enabling studies of the 
structure, kinematics, and chemistry of the disk as well as 
any temporal variations of these properties. 

The atmospheres of low-mass stars retain, to a great extent, the 
chemical composition of the interstellar medium at the time and place 
of their birth. 
The nucleosynthetic yields of the chemical elements 
depend upon stellar mass and metallicity. 
Therefore, measurements 
of metallicity, [Fe/H], and chemical abundance ratios, [X/Fe], 
in stars in open clusters 
offer powerful insight into the formation and evolution of the disk 
\citep{janes79,freeman02,friel02,jbh10,kobayashi11}. 

In recent times there has been considerable effort 
to understand the evolution of the outer Galactic disk 
(e.g., \citealt{hou00,chiappini01,andrievsky02c,daflon04,costa04,cescutti07,magrini09}). 
It appears that the metallicity gradient (i.e., [Fe/H] versus Galactocentric 
distance, \rgc) in the outer disk (\rgc\ $>$ 13 kpc) is flatter than 
the metallicity gradient in the solar neighborhood 
(e.g., \citealt{twarog97,luck03,carraro04,carney05,y05,bragaglia08,sestito08,jacobson09,friel10}). 
Qualitatively similar behavior has now been found in external 
galaxies (e.g., \citealt{worthey05,bresolin09,vlajic09,vlajic11})
suggesting that the processes which 
govern chemical evolution in the outer Galactic disk have also 
operated in the outskirts of other disk galaxies. 

Another intriguing result is evidence for 
enhanced [$\alpha$/Fe] ratios in the outer Galactic disk 
from open clusters \citep{carraro04,y05}, 
field stars \citep{carney05,bensby11}, and 
Cepheids \citep{y06}. Such abundance 
ratios in objects spanning a large range in 
ages can be explained by 
infall of pristine gas in the outer disk and/or 
vigorous star formation. Either explanation requires that the 
outer disk has experienced a significantly different 
star formation history compared to the solar neighborhood 
and this 
has important implications for the formation 
and evolution of the Galactic disk. 

However, not all studies of the outer disk find enhanced [$\alpha$/Fe] 
ratios (e.g., \citealt{bragaglia08}). 
While one explanation for the discrepancy may be 
systematic differences in the analyses, 
studies of additional open clusters in the outer disk are necessary to 
clarify the situation. 
Therefore, to further explore the formation and evolution of 
the outer Galactic disk, we analyze the chemical abundances of 
a new sample of open clusters. 
This is the fourth, and final, paper in our series on elemental 
abundance ratios in stars in the outer Galactic disk. 
Paper I \citep{y05} concentrated upon open clusters, Paper II 
\citep{carney05}
was dedicated to field red giants, and Paper III \citep{y06} 
focused upon Cepheids. 
In this paper, we present radial velocities, metallicities [Fe/H], 
and element abundance ratios [X/Fe] for the open clusters 
Be 18, Be 21, Be 22, Be 32, and PWM 4. 

\section{PROGRAM STARS AND OBSERVATIONS}

\subsection{Target Selection}

\begin{figure}[t!]
\epsscale{1.2}
\plotone{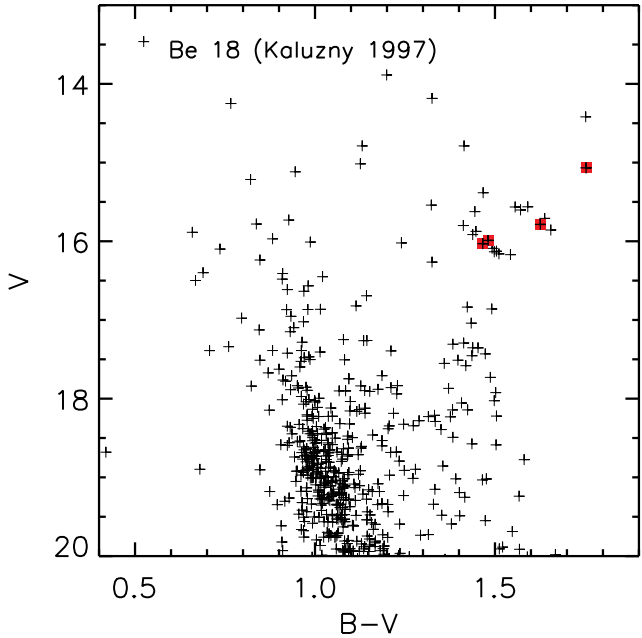}
\caption{Be 18 color-magnitude diagram using the data from 
\citet{kaluzny97}. We distinguish our program 
stars by red squares. 
\label{fig:cmdbe18bv}}
\end{figure}

\begin{figure}[t!]
\epsscale{1.2}
\plotone{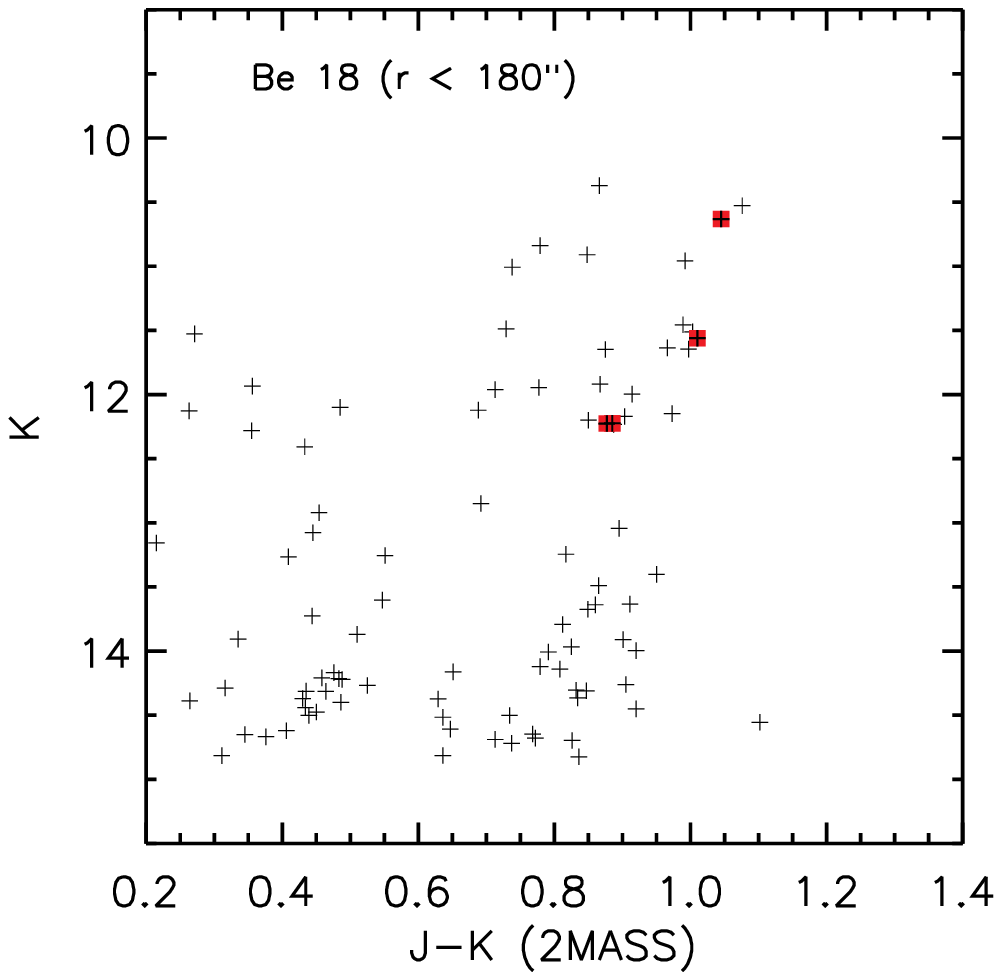}
\caption{Be 18 color-magnitude diagram using the data from 
2MASS \citet{2mass}. We distinguish our program 
stars by red squares. 
\label{fig:cmdbe18jk}}
\end{figure}

\begin{figure}[t!]
\epsscale{1.2}
\plotone{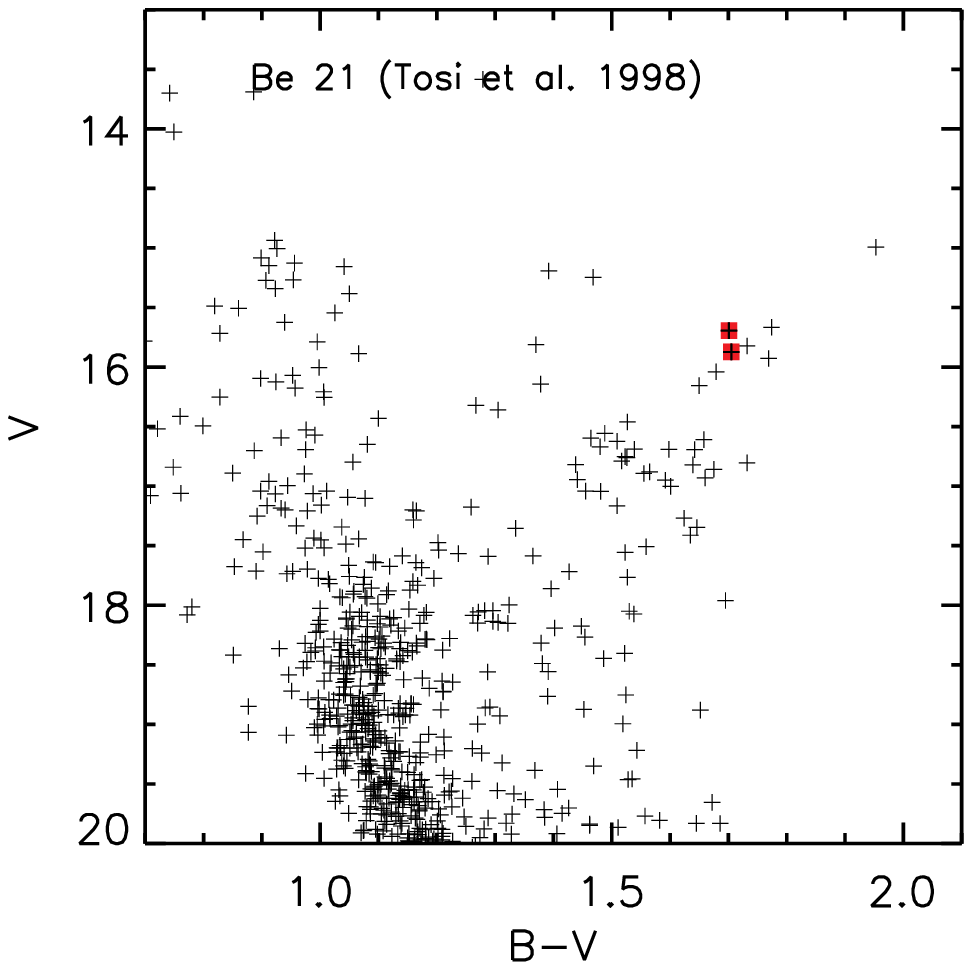}
\caption{Be 21 color-magnitude diagram using the data from 
\citet{tosi98}. We distinguish our program 
stars by red squares. 
\label{fig:cmdbe21bv}}
\end{figure}

\begin{figure}[t!]
\epsscale{1.2}
\plotone{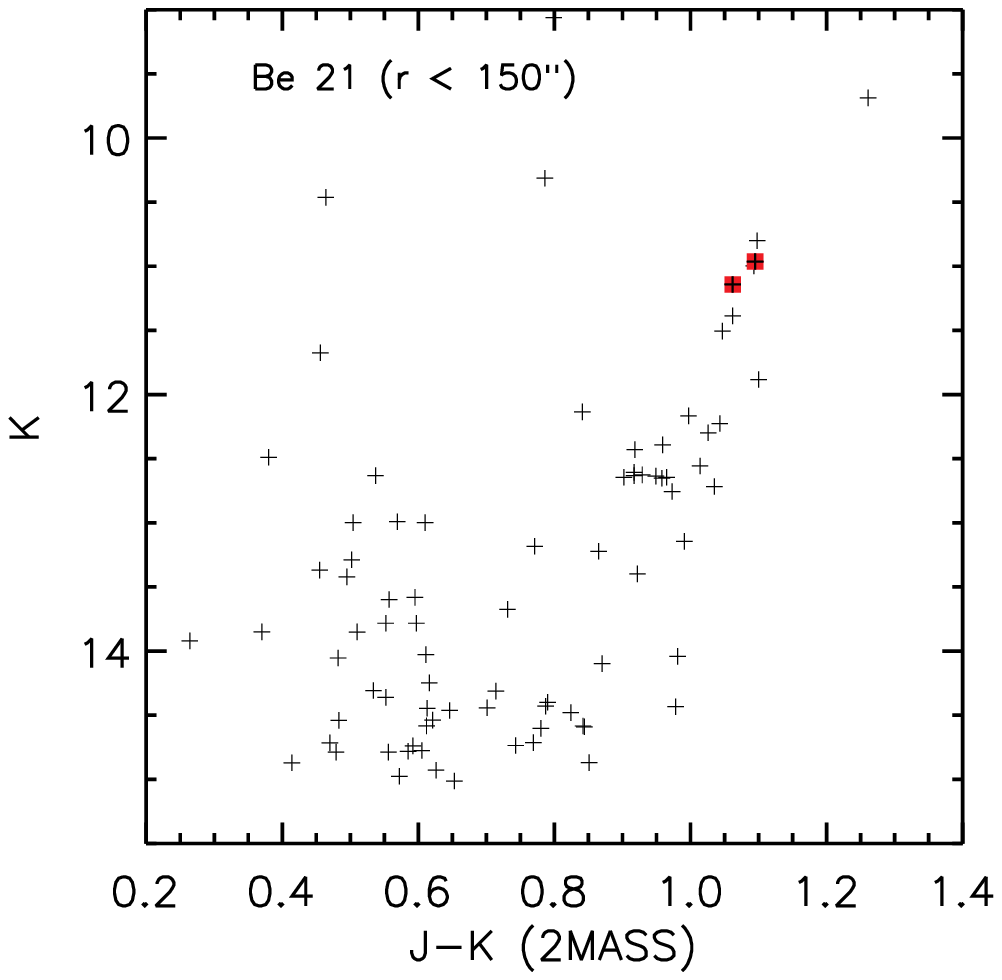}
\caption{Be 21 color-magnitude diagram using the data from 
2MASS \citet{2mass}. We distinguish our program 
stars by red squares. 
\label{fig:cmdbe21jk}}
\end{figure}

\begin{figure}[t!]
\epsscale{1.2}
\plotone{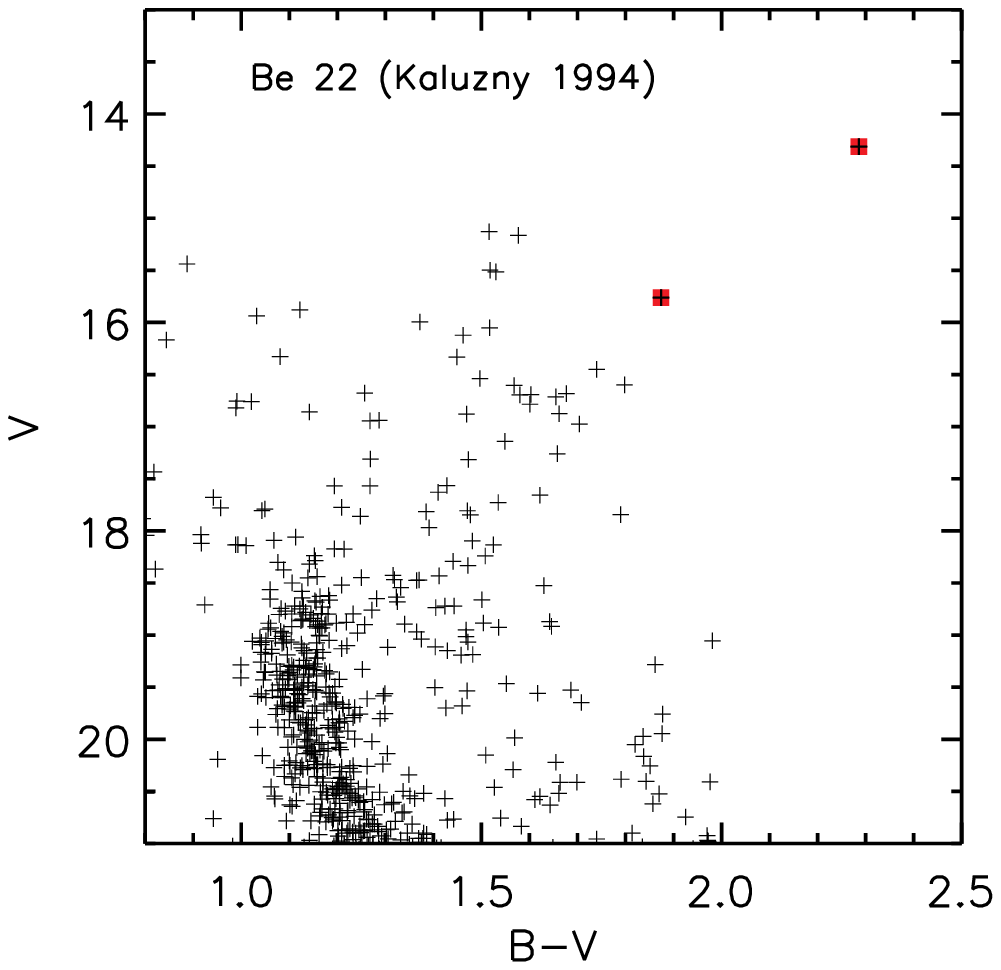}
\caption{Be 22 color-magnitude diagram using the data from 
\citet{kaluzny94}. We distinguish our program 
stars by red squares. 
\label{fig:cmdbe22bv}}
\end{figure}

\begin{figure}[t!]
\epsscale{1.2}
\plotone{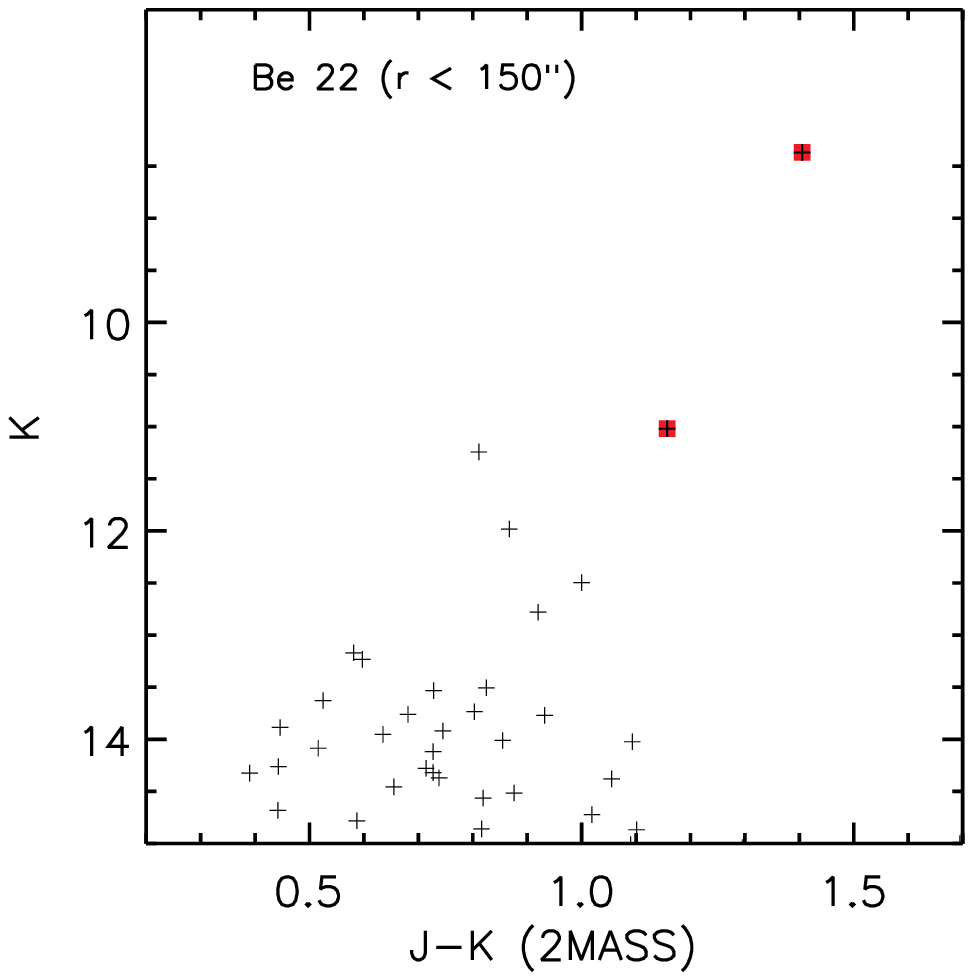}
\caption{Be 22 color-magnitude diagram using the data from 
2MASS \citet{2mass}. We distinguish our program 
stars by red squares. 
\label{fig:cmdbe22jk}}
\end{figure}

\begin{figure}[t!]
\epsscale{1.2}
\plotone{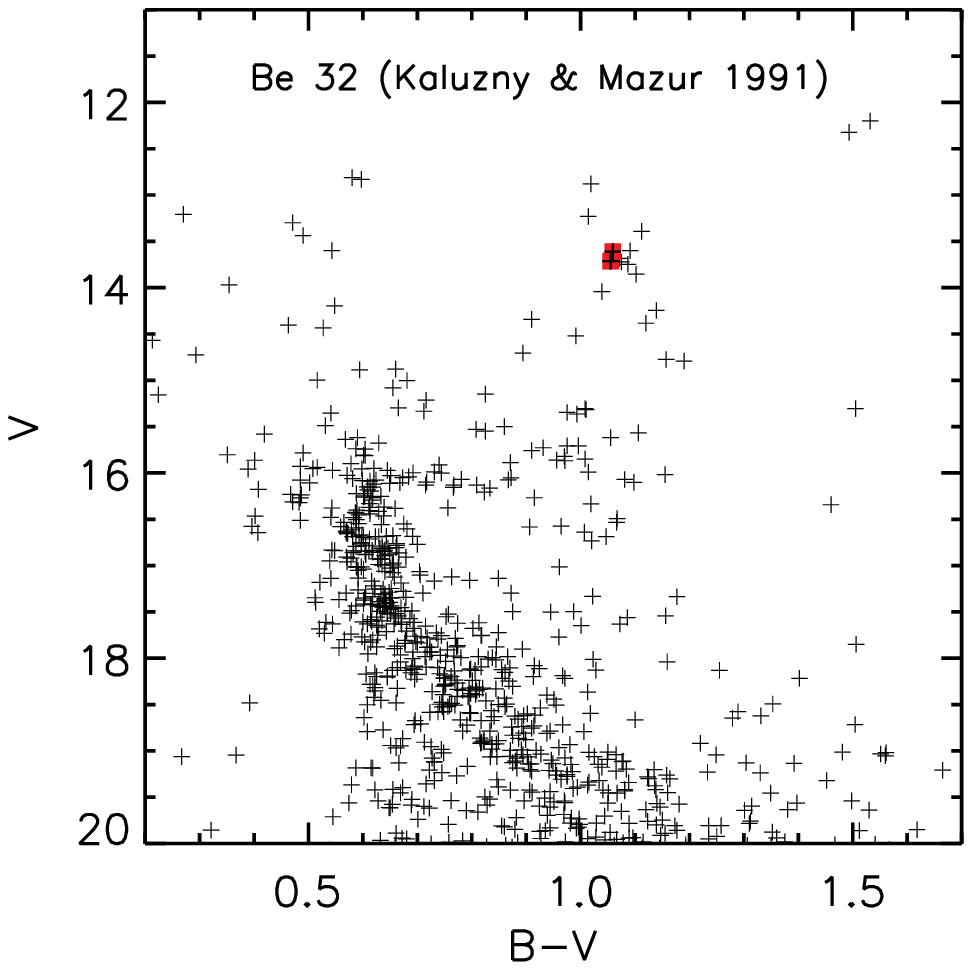}
\caption{Be 32 color-magnitude diagram using the data from 
\citet{kaluzny91}. We distinguish our program 
stars by red squares. 
\label{fig:cmdbe32bv}}
\end{figure}

\begin{figure}[t!]
\epsscale{1.2}
\plotone{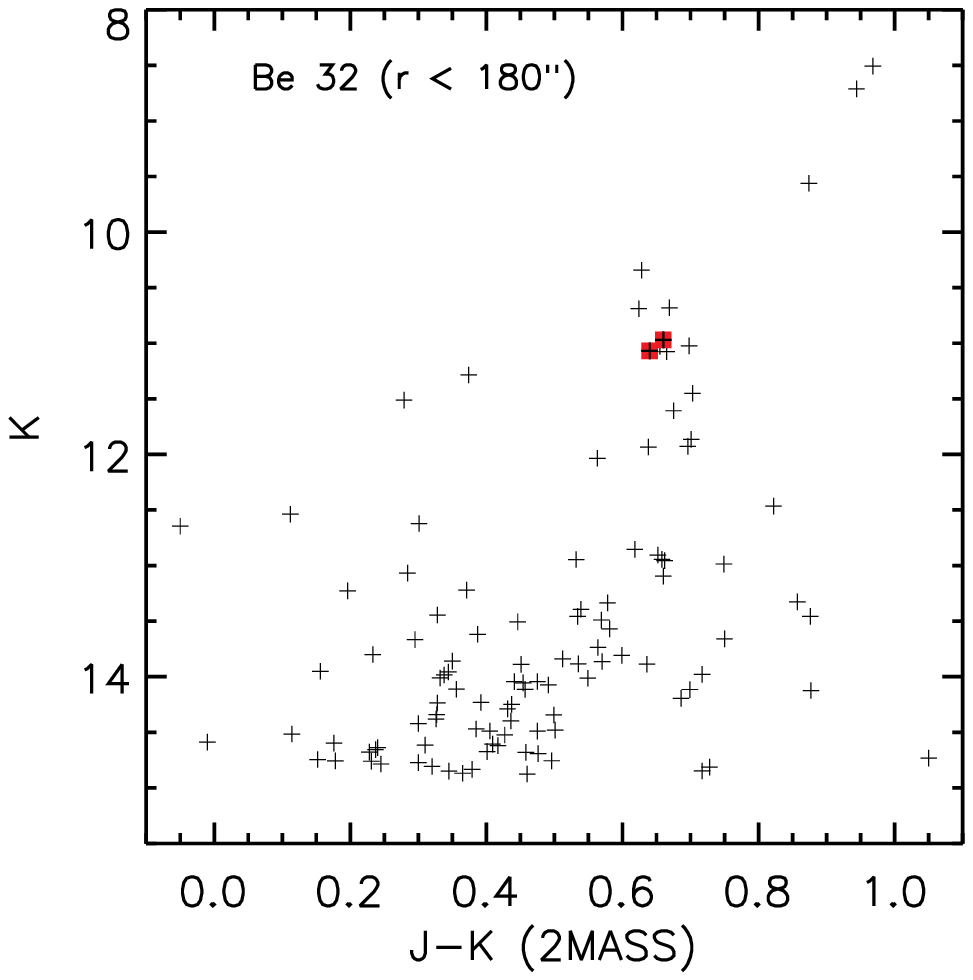}
\caption{Be 32 color-magnitude diagram using the data from 
2MASS \citet{2mass}. We distinguish our program 
stars by red squares. 
\label{fig:cmdbe32jk}}
\end{figure}

\begin{figure}[t!]
\epsscale{1.2}
\plotone{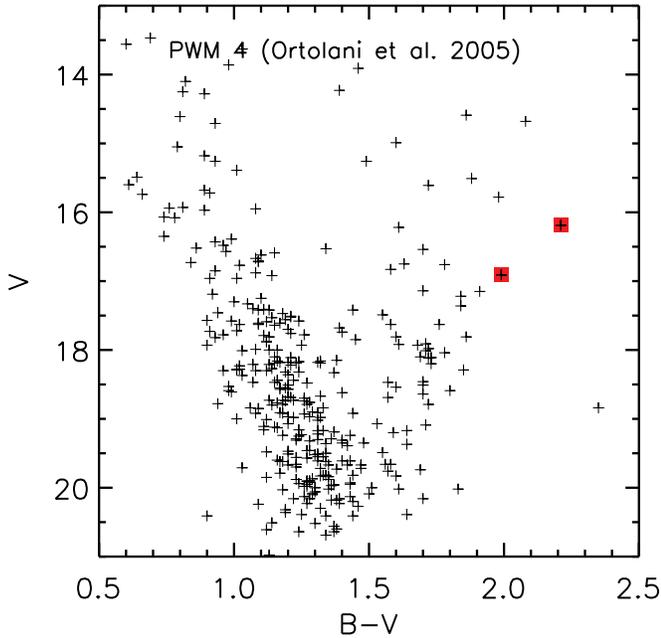}
\caption{PWM 4 color-magnitude diagram using the data from 
\citet{ortolani05}. We distinguish our program 
stars by red squares (the two red clump stars are not 
included in their photometry). 
\label{fig:cmdpwm4vi}}
\end{figure}

\begin{figure}[t!]
\epsscale{1.2}
\plotone{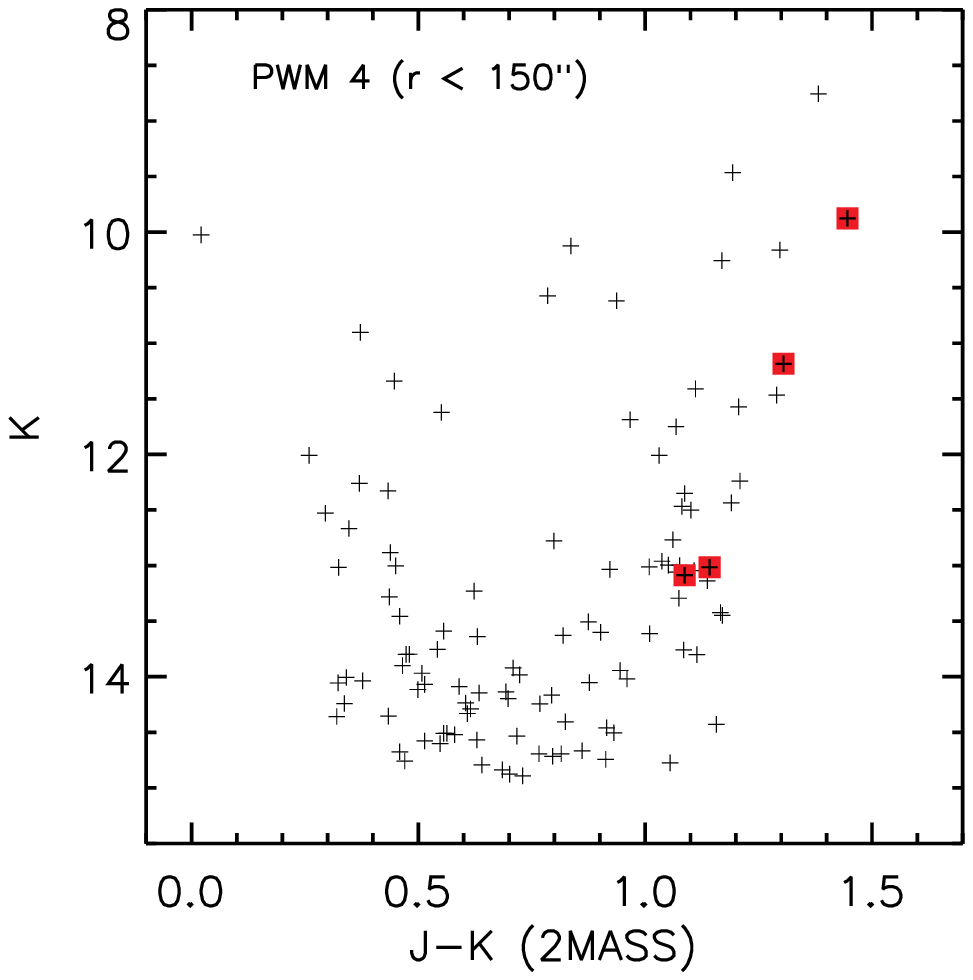}
\caption{PWM 4 color-magnitude diagram using the data from 
2MASS \citet{2mass}. We distinguish our program 
stars by red squares. 
\label{fig:cmdpwm4jk}}
\end{figure}

We searched the WEBDA database\footnote{http://www.univie.ac.at/webda/} 
and the recent literature to 
identify additional old ($>$ 2 Gyr), distant (\rgc\ $>$ 10 kpc) open 
clusters. The following open clusters were identified as suitable 
candidates: Be 18, Be 21, Be 22, Be 32, and PWM 4. 
Following our approach in Paper I, individual stars were 
selected from optical and infra-red color-magnitude diagrams 
(see Figures \ref{fig:cmdbe18bv} to \ref{fig:cmdpwm4jk} 
and Tables \ref{tab:clusters} and \ref{tab:phot}). 
For Be 32, \citet{dorazi06} measured radial velocities for a large 
sample of stars and thus we were able to select likely 
members for this cluster.  

\begin{deluxetable*}{cccccccccc}
\tabletypesize{\scriptsize}
\tablecolumns{10} 
\tablewidth{0pc} 
\tablecaption{Observed Clusters \label{tab:clusters}}
\tablehead{
\colhead{Cluster}  & 
\colhead{R.A.\tablenotemark{a} (J2000)} &
\colhead{Decl.\tablenotemark{a} (J2000)} & 
\colhead{$\ell$\tablenotemark{a}}    & 
\colhead{$b$\tablenotemark{a}} &
\colhead{[Fe/H]\tablenotemark{a}}  &  
\colhead{$\delta V$\tablenotemark{b}} & 
\colhead{MAI\tablenotemark{c}} &
\colhead{d\tablenotemark{d}} & 
\colhead{d\tablenotemark{e}} 
}
\startdata
Berkeley 18 & 05:22:12 & +45:24:00 & 163.63 &   +5.02 &   +0.02 & 2.3 & 5.69 & 5.8 & 5.4 \\
Berkeley 21 & 05:51:42 & +21:47:00 & 186.84 & $-$2.51 & $-$0.83 & 1.6 & 2.18 & 5.0 & 6.2 \\
Berkeley 22 & 05:58:24 & +07:50:00 & 199.88 & $-$8.08 & $-$0.30 & 2.1 & 4.26 & 7.7 & 6.2 \\
Berkeley 32 & 06:58:06 & +06:26:00 & 207.95 &   +4.40 & $-$0.58 & 2.4 & 5.91 & 3.1 & 3.2 \\
PWM 4       & 23:50:55 & +62:19:15 & 115.96 &   +0.27 &  \ldots & 3.0 & 7.00\tablenotemark{f} & 7.9\tablenotemark{f} & 7.2 
\enddata

\tablenotetext{a}{Taken from WEBDA.}
\tablenotetext{b}{Taken from \citet{salaris04}.}
\tablenotetext{c}{Taken from \citet{salaris04}, in Gyrs.}
\tablenotetext{d}{Distance in kpc from \citet{salaris04}.}
\tablenotetext{e}{Distance estimate obtained using red clump stars (this paper).}
\tablenotetext{f}{Taken from \citet{ortolani05}}

\end{deluxetable*}

\begin{deluxetable*}{lccccccrc}
\tabletypesize{\footnotesize}
\tablecolumns{9} 
\tablewidth{0pc} 
\tablecaption{Photometric Data \label{tab:phot}}
\tablehead{ 
\colhead{Star} & 
\colhead{R.A.\tablenotemark{a} (J2000)} &
\colhead{Decl.\tablenotemark{a} (J2000)} & 
\colhead{$V$} &
\colhead{$B-V$} &
\colhead{$V-I_C$} &
\colhead{$K$\tablenotemark{a}} &
\colhead{$J-K$\tablenotemark{a}} &
\colhead{Reference} 
}
\startdata
Be18 532  & 05 22 13.7 & +45 25 46.8 &  15.99 &   1.48 &   1.68 & 12.226 & 0.877 & 1 \\
Be18 1006 & 05 22 16.2 & +45 27 28.5 &  16.03 &   1.47 &   1.70 & 12.224 & 0.885 & 1 \\
Be18 1163 & 05 22 13.8 & +45 27 58.9 &  15.79 &   1.63 &   1.86 & 11.561 & 1.010 & 1 \\
Be18 1383 & 05 22 10.7 & +45 28 49.4 &  15.07 &   1.75 &   1.94 & 10.631 & 1.045 & 1 \\
\\
Be21 T50  & 05 51 42.3 & +21 48 45.4  & 15.87 &   1.71 &   2.07 & 11.142 & 1.062 & 2 \\
Be21 T51  & 05 51 42.0 & +21 48 02.8 &  15.69 &   1.70 &   2.06 & 10.963 & 1.095 & 2 \\
\\
Be22 414  & 05 58 25.9 & +07 46 11.4 &  15.76 &   1.87 &   2.02 & 11.020 & 1.157 & 3 \\
Be22 643  & 05 58 26.9 & +07 45 26.1 &  14.31 &   2.29 &   2.65 &  8.369 & 1.405 & 3 \\
\\
Be32 16   & 06 58 06.9 & +06 25 56.5 &  13.61 &   1.06 & \ldots & 10.970 & 0.660 & 4 \\
Be32 18   & 06 58 13.8 & +06 27 54.9 &  13.71 &   1.06 & \ldots & 11.069 & 0.640 & 4 \\
\\
PWM4 RGB1 & 23 50 57.4 & +62 20 03.2 &  16.19 &   2.21 & \ldots &  9.876 & 1.446 & 5 \\
PWM4 RGB2 & 23 50 58.3 & +62 19 18.0 &  16.91 &   1.99 & \ldots & 11.185 & 1.305 & 5 \\
PWM4 RC1  & 23 50 53.9 & +62 19 07.7 & \ldots & \ldots & \ldots & 13.015 & 1.142 & \ldots \\
PWM4 RC2  & 23 50 53.2 & +62 19 33.8 & \ldots & \ldots & \ldots & 13.087 & 1.087 & \ldots 
\enddata

\tablenotetext{a}{2MASS \citep{2mass} coordinates and photometry.}

\tablerefs{
1 = \citet{kaluzny97}; 
2 = \citet{tosi98}; 
3 = \citet{kaluzny94}; 
4 = \citet{kaluzny91}; 
5 = \citet{ortolani05} 
}

\end{deluxetable*}

\subsection{Observations and Data Reduction}

We used the HIRES \citep{vogt94} spectrograph on the Keck I Telescope 
on 12 November 2006 and 27 December 2006. 
On both occasions we used the red cross-disperser 
and the B5 decker which has a slit width of 0\farcs86 and a slit 
length of 3\farcs5. On the November run, we binned the CCD 
3 $\times$ 1 (spatial $\times$ spectral) and on the December 
run we binned the CCD 
2 $\times$ 1. 
The observing routine included 20 quartz lamp 
exposures for flat fielding as well as 20 zero second exposures for 
``bias'' frames. ThAr frames were taken during each night 
(three on 12 November and eight on 27 December) 
and two radial velocity standards were observed 
on each night. 
The spectroscopic data were reduced using the 
IRAF\footnote{IRAF (Image Reduction and Analysis
Facility) is distributed by the National Optical Astronomy
Observatory, which is operated by the Association of Universities
for Research in Astronomy, Inc., under cooperative agreement with the National
Science Foundation.} packages as described in Paper I. 
From the reduced ThAr frames, we 
measured a spectral resolution of $R$ = 47,000 ($\sigma$ = 3000) 
on each of the 
three detectors, and this value is close to 
the expected value of $R$ = 48,000. Since we did not bin the CCD 
in the spectral dimension, the data are oversampled with approximately 
4.8 pixels per resolution element. The wavelength coverage was 
from 4000\AA\ to 8500\AA. There are gaps in the spectra between the 
blue and green CCDs ($\sim$ 5400\AA\ to 5440\AA) and between the 
green and red CCDs ($\sim$ 7000\AA\ to 7075\AA). Furthermore, the 
free spectral range exceeds the CCD coverage beyond $\sim$ 6400\AA. 

\begin{figure}[t!] 
\epsscale{1.2}
\plotone{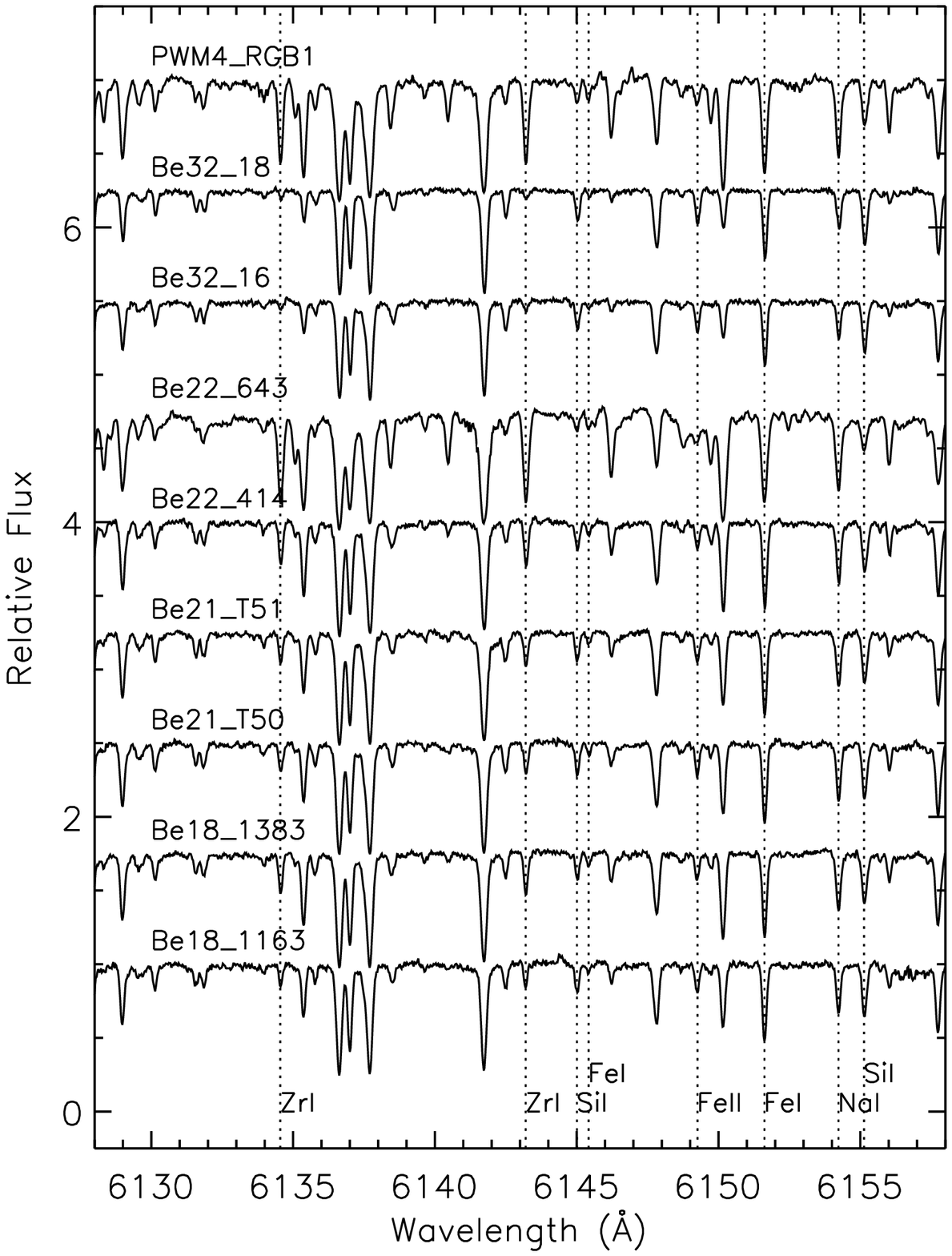}
\caption{Sample region of spectra of the program stars. Lines used in the analysis 
are marked. 
\label{fig:redspectra}}
\end{figure}

Exposure times and resulting signal-to-noise ratios (S/N) 
for the program stars are given in Table \ref{tab:specobs}. 
(Note that the S/N is given per pixel at 6050\AA\ and that we have 4.8 pixels 
per resolution element. To obtain S/N ratios per resolution element, 
multiply our values by 2.19.) 
Our observing program consisted of relatively short exposures of 
some stars in order to measure their radial velocities as 
well as (multiple) longer exposures of other stars to obtain high S/N 
ratios for detailed chemical abundance analysis. 
For the longer exposures, preliminary reduction was performed in real-time 
in order to estimate the radial velocities. 
Shorter exposures were taken of four candidate red clump giants 
(two stars in each of Be 18 and PWM4) and one red giant 
in PWM4, and the radial velocities were 
measured during the analysis. 
Examples of reduced spectra are shown in Figure \ref{fig:redspectra}. 

\begin{deluxetable}{lcccc}
\tablecolumns{9} 
\tablewidth{0pc} 
\tablecaption{Spectroscopic Observations\label{tab:specobs}}
\tablehead{ 
\colhead{Star} & 
\colhead{Exposure Time (s)} & 
\colhead{HJD $-$ 2,450,000} & 
\colhead{S/N\tablenotemark{a}} & 
\colhead{$V_{\rm rad}$} } 
\startdata
Be18-532 & 600 & 4052.9209 & 22 & $-$3.6 \\
Be18-1006 & 500 & 4052.9325 & 20 & $-$8.3 \\
Be18-1163 & 500 & 4052.9414 & 24 & $-$3.1 \\
Be18-1163 & 2400 & 4097.7714 & 48 & $-$4.1 \\
Be18-1163 & 2401 & 4097.7998 & 51 & $-$4.4 \\
Be18-1163 & 2400 & 4097.8282 & 50 & $-$4.4 \\
Be18-1383 & 1200 & 4052.9535 & 48 & $-$6.5 \\
Be18-1383 & 1800 & 4052.9717 & 58 & $-$6.1 \\
Be18-1383 & 1500 & 4052.9916 & 54 & $-$6.1 \\
Be21-T50 & 2400 & 4097.9890 & 53 & $-$1.3 \\
Be21-T50 & 2401 & 4098.0174 & 54 & $-$1.6 \\
Be21-T50 & 2401 & 4098.0458 & 51 & $-$1.2 \\
Be21-T51 & 2064 & 4053.0212 & 53 & +0.9 \\
Be21-T51 & 2400 & 4053.0535 & 57 & $-$1.1 \\
Be21-T51 & 2400 & 4053.0818 & 58 & $-$0.9 \\
Be21-T51 & 600 & 4053.0998 & 28 & $-$0.9 \\
Be22-414 & 2400 & 4097.8690 & 51 & +93.1 \\
Be22-414 & 983 & 4097.8893 & 31 & +93.0 \\
Be22-414 & 2400 & 4097.9184 & 51 & +93.4 \\
Be22-414 & 2533 & 4097.9487 & 52 & +93.7 \\
Be22-643 & 1200 & 4053.1184 & 72 & +90.4 \\
Be22-643 & 1200 & 4053.1328 & 69 & +90.4 \\
Be32-16 & 600 & 4098.0674 & 61 & +105.9 \\
Be32-16 & 600 & 4098.0751 & 61 & +105.9 \\
Be32-16 & 600 & 4098.0827 & 62 & +105.8 \\
Be32-18 & 600 & 4098.0910 & 58 & +105.1 \\
Be32-18 & 600 & 4098.0985 & 54 & +105.1 \\
Be32-18 & 901 & 4098.1080 & 64 & +105.1 \\
PWM4-RGB1 & 2002 & 4052.7283 & 42 & $-$125.4 \\
PWM4-RGB1 & 2700 & 4052.7631 & 48 & $-$124.8 \\
PWM4-RGB1 & 2700 & 4052.7953 & 51 & $-$124.8 \\
PWM4-RGB1 & 2700 & 4052.8276 & 51 & $-$124.6 \\
PWM4-RGB2 & 600 & 4052.8527 & 17 & $-$128.8 \\
PWM4-RC1 & 1800 & 4052.8709 & 17 & $-$125.4 \\
PWM4-RC2 & 1800 & 4052.8963 & 16 & $-$124.3 
\enddata

\tablenotetext{a}{Signal-to-noise ratio per pixel near 6050\AA.}

\end{deluxetable}

\subsection{Radial Velocities} 

Radial velocities were measured by cross-correlating the program 
stars with radial velocity standards. For 12 November 2006, 
our radial velocity standard was HD 90861 with a 
heliocentric radial velocity of 36.3 \kms\ 
(this value was used in Paper I and was taken from the  
Astronomical Almanac 1999). 
For 27 December 2006, our radial velocity standard was HD 110885 
with a heliocentric radial velocity of $-$47.89 \kms\ 
\citep{carney03}. 

The cross-correlation typically involved some 30 orders per star 
and the dispersion about the mean velocity 
ranged from 0.1 to 0.7 \kms. However, we found that on 12 November 2006, 
the velocity "jumps" by $\sim$ 2 \kms\ within the 
consecutive series of spectra 
taken for Be21-T51. On closer inspection of 
the data, we found that one of the four spectra obtained has an anomalous 
radial velocity. The wavelength solution for this spectrum comes from a 
different ThAr frame than for the other three spectra. On this observing run, 
we only took three ThAr frames, one each at the start and end of the night 
and only one in the middle of the night. At face value, it would appear that 
there has been a 
shift during the course of the night and 
that the radial velocities on this night probably have a systematic 
uncertainty of $\sim$ 2 \kms. This issue was noted shortly after the 
November run and more frequent ThAr frames, eight in total, were taken on the 
December run. 
For Be21-T51, measurement 
of the telluric absorption line wavelengths suggests that 
the three spectra with consistent radial velocities give the correct value. 
Another point worth noting is that 
Be18-1163 was observed on both runs. For this star, we find that the 
radial velocity from 12 November 2006 is $\sim$1 \kms\ higher than 
the values obtained from 27 December 2006. This could be explained 
by either radial velocity variation (and/or jitter -- e.g., see \citealt{carney08} 
and references therein) or 
systematic errors in the radial velocity standards. 
The bottom line is that our 
radial velocities may be uncertain by $\sim$1 \kms\ between the 
two observing runs. 

For Be18 and PWM4, there are no previous radial velocity studies 
to our knowledge. As noted, Be18-1163 was observed on both 
runs and there was a 1.0 \kms\ radial velocity offset between the 
two runs. Without knowing which of the two values may be spurious, 
we include all values for all stars when computing the average cluster 
radial velocity for Be18 of $-$5.5 $\pm$ 1.1 \kms\ 
($\sigma$ = 2.2 \kms). For PWM4, we find an average cluster 
radial velocity of $-$125.9 $\pm$ 1.0 \kms\ 
($\sigma$ = 2.0 \kms). For both clusters we studied four stars. 

For Be21 T50, our average radial velocity is $-$1.4 \kms\ 
($\sigma_{\rm internal}$ = 0.2 \kms)  and this value agrees well 
with the average cluster value of $-$0.6 \kms\ ($\sigma$ = 2.9 \kms) 
from Paper I. For Be21 T51, our average radial velocity from all four 
spectra is $-$0.5 \kms\ ($\sigma_{\rm internal}$ = 0.9 \kms). Had 
we excluded the apparently anomalous +0.9 \kms\ value, we would have 
obtained an average value of $-$1.0 \kms\ 
($\sigma_{\rm internal}$ = 0.1 \kms). Either value is in good agreement 
with the value measured for this star in Paper I, $-$1.0 $\pm$ 1.0 \kms. 
However, as noted in Paper I, our radial velocities for this cluster 
differ from those of \citet{hill99} who obtained +12.35 $\pm$ 0.6 \kms\ 
based on four stars. 

For Be22, we find an average cluster radial velocity of 
+91.9 $\pm$ 1.5 \kms\ ($\sigma_{\rm internal}$ = 2.1 \kms) based 
on two stars. 
This value is in fair agreement with the value from 
\citet{villanova05} of +95.3 $\pm$ 2.0 \kms\ ($\sigma$ = 2.8 \kms). 

For Be32 18, our average radial velocity of +105.1 \kms\ is in 
good agreement with the values from \citet{dorazi06} and \citet{sestito06} 
of +108.07 \kms\ ($\sigma$ = 4.12 \kms) and +105.5 \kms, respectively. 
We find an average cluster radial velocity of 
+105.5 $\pm$ 0.4 \kms\ ($\sigma_{\rm internal}$ = 0.5 \kms) based 
on two stars. Our value is in good agreement with 
previous studies of this cluster: 
+101 $\pm$ 3 \kms\ ($\sigma$ = 10 \kms) \citep{scott95}; 
+106.7 ($\sigma$ = 8.5 \kms) \citep{dorazi06}; 
+106.0 $\pm$ 1.4 \kms\ \citep{sestito06}; 
+105.4 $\pm$ 0.4 \kms\ \citep{friel10}. 

\subsection{Distance Estimations}

We determined distances and interstellar reddening for the clusters 
following the prescription outlined in Paper I. From the 2MASS 
color-magnitude diagrams, we estimated the location of the red 
clump. \citet{carney05b} derived a relation between the 
mean $J-K$ color of the red clump and metallicity. 
Using our spectroscopic measurement for metallicity, this 
relation is used to derive the reddening $E(J-K)$ [= 0.52$E(B-V)$ 
\citep{rieke85}]. 
The luminosity of the red clump is $M_K$ $\simeq$ $-$1.61 mag 
and is insensitive to metallicity in the regime [Fe/H] $>$ $-$1.0 
\citep{alves00}. 
We present our distances and reddening in Table \ref{tab:rc}. 
Our distance estimates are, in general, in good agreement with the 
WEBDA literature values. Such agreement is encouraging given that 
in some cases, identifying the clump from 2MASS CMDs can 
be difficult.  

\begin{deluxetable*}{lccccccrcccc}
\tabletypesize{\footnotesize}
\tablecolumns{9} 
\tablewidth{0pc} 
\tablecaption{Red Clump Data \label{tab:rc}}
\tablehead{ 
\colhead{Cluster} & 
\colhead{$<K>$} &
\colhead{$<J-K>$} &
\colhead{[Fe/H]} &
\colhead{$(J-K)_0$} &
\colhead{$E(J-K)$} &
\colhead{$E(B-V)$} &
\colhead{$(m-M)_0$} &
\colhead{$d$ (kpc)} &
\colhead{\rgc} &
\colhead{$E(B-V)$\tablenotemark{a}} &
\colhead{$d$\tablenotemark{a} (kpc)} 
}
\startdata
Berkeley 18 & 12.21 & 0.87 & $-$0.4 & 0.53 & 0.34 & 0.66 & 13.60 & 5.2 & 13.10 & 0.46 & 5.8 \\
Berkeley 21 & 12.62 & 0.93 & $-$0.3 & 0.55 & 0.39 & 0.74 & 13.98 & 6.2 & 14.22 & 0.76 & 5.0 \\
Berkeley 22 & 12.57 & 0.88 & $-$0.4 & 0.53 & 0.35 & 0.68 & 13.95 & 6.2 & 13.92 & 0.70 & 7.7 \\
Berkeley 32 & 11.02 & 0.65 & $-$0.3 & 0.55 & 0.11 & 0.20 & 12.56 & 3.3 & 10.97 & 0.16 & 3.1 \\
PWM 4       & 13.02 & 1.06 & $-$0.3 & 0.55 & 0.52 & 0.99 & 14.29 & 7.2 & 12.91 & 0.62 & 7.9 
\enddata

\tablenotetext{a}{Taken from WEBDA.} 

\end{deluxetable*}

\section{ELEMENTAL ABUNDANCES}

\subsection{Stellar Parameters}

Stellar parameters are required to conduct a chemical abundance analysis 
and we obtained the values in the following manner. Following Paper I, 
initial estimates of \teff\ come from the infra-red flux method 
metallicity-dependent color-temperature relations for giant stars 
by \citet{alonso99} and \citet{ramirez05}. 
With distances, temperatures, and bolometric 
corrections, we can calculate $\log g$ values if we assume a 
mass for the red giants. As in Paper I, we adopt a mass of $1 M_\odot$ 
and note that the surface gravity is not especially sensitive to 
the assumed value. 

These initial estimates for \teff\ and $\log g$ were then used 
as starting values for the spectroscopic stellar parameters. 
Equivalent widths (EW) were measured for a selection of \fei\ and \feii\ 
lines using routines in IRAF. We used the same line list as 
in Paper I. Model atmospheres were computed using the ATLAS9 
program \citep{kurucz93}. Using the 
LTE stellar line analysis program {\sc Moog} \citep{moog}, 
we computed abundances for each line based on the measured EW. 
(As in Papers I, II, and III, we used the 2002 version of {\sc Moog}.) 
We set the effective temperature, \teff, 
by requiring that the abundances from \fei\ lines 
showed no trend with lower excitation potential, i.e., excitation 
equilibrium. The microturbulent velocity, $\xi_t$, was determined 
by ensuring that the abundance from \fei\ lines showed no trend 
with reduced equivalent width, $\log (W_\lambda/\lambda)$. 
We adjusted the surface gravity, $\log g$, until the 
abundance from \fei\ lines agreed with the abundance from \feii\ 
lines, i.e., ionization equilibrium. 
We also required that the derived metallicity, [Fe/H], was within 0.1 
dex of the metallicity of the model, [m/H], otherwise 
the model atmosphere was re-computed with the appropriate metallicity. 
The initial photometric stellar parameters and the final spectroscopic 
stellar parameters are given in Table \ref{tab:param}. 

\begin{deluxetable*}{lccccccccc}
\tabletypesize{\footnotesize}
\tablecolumns{9} 
\tablewidth{0pc} 
\tablecaption{Atmospheric Parameters\label{tab:param}}
\tablehead{ 
\colhead{Star} & 
\colhead{\teff} & 
\colhead{$\log g$} &
\colhead{\teff} & 
\colhead{$\log g$} &
\colhead{\teff} & 
\colhead{$\log g$} &
\colhead{[m/H]} & 
\colhead{$\xi_t$} & 
\colhead{[Fe/H]} 
\\
\colhead{} &
\multicolumn{2}{c}{Photometric\tablenotemark{a}} &
\multicolumn{2}{c}{Photometric\tablenotemark{b}} &
\multicolumn{5}{c}{Spectroscopic\tablenotemark{c}}
} 
\startdata
Be18-1163 & 4666 & 2.1 & 4655 & 2.1 & 4500 & 2.2 & $-$0.5 & 1.20 & $-$0.46 \\
Be18 1383 & 4489 & 1.7 & 4495 & 1.7 & 4400 & 1.9 & $-$0.5 & 1.27 & $-$0.41 \\
\\ 
Be21 T50 & 4464 & 1.7 & 4468 & 1.7 & 4625 & 1.9 & $-$0.3 & 1.32 & $-$0.26 \\
Be21 T51 & 4481 & 1.7 & 4452 & 1.6 & 4500 & 1.7 & $-$0.3 & 1.29 & $-$0.35 \\
\\ 
Be22 414 & 4340 & 1.7 & 4297 & 1.7 & 4350 & 1.7 & $-$0.5 & 1.18 & $-$0.41 \\
Be22 643 & 3791 & 0.6 & 3752 & 0.5 & 3850 & 0.2 & $-$0.5 & 1.55 & $-$0.49 \\
\\ 
Be32 16 & 4977 & 2.4 & 5000 & 2.4 & 4875 & 2.4 & $-$0.3 & 1.07 & $-$0.39 \\
Be32 18 & 4962 & 2.4 & 5007 & 2.4 & 4950 & 2.7 & $-$0.3 & 1.46 & $-$0.37 \\
\\
PWM4 RGB1 & 4368 & 1.4 & 3942 & 1.0 & 3950 & 0.5 & $-$0.3 & 1.28 & $-$0.33 
\enddata

\tablenotetext{a}{Estimates obtained using (i) the photometric data in 
Table \ref{tab:phot}, (ii) the distance, reddening, and metallicity 
from Table \ref{tab:rc}, and (iii) the \citet{ramirez05} IRFM calibration.}

\tablenotetext{b}{Estimates obtained using (i) the photometric data in 
Table \ref{tab:phot}, (ii) the distance, reddening, and metallicity 
from Table \ref{tab:rc}, and (iii) the \citet{alonso99} IRFM calibration.}

\tablenotetext{c}{Quantities are derived using the spectroscopic 
methods described in the text.}

\end{deluxetable*}

We note that there is good agreement between the photometric and 
spectroscopic values. 
Given the large reddening for the program clusters, the infra-red 
flux method temperatures may be regarded as being less reliable 
than spectroscopic values. Furthermore, while uncertainties in the distances 
estimates affect the photometric temperatures, the spectroscopic 
values are not affected. For PWM4-RGB1, the photometric parameters, 
using the \citet{ramirez05} calibration, differ from our spectroscopic 
values. Such a discrepancy may be attributed to the uncertainty in the 
reddening (we obtain $E(B-V)$ = 0.99 compared to the WEBDA value of 0.62). 
With the exception of PWM4, 
that the two sets of temperatures are in good 
agreement, in general, probably suggests that the reddening 
and distance estimates are reasonable.  

In practice, we explored discrete values for \teff\ 
(every 25 K, i.e., 4525K, 4550K, etc) 
and $\log g$ (every 0.05 dex, i.e., 1.05 dex, 1.10 dex, etc). We assumed that 
excitation equilibrium was satisfied when the slope between 
$\log\epsilon$(\fei) and lower excitation potential was 
$<$ 0.004 and that ionization equilibrium 
was satisfied when 
$| \log\epsilon$(\fei) $-$ 
$\log\epsilon$(\feii) $| < 0.05$ dex. 
The microturbulent velocity was considered satisfactory when the slope 
between $\log\epsilon$(\fei) and reduced equivalent width was $<$ 0.004. 
We iterated until the criteria were simultaneously satisfied. 
During this process, we estimate that the internal uncertainties are 
\teff\ $\pm$ 100K, $\log g$ $\pm$ 0.3 dex, and $\xi_t$ $\pm$ 0.2 \kms. 

\subsection{Elemental Abundance Analysis}

In Paper I, we relied upon spectrum synthesis to measure chemical abundances 
for individual elements. In that paper, the spectral resolution was 
$R$ = 28,000 and we were concerned that blends might affect the 
abundances derived from an equivalent width analysis. 
In this paper, the spectral resolution $R$ = 47,000 was sufficiently 
high that abundance measurements 
based 
on EW measurements from our high S/N spectra were regarded 
to be reliable. 
Therefore, we measured chemical abundances 
from spectrum synthesis as well as from equivalent width analysis, when 
possible, using 
{\sc Moog} in both cases. 
For both approaches, we took into account the effects of hyperfine 
splitting and/or isotopic splitting when necessary 
(e.g., Mn, Co, Rb, Ba, La, and Eu) using the 
same approach as in Paper I. 
The line list and solar abundances 
were the same as in Paper I. 
(See Table \ref{tab:ews} for the atomic data and EW measurements.) 
In Tables \ref{tab:abund.be18} to \ref{tab:abund.pwm4} 
we present the chemical abundances, [X/Fe], for the 
program stars where these values represent the average abundances 
from EW analysis and from spectrum synthesis.  
We note that the two approaches gave quite similar results with an 
average difference in abundance ratio $\Delta$[X/Fe] (EW $-$ Synth) 
of 0.06 $\pm$ 0.01 dex ($\sigma$ = 0.14). 
We did not find any significant global differences for a given line 
or a given species between the two approaches. 
(The 
[X/Fe] ratios are based on the individual stellar [Fe/H] rather than 
the cluster mean [Fe/H].)
In Table \ref{tab:parvar} we show the abundance dependences upon 
the model parameters for a representative star, Be21 T51, assuming 
that the errors are symmetric for positive and negative changes. 

\begin{deluxetable*}{lllrrrrrrrr}
\tabletypesize{\tiny}
\tablecolumns{9} 
\tablewidth{0pc} 
\tablecaption{Equivalent Widths for Program Stars \label{tab:ews}}
\tablehead{ 
\colhead{Wavelength (\AA)} &
\colhead{Species} & 
\colhead{LEP (eV)} &
\colhead{$\log gf$} &
\colhead{Be18 1163} &
\colhead{Be18 1383} &
\colhead{Be21 T50} &
\colhead{Be21 T51} &
\colhead{Be22 414} &
\colhead{Be22 643} &
\colhead{Be32 16} 
}
\startdata
 6300.30 &  8.0 &     0.00 &  $-$9.72 &  \ldots &    38.7 &  \ldots &    38.9 &    48.4 &   114.0 &    24.9  \\
 6363.78 &  8.0 &     0.02 & $-$10.19 &  \ldots &    20.8 &  \ldots &    20.3 &    27.3 &    83.2 &    16.3  \\
 5688.19 & 11.0 &     2.11 &  $-$0.42 &  \ldots &  \ldots &  \ldots &  \ldots &  \ldots &  \ldots &   120.7  \\
 6154.23 & 11.0 &     2.10 &  $-$1.53 &    60.8 &    75.0 &    74.6 &    73.2 &    77.8 &  \ldots &    44.3  \\
 6160.75 & 11.0 &     2.10 &  $-$1.23 &    83.8 &    97.6 &    97.9 &    96.0 &   103.6 &  \ldots &    68.1  \\
 5711.09 & 12.0 &     4.35 &  $-$1.83 &   122.5 &  \ldots &  \ldots &  \ldots &  \ldots &  \ldots &   105.5  
\enddata

\tablerefs{
Note. Table 6 is published in its entirety in the electronic edition of the Astronomical Journal. A portion is shown here for guidance regarding its form and content.
}

\end{deluxetable*}

\begin{deluxetable}{lrcccrcc}
\tablecolumns{9} 
\tablewidth{0pc} 
\tablecaption{Chemical Abundances for Berkeley 18 
\label{tab:abund.be18}}
\tablehead{ 
\colhead{Species} & 
\colhead{Abundance} & 
\colhead{$\sigma$} &
\colhead{$N$} & 
\colhead{} & 
\colhead{Abundance} & 
\colhead{$\sigma$} &
\colhead{$N$} 
}
\startdata
 & \multicolumn{3}{c}{Be18-1163} & & \multicolumn{3}{c}{Be18-1383} \\
 \noalign{\vskip  .8ex} \hline
 \noalign{\vskip -2ex}\\
{\rm [O/Fe]}  & \ldots & \ldots & \ldots &  & 0.11 & 0.07 & 2 \\
{\rm [Na/Fe]}  & 0.10 & 0.05 & 2 &  & 0.32 & 0.10 & 3 \\
{\rm [Mg/Fe]}  & 0.25 & 0.10 & 4 &  & 0.10 & 0.14 & 4 \\
{\rm [Al/Fe]}  & 0.15 & 0.04 & 2 &  & 0.22 & 0.00 & 2 \\
{\rm [Si/Fe]}  & 0.15 & 0.18 & 7 &  & 0.19 & 0.04 & 10 \\
{\rm [Ca/Fe]}  & $-$0.01 & 0.18 & 11 &  & 0.25 & 0.09 & 6 \\
{\rm [Ti/Fe]}  & 0.10 & 0.17 & 41 &  & 0.27 & 0.09 & 30 \\
{\rm [Mn/Fe]}  & $-$0.27 & 0.12 & 3 &  & 0.08 & 0.08 & 3 \\
{\rm [\fei/H]}  & $-$0.47 & 0.10 & 46 &  & $-$0.41 & 0.12 & 46 \\
{\rm [\feii/H]}  & $-$0.44 & 0.17 & 11 &  & $-$0.41 & 0.18 & 8 \\
{\rm [Co/Fe]}  & 0.12 & 0.06 & 3 &  & 0.25 & 0.08 & 3 \\
{\rm [Ni/Fe]}  & $-$0.07 & 0.15 & 6 &  & 0.04 & 0.12 & 6 \\
{\rm [Rb/Fe]}  & $-$0.31 & \ldots & 1 &  & $-$0.15 & \ldots & 1 \\
{\rm [Zr/Fe]}  & $-$0.30 & 0.09 & 3 &  & $-$0.11 & 0.03 & 3 \\
{\rm [Ba/Fe]}  & 0.25 & \ldots & 1 &  & 0.34 & \ldots & 1 \\
{\rm [La/Fe]}  & 0.28 & 0.07 & 2 &  & 0.39 & 0.00 & 2 \\
{\rm [Eu/Fe]}  & 0.25 & \ldots & 1 &  & 0.34 & \ldots & 1 
\enddata
\end{deluxetable} 

\begin{deluxetable}{lrcccrcc}
\tablecolumns{9} 
\tablewidth{0pc} 
\tablecaption{Chemical Abundances for Berkeley 21 
\label{tab:abund.be21}}
\tablehead{ 
\colhead{Species} & 
\colhead{Abundance} & 
\colhead{$\sigma$} &
\colhead{$N$} & 
\colhead{} & 
\colhead{Abundance} & 
\colhead{$\sigma$} &
\colhead{$N$} 
}
\startdata
 & \multicolumn{3}{c}{Be21-T50} & & \multicolumn{3}{c}{Be21-T51} \\
 \noalign{\vskip  .8ex} \hline
 \noalign{\vskip -2ex}\\
{\rm [O/Fe]}  &  \ldots & \ldots & \ldots &  & 0.18 & 0.04 & 2 \\
{\rm [Na/Fe]}  & 0.39 & 0.30 & 3 &  & 0.32 & 0.18 & 3 \\
{\rm [Mg/Fe]}  & 0.17 & 0.13 & 4 &  & 0.20 & 0.15 & 4 \\
{\rm [Al/Fe]}  & 0.15 & 0.00 & 2 &  & 0.19 & 0.07 & 2 \\
{\rm [Si/Fe]}  & 0.18 & 0.08 & 10 &  & 0.24 & 0.08 & 12 \\
{\rm [Ca/Fe]}  & 0.21 & 0.12 & 4 &  & 0.19 & 0.11 & 6 \\
{\rm [Ti/Fe]}  & 0.22 & 0.05 & 27 &  & 0.17 & 0.04 & 30 \\
{\rm [Mn/Fe]}  & 0.18 & 0.08 & 3 &  & 0.03 & 0.13 & 3 \\
{\rm [\fei/H]}  & $-$0.26 & 0.13 & 44 &  & $-$0.35 & 0.14 & 31 \\
{\rm [\feii/H]}  & $-$0.26 & 0.16 & 12 &  & $-$0.33 & 0.18 & 6 \\
{\rm [Co/Fe]}  & 0.14 & 0.08 & 3 &  & 0.15 & 0.10 & 3 \\
{\rm [Ni/Fe]}  & 0.07 & 0.15 & 7 &  & 0.06 & 0.15 & 6 \\
{\rm [Rb/Fe]}  & 0.05 & \ldots & 1 &  & $-$0.20 & \ldots & 1 \\
{\rm [Zr/Fe]}  & $-$0.05 & 0.05 & 3 &  & $-$0.16 & 0.06 & 3 \\
{\rm [Ba/Fe]}  & 0.59 & \ldots & 1 &  & 0.58 & \ldots & 1 \\
{\rm [La/Fe]}  & 0.55 & 0.07 & 2 &  & 0.57 & 0.07 & 2 \\
{\rm [Eu/Fe]}  & 0.25 & \ldots & 1 &  & 0.36 & \ldots & 1 
\enddata
\end{deluxetable} 

\begin{deluxetable}{lrcccrcc}
\tablecolumns{9} 
\tablewidth{0pc} 
\tablecaption{Chemical Abundances for Berkeley 22 
\label{tab:abund.be22}}
\tablehead{ 
\colhead{Species} & 
\colhead{Abundance} & 
\colhead{$\sigma$} &
\colhead{$N$} & 
\colhead{} & 
\colhead{Abundance} & 
\colhead{$\sigma$} &
\colhead{$N$} 
}
\startdata
 & \multicolumn{3}{c}{Be22-414} & & \multicolumn{3}{c}{Be22-643} \\
 \noalign{\vskip  .8ex} \hline
 \noalign{\vskip -2ex}\\
{\rm [O/Fe]}  & 0.12 & 0.04 & 2 &  & 0.21 & 0.04 & 2 \\
{\rm [Na/Fe]}  & 0.42 & 0.23 & 3 &  & \ldots & \ldots & \ldots \\
{\rm [Mg/Fe]}  & 0.16 & 0.12 & 3 &  & 0.19 & 0.16 & 4 \\
{\rm [Al/Fe]}  & 0.28 & 0.00 & 2 &  & 0.49 & 0.00 & 2 \\
{\rm [Si/Fe]}  & 0.19 & 0.07 & 11 &  & 0.19 & 0.17 & 5 \\
{\rm [Ca/Fe]}  & 0.23 & 0.12 & 6 &  & 0.27 & 0.27 & 4 \\
{\rm [Ti/Fe]}  & 0.31 & 0.11 & 17 &  & 0.38 & 0.19 & 6 \\
{\rm [Mn/Fe]}  & 0.00 & 0.10 & 3 &  & $-$0.45 & 0.09 & 3 \\
{\rm [\fei/H]}  & $-$0.40 & 0.11 & 39 &  & $-$0.49 & 0.22 & 23 \\
{\rm [\feii/H]}  & $-$0.44 & 0.09 & 10 &  & $-$0.50 & 0.33 & 5 \\
{\rm [Co/Fe]}  & 0.25 & 0.10 & 3 &  & 0.12 & 0.03 & 3 \\
{\rm [Ni/Fe]}  & 0.06 & 0.16 & 6 &  & 0.14 & 0.26 & 6 \\
{\rm [Rb/Fe]}  & $-$0.20 & \ldots & 1 &  & $-$0.35 & \ldots & 1 \\
{\rm [Zr/Fe]}  & $-$0.14 & 0.08 & 3 &  & \ldots & \ldots & \ldots \\
{\rm [Ba/Fe]}  & 0.64 & \ldots & 1 &  & 0.57 & \ldots & 1 \\
{\rm [La/Fe]}  & 0.39 & 0.07 & 2 &  & 0.36 & 0.00 & 2 \\
{\rm [Eu/Fe]}  & 0.33 & \ldots & 1 &  & 0.19 & \ldots & 1 
\enddata
\end{deluxetable} 

\begin{deluxetable}{lrcccrcc}
\tablecolumns{9} 
\tablewidth{0pc} 
\tablecaption{Chemical Abundances for Berkeley 32 
\label{tab:abund.be32}}
\tablehead{ 
\colhead{Species} & 
\colhead{Abundance} & 
\colhead{$\sigma$} &
\colhead{$N$} & 
\colhead{} & 
\colhead{Abundance} & 
\colhead{$\sigma$} &
\colhead{$N$} 
}
\startdata
 & \multicolumn{3}{c}{Be32-16} & & \multicolumn{3}{c}{Be32-18} \\
 \noalign{\vskip  .8ex} \hline
 \noalign{\vskip -2ex}\\
{\rm [O/Fe]}  & 0.15 & 0.04 & 2 &  & 0.30 & 0.07 & 2 \\
{\rm [Na/Fe]}  & 0.22 & 0.21 & 3 &  & 0.25 & 0.28 & 3 \\
{\rm [Mg/Fe]}  & 0.22 & 0.08 & 4 &  & 0.20 & 0.08 & 3 \\
{\rm [Al/Fe]}  & 0.16 & 0.00 & 2 &  & 0.19 & 0.04 & 2 \\
{\rm [Si/Fe]}  & 0.18 & 0.10 & 13 &  & 0.22 & 0.05 & 13 \\
{\rm [Ca/Fe]}  & 0.16 & 0.20 & 11 &  & 0.14 & 0.12 & 12 \\
{\rm [Ti/Fe]}  & 0.23 & 0.07 & 26 &  & 0.29 & 0.08 & 30 \\
{\rm [Mn/Fe]}  & $-$0.13 & 0.15 & 3 &  & 0.03 & 0.12 & 3 \\
{\rm [\fei/H]}  & $-$0.38 & 0.11 & 59 &  & $-$0.37 & 0.10 & 52 \\
{\rm [\feii/H]}  & $-$0.42 & 0.17 & 14 &  & $-$0.36 & 0.13 & 16 \\
{\rm [Co/Fe]}  & 0.12 & 0.10 & 3 &  & 0.18 & 0.15 & 3 \\
{\rm [Ni/Fe]}  & 0.04 & 0.12 & 7 &  & 0.03 & 0.12 & 7 \\
{\rm [Rb/Fe]}  & $-$0.25 & \ldots & 1 &  & $-$0.10 & \ldots & 1 \\
{\rm [Zr/Fe]}  & $-$0.11 & 0.03 & 3 &  & $-$0.01 & 0.03 & 3 \\
{\rm [Ba/Fe]}  & 0.37 & \ldots & 1 &  & 0.20 & \ldots & 1 \\
{\rm [La/Fe]}  & 0.42 & 0.07 & 2 &  & 0.45 & 0.07 & 2 \\
{\rm [Eu/Fe]}  & 0.23 & \ldots & 1 &  & 0.38 & \ldots & 1 
\enddata
\end{deluxetable} 

\begin{deluxetable}{lrcc}
\tablecolumns{9} 
\tablewidth{0pc} 
\tablecaption{Chemical Abundances for PWM4 
\label{tab:abund.pwm4}}
\tablehead{ 
\colhead{Species} & 
\colhead{Abundance} & 
\colhead{$\sigma$} &
\colhead{$N$} 
}
\startdata
 & \multicolumn{3}{c}{PWM4-RGB1} \\
 \noalign{\vskip  .8ex} \hline
 \noalign{\vskip -2ex}\\
{\rm [O/Fe]}  & 0.18 & \ldots & 2 \\
{\rm [Na/Fe]}  & 0.42 & 0.12 & 3 \\
{\rm [Mg/Fe]}  & 0.23 & 0.11 & 4 \\
{\rm [Al/Fe]}  & 0.26 & 0.04 & 2 \\
{\rm [Si/Fe]}  & 0.10 & 0.05 & 7 \\
{\rm [Ca/Fe]}  & 0.19 & 0.09 & 4 \\
{\rm [Ti/Fe]}  & 0.14 & 0.05 & 6 \\
{\rm [Mn/Fe]}  & $-$0.28 & 0.13 & 3 \\
{\rm [\fei/H]}  & $-$0.34 & 0.15 & 33 \\
{\rm [\feii/H]}  & $-$0.29 & 0.12 & 8 \\
{\rm [Co/Fe]}  & 0.13 & 0.06 & 3 \\
{\rm [Ni/Fe]}  & $-$0.03 & 0.11 & 6 \\
{\rm [Rb/Fe]}  & $-$0.30 & \ldots & 1 \\
{\rm [Zr/Fe]}  & 0.11 & 0.06 & 3 \\
{\rm [Ba/Fe]}  & 0.46 & \ldots & 1 \\
{\rm [La/Fe]}  & 0.22 & 0.07 & 2 \\
{\rm [Eu/Fe]}  & 0.12 & \ldots & 1 
\enddata
\end{deluxetable} 

\begin{deluxetable}{lrrrr}
\tabletypesize{\scriptsize}
\tablecolumns{9} 
\tablewidth{0pc} 
\tablecaption{Abundance Dependences on Model Parameters for Be 21 T51 \label{tab:parvar}}
\tablehead{ 
\colhead{Species} & 
\colhead{\teff\ + 100K} & 
\colhead{$\log g$ + 0.3 dex} &
\colhead{$\xi_t$ + 0.2 \kms} & 
\colhead{Total} 
}
\startdata
{\rm [O/H]}  & $-$0.01 &  0.11 & $-$0.01 &  0.11  \\
{\rm [Na/H]}  &    0.09 &  0.00 & $-$0.05 &  0.10  \\
{\rm [Mg/H]}  &    0.03 &  0.00 & $-$0.02 &  0.04  \\
{\rm [Al/H]}  &    0.08 & -0.01 & $-$0.04 &  0.09  \\
{\rm [Si/H]}  & $-$0.04 &  0.06 & $-$0.04 &  0.08  \\
{\rm [Ca/H]}  &    0.10 & -0.02 & $-$0.10 &  0.14  \\
{\rm [Ti/H]}  &    0.17 &  0.01 & $-$0.10 &  0.20  \\
{\rm [Mn/H]}  &    0.10 &  0.02 & $-$0.17 &  0.20  \\
{\rm [\fei/H]}  &    0.06 &  0.03 & $-$0.09 &  0.11  \\
{\rm [\feii/H]}  & $-$0.09 &  0.15 & $-$0.07 &  0.19  \\
{\rm [Co/H]}  &    0.05 &  0.06 & $-$0.06 &  0.10  \\
{\rm [Ni/H]}  &    0.03 &  0.08 & $-$0.07 &  0.11  \\
{\rm [Rb/H]}  &    0.12 &  0.00 & $-$0.02 &  0.12  \\
{\rm [Zr/H]}  &    0.21 &  0.01 & $-$0.03 &  0.21  \\
{\rm [Ba/H]}  &    0.03 &  0.07 & $-$0.17 &  0.19  \\
{\rm [La/H]}  &    0.02 &  0.12 & $-$0.06 &  0.14  \\
{\rm [Eu/H]}  & $-$0.01 &  0.12 & $-$0.04 &  0.13  \\
\\
\hline
\\
{\rm [O/Fe]} & $-$0.07 & 0.08 & 0.08 & 0.13 \\
{\rm [Na/Fe]} & 0.03 & $-$0.03 & 0.04 & 0.06 \\
{\rm [Mg/Fe]} & $-$0.03 & $-$0.03 & 0.07 & 0.08 \\
{\rm [Al/Fe]} & 0.02 & $-$0.04 & 0.05 & 0.07 \\
{\rm [Si/Fe]} & $-$0.10 & 0.03 & 0.05 & 0.12 \\
{\rm [Ca/Fe]} & 0.04 & $-$0.04 & $-$0.01 & 0.06 \\
{\rm [Ti/Fe]} & 0.11 & $-$0.01 & $-$0.01 & 0.11 \\
{\rm [Mn/Fe]} & 0.04 & $-$0.01 & $-$0.08 & 0.09 \\
{\rm [Co/Fe]} & $-$0.01 & 0.03 & 0.03 & 0.04 \\
{\rm [Ni/Fe]} & $-$0.03 & 0.05 & 0.02 & 0.06 \\
{\rm [Rb/Fe]} & 0.06 & $-$0.03 & 0.07 & 0.10 \\
{\rm [Zr/Fe]} & 0.15 & $-$0.01 & 0.06 & 0.16 \\
{\rm [Ba/Fe]} & $-$0.03 & 0.05 & $-$0.08 & 0.10 \\
{\rm [La/Fe]} & $-$0.04 & 0.09 & 0.03 & 0.10 \\
{\rm [Eu/Fe]} & $-$0.07 & 0.09 & 0.05 & 0.12 
\enddata
\end{deluxetable}

\subsection{Comparison With Literature}
\label{sec:luckcomp}

\citet{friel10} conducted an examination of the 
various systematic differences between the work by 
(a) \citet{sestito08} and \citet{bragaglia08} (SB using their notation), 
(b) preliminary results from This Study (CY using their 
notation\footnote{The results presented here are very similar to those 
used in \citet{friel10}. The average difference in [X/Fe] is only 
0.03 dex $\pm$ 0.01 dex ($\sigma$ = 0.06 dex). The largest 
differences were for Na (our [Na/Fe] values are lower by 0.14 dex) 
and Ca (our [Ca/Fe] values are lower by 0.18 dex). All other 
elements are within 0.1 dex.}), and  
(c) re-analysis by \citet{friel10} using the spectra employed by SB and CY. 
We refer the reader to their careful analysis that takes into account 
solar abundances, EW measurements, $\log gf$ values and 
atmospheric parameters. \citet{friel10} identify systematic 
abundance differences between the various studies, but note that 
``it is difficult 
to identify the component sources of systematic differences''. 

\begin{figure}[t!]
\epsscale{1.2}
\plotone{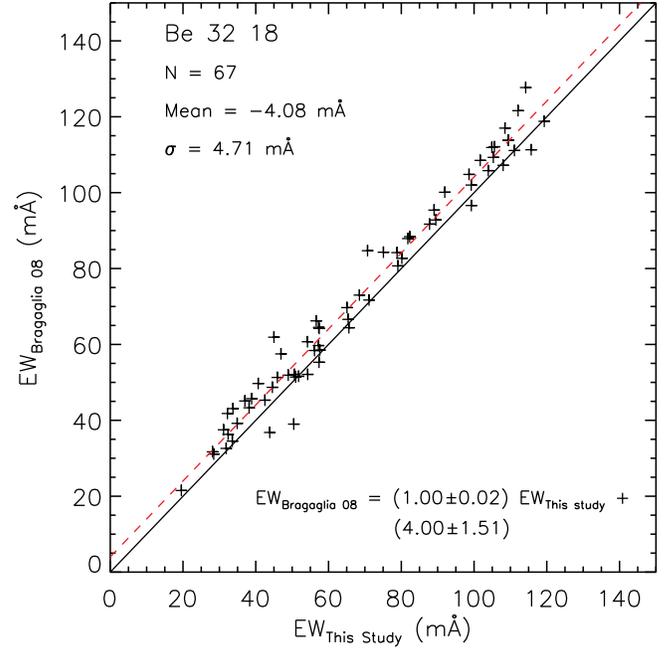}
\caption{Comparison of equivalent width measurements for Be 32 18 
between this study and \citet{bragaglia08}. The one-to-one relation 
is shown (solid line) along with the linear fit to the data (dashed 
line).  
\label{fig:spectra}}
\end{figure}

\begin{figure}[t!]
\epsscale{1.2}
\plotone{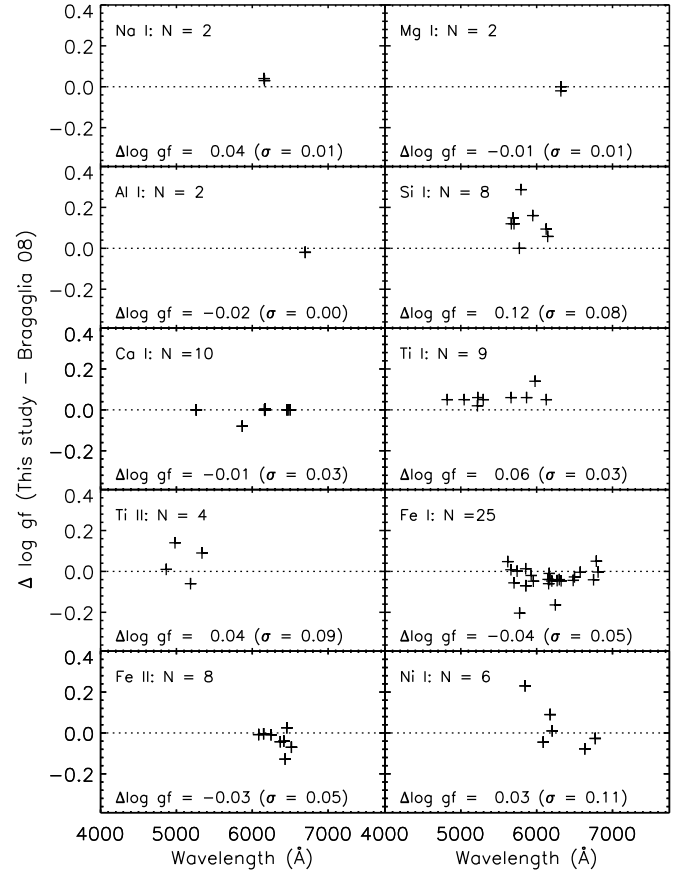}
\caption{Differences in $\log gf$ values for various elements  
between this study and 
\citet{bragaglia08} versus wavelength. 
\label{fig:spectra2}}
\end{figure}

For completeness, we also include a comparison of measured equivalent  
widths for Be32-18. For 67 lines in common with \citet{bragaglia08}, 
we find a mean difference (This Study $-$ Literature) of $-$4.08 m\AA\ 
($\sigma$ = 4.71 m\AA) (see Figure \ref{fig:spectra}). 
While the agreement is reasonable, there is a small systematic difference. 
(We note that the \citet{bragaglia08} spectra are of comparable 
spectral resolution, R = 45,000, and similar S/N.) 
For Be32-18, uniformly increasing the EWs by 4.0 m\AA, while keeping all 
stellar parameters unchanged, would increase [Fe/H] by 0.09 dex. 
We also compare our $\log gf$ values for individual elements 
with \citet{bragaglia08} (see Figure \ref{fig:spectra2}). 
The largest difference is for Si, $\Delta \log gf$ = 0.12 dex. For 
all other elements, the average differences in $\log gf$ values 
for a given element are 0.06 dex or smaller. 

\begin{figure}[t!]
\epsscale{1.2}
\plotone{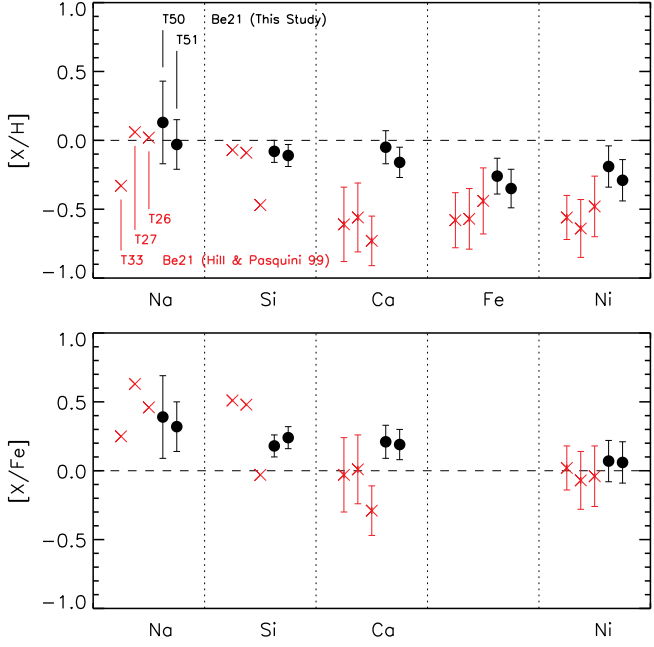}
\caption{Comparison of the abundance ratios 
[X/H] ($upper$) and [X/Fe] ($lower$) in Be 21 between this 
study (filled circles) and \citet{hill99} (crosses). 
The individual stars are marked and their relative positions are the same 
in each panel. 
\label{fig:compbe21}}
\end{figure}

Next, we compare our abundances [X/H] and [X/Fe] 
for a given cluster with literature 
values. 
In Figure \ref{fig:compbe21} we compare our abundances for 
stars Be 21 T50 and Be 21 T51 with the three Be 21 
stars analyzed by \citet{hill99}. 
Our mean metallicity is [Fe/H] = $-$0.31 
while their value is [Fe/H] = $-$0.54. 
For the elements in common, our [X/Fe] ratios are on average higher 
by 0.05 dex ($\sigma$ = 0.20 dex). 
For Ca, there is a large difference between the two studies, 
$\Delta$[Ca/Fe] = 0.30 dex. For all other elements, the maximum 
difference in [X/Fe] is 0.11 dex. 

\begin{figure}[t!]
\epsscale{1.2}
\plotone{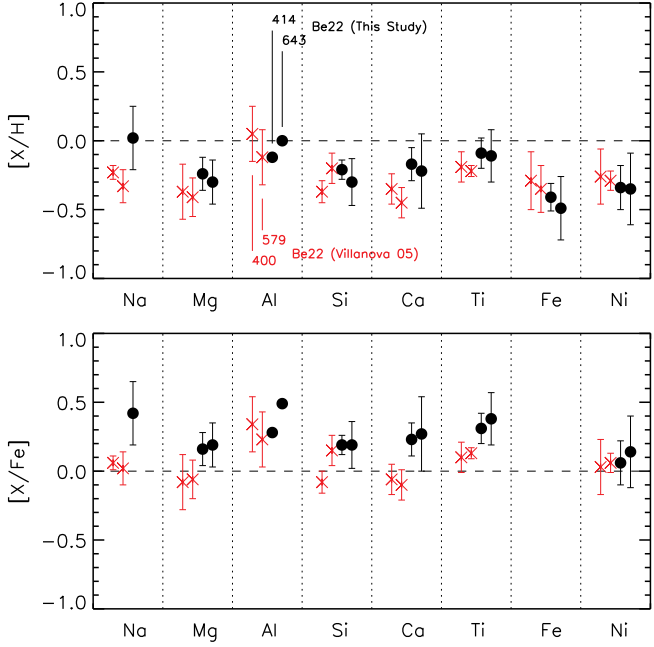}
\caption{Comparison of the abundance ratios 
[X/H] ($upper$) and [X/Fe] ($lower$) in Be 22 between this 
study (filled circles) and \citet{villanova05} (crosses). 
The individual stars are marked and their relative positions are the same 
in each panel. 
\label{fig:compbe22}}
\end{figure}

In Figure \ref{fig:compbe22} we compare our Be 22 abundances for two stars 
with the two different stars analyzed by \citet{villanova05}. 
Our mean metallicity is [Fe/H] = $-$0.45 
and their value is [Fe/H] = $-$0.32. 
For the elements in common, our [X/Fe] ratios are on average higher 
by 0.21 dex ($\sigma$ = 0.12 dex).  

\begin{figure}[t!]
\epsscale{1.2}
\plotone{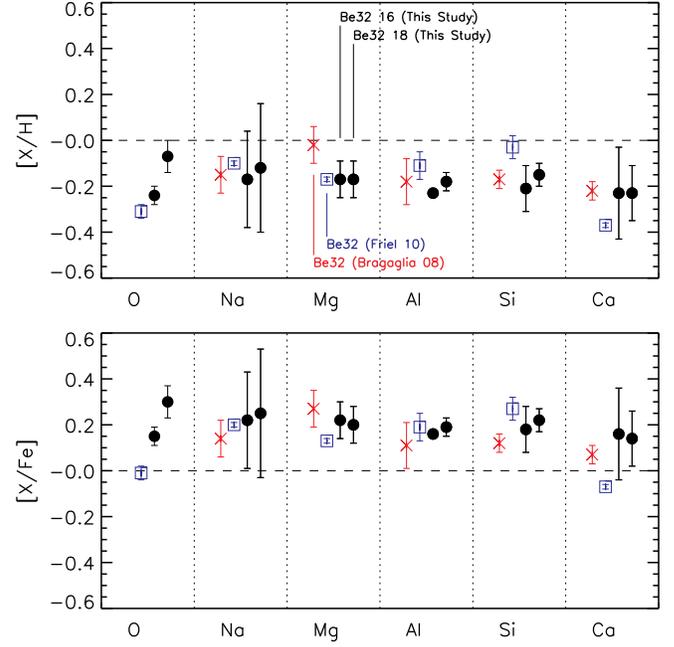}
\caption{Comparison of the abundance ratios 
[X/H] ($upper$) and [X/Fe] ($lower$) 
for O, Na, Mg, Al, Si, and Ca 
in Be 32 between the 
individual stars from study (filled circles) and 
the mean cluster values from \citet{bragaglia08} (crosses) and 
\citet{friel10} (squares). 
\label{fig:compbe32m1}}
\end{figure}

\begin{figure}[t!]
\epsscale{1.2}
\plotone{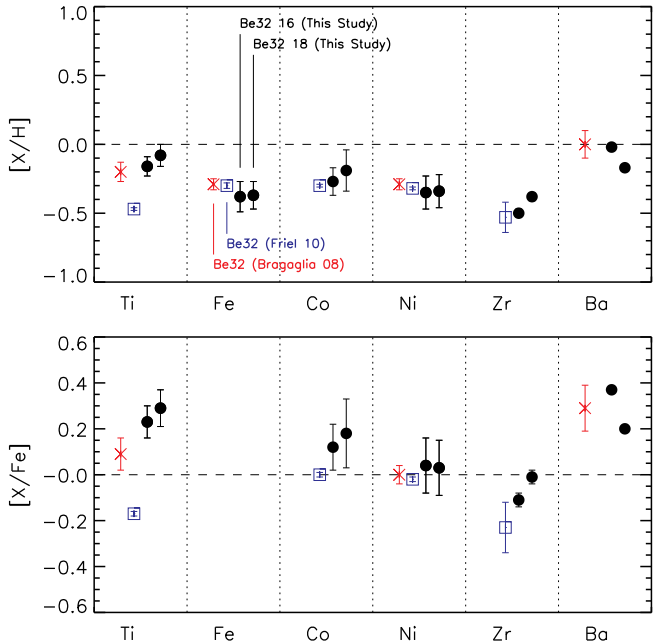}
\caption{Same as Figure \ref{fig:compbe32m1} but for 
Ti, Fe, Co, Ni, Zr, and Ba.
\label{fig:compbe32m2}}
\end{figure}

In Figures \ref{fig:compbe32m1} and \ref{fig:compbe32m2} 
we compare our abundances for our two stars in Be32 with the mean 
cluster values from the analyses of \citet{bragaglia08} and \citet{friel10}. 
Our mean metallicity, [Fe/H] = $-$0.38, is 0.09 dex lower than \citet{bragaglia08}, 
[Fe/H] = $-$0.29. 
For elements in common, our [X/Fe] ratios are on average higher than 
those of \citet{bragaglia08} by 0.06 dex ($\sigma$ = 0.07 dex). 
Our mean metallicity, [Fe/H] = $-$0.38, is 0.08 dex lower than \citet{friel10}, 
[Fe/H] = $-$0.30. 
For elements in common, our [X/Fe] ratios are on average higher than 
those of \citet{friel10} by 0.13 dex ($\sigma$ = 0.15 dex). 
This difference is driven primarily by Ti where $\Delta$[Ti/Fe] = 0.43 dex. 
Such a large abundance difference is of concern, but as noted in 
\citet{friel10} and apparent later in Figure \ref{fig:alpha1}, [Ti/Fe] 
abundances appear to be especially vulnerable to at times large systematic 
differences between studies. For example, as we shall note in Section 4.4.1, 
M67 is in common to \citet{friel10} and Paper I, and while the two studies 
obtain very similar metallicities, $\Delta$[Fe/H] = $-$0.01 dex, 
a large abundance difference exists for Ti, $\Delta$[Ti/Fe] = 0.26 dex. 
The origin of these differences is not altogether clear, and does not appear 
to be due to simple effects of a single source, but is likely driven by a 
complex combination of factors, such as EWs, stellar parameters, $\log gf$ values, 
specific lines utilized, reference abundances, analysis techniques, etc. 

\begin{figure}[t!]
\epsscale{1.2}
\plotone{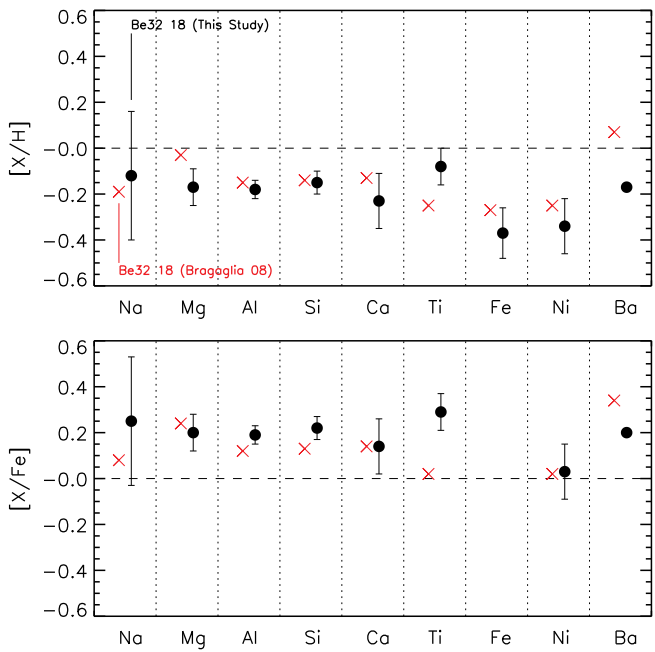}
\caption{Comparison of the abundance ratios 
[X/H] ($upper$) and [X/Fe] ($lower$) in Be 32 18 between this 
study (filled circles) and \citet{bragaglia08} (crosses). 
\label{fig:compbe32}}
\end{figure}

In Figure \ref{fig:compbe32}, we compare our abundances for 
the star Be32 18 with the analysis of \citet{bragaglia08}. 
We derive a metallicity of [Fe/H] = $-$0.37 
compared to their value of [Fe/H] = $-$0.27. 
The elements in common are 
Na, Mg, Al, Si, Ca, Ti, Ni, Ba, and Fe. For [X/H] we find a mean 
difference (This Study $-$ Literature) of 
$-$0.05 dex $\pm$ 0.04 dex ($\sigma$ = 0.12 dex). 
For [X/Fe] we find a mean difference (This Study $-$ Literature) 
of +0.05 dex $\pm$ 0.04 dex ($\sigma$ = 0.13 dex). 
We note that, as with the comparison to \citet{friel10}, the agreement 
is especially poor for Ti, with $\Delta$[Ti/Fe] = 0.27.

From these comparisons, it is clear that there are systematic differences 
at the $\sim$0.1 dex level between various studies. For some 
elements, the abundance differences can be as large as 0.4 
dex. Such differences should be 
borne in mind in the 
comparisons that follow. 
Of particular interest is that our metallicities are, with the 
exception of Be 21, lower than the literature values 
by $\sim$0.1 dex which may 
potentially lead to larger [X/Fe] ratios than compared 
to the literature values. In the subsequent discussion, we will (at times) 
re-evaluate our conclusions by arbitrarily shifting our 
metallicities, [Fe/H], by +0.1 dex and decreasing 
our abundance ratios, [X/Fe], by 0.1 dex. 

\section{DISCUSSION} 

\subsection{Trends With Galactocentric Distance}

\begin{figure}[t!] 
\epsscale{1.2}
\plotone{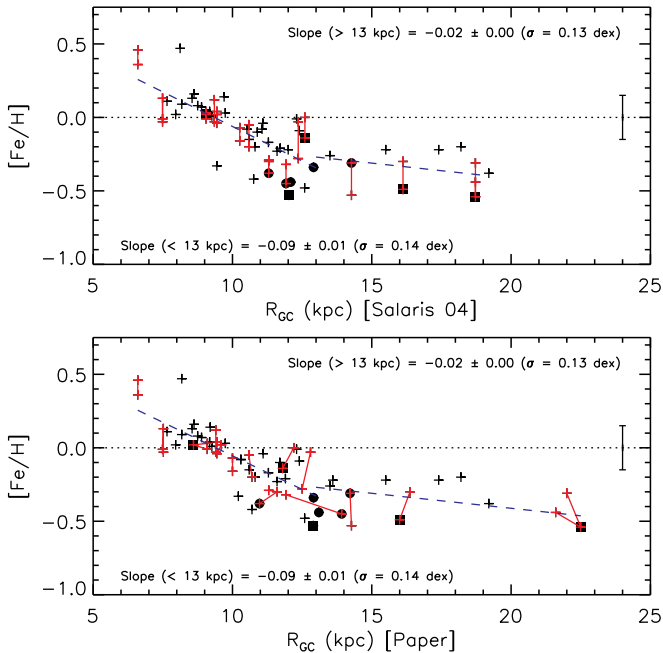}
\caption{[Fe/H] vs.\ Galactocentric distance. In the upper panel, 
the distances are taken from \citet{salaris04} while in the lower panel, 
the distances are from the literature papers. The data from this study 
are filled circles, the Paper I data are filled squares, and the 
literature data are plus signs. The connected red symbols represent 
multiple measurements of the same cluster. In both panels, we 
show the linear fit to the data with \rgc\ $<$ 13 kpc and \rgc\ $>$ 13 kpc. 
A representative error bar is shown. 
\label{fig:fehab}}
\end{figure}

In Figure \ref{fig:fehab} we plot the metallicity, [Fe/H], 
versus Galactocentric distance. This figure includes measurements 
of open clusters 
from this paper (filled circles), Paper I (filled squares), and 
literature sources (plus signs). 
(Although membership of Be 31 remains unclear \citep{friel10}, 
we retain this key cluster, \rgc\ = 12 kpc, in subsequent figures 
and discussion.) 
The compilation of literature data 
(LTE abundances) and 
sources used to generate this figure can be found in Table \ref{tab:compile}. 
The literature search was not exhaustive.  
We have not made a critical assessment of the literature sample 
and note that some of the abundance measurements were based on data that we 
would regard to have unacceptably low S/N. (In the Appendix A.1, we 
explore how selections based on S/N may affect the results.) 
This combined sample consists of some 68 chemical abundance measurements 
in 49 unique clusters. 
(We note that \citet{pancino10} assembled a compilation of 
some 57 open clusters. They have more clusters in the 
solar vicinity but fewer in 
the outer disk, which is the focus of this series of papers.) 
One issue is that different authors 
may adopt independent distances. Therefore, this figure 
shows the distances from \citet{salaris04} (upper panel) as well as the 
distances from the individual literature sources 
(lower panel)\footnote{When distances 
from \citet{salaris04} are not available, the 
values from the individual literature sources are used in the 
upper panel. Similarly, when distances are not given in the literature 
sources, the values from \citet{salaris04} are employed. 
The \citet{salaris04} distances are based on main sequence fitting and 
are therefore sensitive to extinction and metallicity. 
The literature distances come from a variety of methods. As noted in Section
2.4, we use the red clump luminosity. Distances based on this approach are
insensitive to metallicity, but identifying 
the red clump location and extinction affect our results.}. 
In both panels, we join individual clusters with multiple measurements 
using red lines. That is, in the event that a given cluster has been 
studied by two or more authors, we treat all studies as 
independent measurements regardless of the sample size, 
spectral resolution, and S/N of each study. 
Following \citet{pancino10}, rather than averaging multiple 
measurements into a single value for a given cluster, we show all 
measurements to provide a ``realistic idea of the uncertainties 
involved in the compilation.'' 
We also measure the linear\footnote{While we determine 
linear fits to the data throughout the paper, 
we are not suggesting that this is the correct function to use. 
Rather, we consider this a first pass to begin to understand 
the behavior of abundance ratios with distance, age, and/or 
metallicity, [Fe/H].} 
fit to the 
local (\rgc\ $<$ 13 kpc) and distant (\rgc\ $>$ 13 kpc) samples. 
The 13 kpc boundary dividing the local and distant clusters was 
arbitrarily chosen (see Appendix A.2 for a more thorough examination 
of the division between the inner and outer disk and we also 
refer the reader to \citealt{jacobsonphd} and 
\citealt{jacobson11b} for further discussion on this issue). 
As noted previously, the metallicity gradients 
in the two regions are considerably different. The inner region shows a 
steeper gradient than in the outer disk. Interestingly, the formal slopes, 
uncertainties, and dispersion about the linear fit are virtually 
identical in both panels, that is, these values do not depend upon 
the distance estimate, \citet{salaris04} vs.\ literature. 
Our conclusions would not change had we selected a boundary value 
of 12 kpc or 14 kpc rather than 13 kpc. 
For completeness, we note that the linear fit includes multiple 
measurements of a given cluster such that those clusters in effect 
are given more weight than a cluster with a single measurement. 
In subsequent figures, we use the literature distance estimates. 

\begin{deluxetable*}{llrrrrrrrrrr}
\tablecolumns{12} 
\tablewidth{0pc} 
\tablecaption{Compilation of (LTE) Abundance Ratios for Open Clusters \label{tab:compile}}
\tablehead{ 
\colhead{} & 
\colhead{} &
\colhead{Age\tablenotemark{a}} &
\colhead{Distance\tablenotemark{a}} &
\colhead{Age\tablenotemark{b}} &
\colhead{Distance\tablenotemark{b}} &
\colhead{} &
\colhead{} &
\colhead{} &
\colhead{} &
\colhead{} &
\colhead{} 
%\colhead{} &
%\colhead{} &
%\colhead{} &
%\colhead{} &
%\colhead{} &
%\colhead{} &
%\colhead{} &
%\colhead{} &
%\colhead{} \\
\\
\colhead{Cluster} & 
\colhead{Source} &
\colhead{(Gyr)} &
\colhead{(kpc)} &
\colhead{(Gyr)} &
\colhead{(kpc)} &
\colhead{[Fe/H]} &
\colhead{[O/Fe]} &
\colhead{[Na/Fe]} &
\colhead{[Mg/Fe]} &
\colhead{[Al/Fe]} &
\colhead{[Si/Fe]} 
%\colhead{[Ca/Fe]} &
%\colhead{[Ti/Fe]} &
%\colhead{[Mn/Fe]} &
%\colhead{[Co/Fe]} &
%\colhead{[Ni/Fe]} &
%\colhead{[Zr/Fe]} &
%\colhead{[Ba/Fe]} &
%\colhead{[La/Fe]} &
%\colhead{[Eu/Fe]} 
}
\startdata
Be 17 & Friel10 & 10.10 & 11.70 & 10.06 & 10.89 & $-$0.10 & 0.00 & 0.33 & 0.12 & 0.25 & 0.30 \\ 
Be 18 & This Study & 5.69 & 13.10 & 5.69 & 12.09 & $-$0.44 & 0.11 & 0.21 & 0.18 & 0.19 & 0.17 \\ 
Be 20 & Sestito08 & 6.00 & 16.37 & 4.05 & 16.12 & $-$0.30 & \ldots & 0.17 & 0.28 & 0.17 & 0.05 \\ 
Be 20 & Y05 & 4.10 & 16.00 & 4.05 & 16.12 & $-$0.49 & 0.18 & 0.21 & 0.24 & 0.18 & 0.06 \\ 
Be 21 & This Study & 2.18 & 14.27 & 2.18 & 14.27 & $-$0.31 & 0.18 & 0.36 & 0.19 & 0.17 & 0.21 
\enddata

\tablenotetext{a}{Ages and distances taken from the individual papers}
\tablenotetext{b}{Ages and distances taken from \citet{salaris04}}
\tablenotetext{c}{Ages and distances taken from \citet{bragaglia06}}

\tablerefs{
Bragaglia01 = \citet{bragaglia01}; 
Bragaglia08 = \citet{bragaglia08};
Carraro04 = \citet{carraro04}; 
Carrera11 = \citet{carrera11}; \\
Carretta04 = \citet{carretta04};
Carretta05 = \citet{carretta05};
Carretta07 = \citet{carretta07};
de Silva06 = \citet{desilva06}; \\
de Silva07 = \citet{desilva07}; 
Friel10 = \citet{friel05,jacobson08,jacobson09,friel10}; 
Hill99 = \citet{hill99}; 
Jacobson11a = \citet{jacobson11a};\\
Jocobson11b = \citet{jacobson11b}; 
Pancino10 = \citet{pancino10}; 
Paulson03 = \citet{paulson03};
Sestito08 = \citet{sestito06,sestito07,sestito08,bragaglia08}; 
Villanova05 = \citet{villanova05}; 
Y05 = \citet{y05} \\ 
\\
Note. Table 13 is published in its entirety in the electronic edition of the Astronomical Journal. A portion is shown here for guidance regarding its form and content.
\\
}

\end{deluxetable*}

The five clusters studied in the present paper lie on, or slightly below, the 
mean trends. (If we adopt the \citet{salaris04} distances, then the 
two most distant clusters in this paper are more metal-rich 
than the other three clusters. Although we note that this is not the case
for our distance estimates.) 
Given the inhomogeneous nature of this 
figure, it is difficult to ascertain whether or not the dispersion about 
the linear trend represents a real spread in metallicity or systematic 
offsets between the various studies. Our typical measurement uncertainty 
for [Fe/H] is 0.15 dex, a value comparable with the dispersion about 
the linear fit which may indicate that the dispersion about the 
mean trend may be almost entirely attributable to measurement uncertainties. 
Indeed, this issue of inhomogeneous 
comparisons is a recurring theme throughout the rest of the paper. 
We note that in this figure, and subsequent figures, no attempt has been 
made to normalize the abundances onto a common scale. It would be possible 
to make adjustments for the adopted solar abundances. However, 
quantifying systematic differences arising from equivalent widths, 
$\log gf$ values, stellar parameters, 
model atmospheres, spectrum synthesis codes, and so on 
would require an independent re-analysis of the entire data set 
(a substantial, but feasible analysis that should be conducted). 
Such a re-analysis, which should also include non-LTE and/or 3D effects 
when possible (e.g., \citealt{bergemann08,lind09}), 
is beyond the scope of the present paper. 

Had we shifted our metallicities (i.e., only our measurements 
in the present paper and in Paper I), [Fe/H], by +0.1 dex, we would have 
obtained identical slopes for all subsamples considered in Figure 
\ref{fig:fehab}. For the distant samples, the dispersion about the 
linear fit would decrease by 0.01 to 0.02 dex depending on the 
adopted distance scale.

\begin{figure}[t!] 
\epsscale{1.2}
\plotone{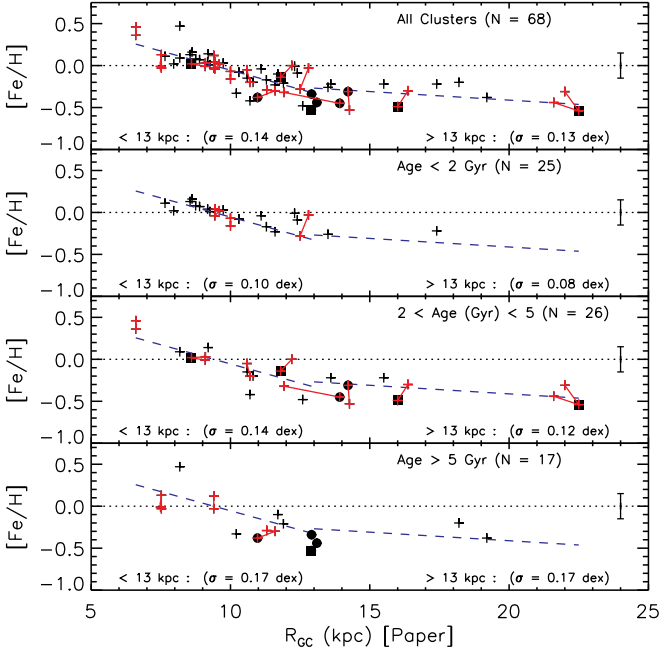}
\caption{[Fe/H] vs.\ Galactocentric distance using the 
same symbols as in Figure \ref{fig:fehab}. The upper panel 
shows all stars. The lower panels show stars in various 
age groups: $<$ 2 Gyr, 2 $<$ Age $<$ 5, and $>$ 5 Gyr. 
The linear fit to the data with \rgc\ $<$ 13 kpc 
and \rgc\ $>$ 13 kpc is shown in the upper panel. We superimpose 
these fits to the lower three panels but remeasure the dispersion 
about the linear fit. 
A representative error bar is shown. 
\label{fig:fehage}}
\end{figure}

Although we explicitly consider abundance trends with age 
in the following subsection, we now consider whether the trends 
and dispersions 
between metallicity and Galactocentric distance differ when 
considering open clusters of different age ranges. 
In Figure \ref{fig:fehage}, we again plot metallicity versus 
Galactocentric distance, but the sample is split into various 
age groups: $<$ 2 Gyr, 2 $<$ Age $<$ 5, and $>$ 5 Gyr. These 
bins were arbitrarily chosen to contain roughly equal numbers of clusters 
and none of our conclusions would change had we changed the age 
groups by $\pm$ 0.5 Gyr. 
The upper panel of this figure shows the full sample along with the 
linear fit to the data for the local (\rgc\ $<$ 13 kpc) and the 
distant (\rgc\ $>$ 13 kpc) samples. In the lower panels, 
each of which covers a different age range, we superimpose the 
linear fit to the complete sample 
and we re-measure the dispersions about the linear fit for each 
subsample. The most notable feature of this dissection is that 
for the local sample, \rgc\ $<$ 13 kpc, the scatter about 
the mean relation is largest for the oldest clusters ($>$ 5 Gyr) and smallest 
for the youngest clusters ($<$ 2 Gyr). This result would remain unchanged 
had we 
(i) shifted our metallicities by +0.1 dex or 
(ii) excluded the very old (10.2 Gyr) and metal-rich ([Fe/H] = +0.47) 
cluster NGC 6791 \citep{carretta07}. 

Clearly, it would be interesting to study whether the 
transition radius between the inner and outer disk 
changes as a function of age. To this end, \citet{jacobson11b} 
explored this issue and found a suggestion of some variation. 

\begin{figure}[t!] 
\epsscale{1.2}
\plotone{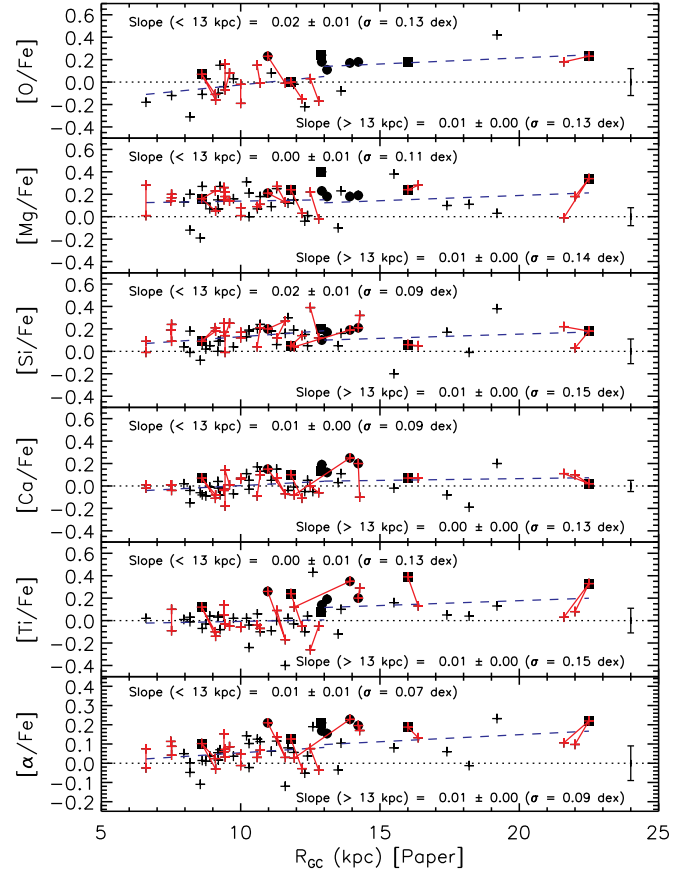}
\caption{Abundance ratios [X/Fe] vs.\ Galactocentric distance 
for the $\alpha$ elements. Symbols are the same as in Figure \ref{fig:fehab}. 
In all panels, we 
show the linear fit to the data with \rgc\ $<$ 13 kpc 
and \rgc\ $>$ 13 kpc. 
A representative error bar is shown. 
\label{fig:alpha1}}
\end{figure}

In Figure \ref{fig:alpha1}, we plot [X/Fe] for the $\alpha$ 
elements versus Galactocentric distance. 
In the bottom panel, $\alpha$ represents the direct average of 
O, Mg, Si, Ca, and Ti. As in previous figures, we compute and 
plot the fit to the data for \rgc\ $<$ 13 kpc and \rgc\ $>$ 13 kpc. 
For all elements, there is no statistically significant 
difference between the gradient for the local and distant samples. 
While combining the $\alpha$ elements into a single 
measure may be a convenient way to examine their behavior and to 
facilitate comparisons with model predictions, we note that 
the individual elements do not appear to follow identical patterns: 
[Ca/Fe] is roughly solar at all distances, [Si/Fe] and [Mg/Fe] 
are uniformly enhanced at all distances, and [O/Fe] and [Ti/Fe] show 
considerably higher ratios in the outer disk compared to the inner disk. 
Thus the behavior of [$\alpha$/Fe] would appear to be driven primarily by 
O and Ti. 
As noted in Paper I, the outer disk open clusters are probably not members of
an older population such as the thick disk (e.g., \citealt{bensby04}) despite some
similarities in [X/Fe] abundance ratios.  

\begin{figure}[t!]
\epsscale{1.2}
\plotone{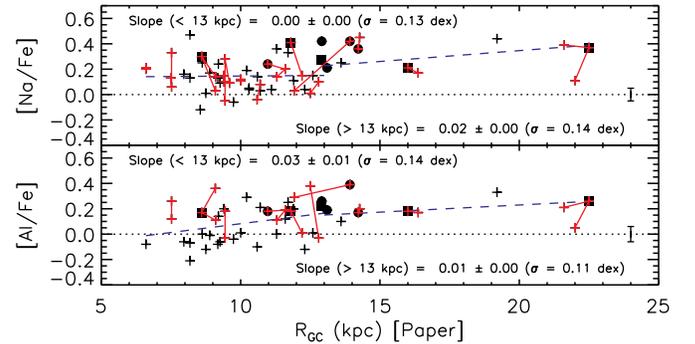}
\caption{Same as Figure \ref{fig:alpha1} but for the light 
elements [Na/Fe] and [Al/Fe]. 
\label{fig:nafe1}}
\end{figure}

\begin{figure}[t!]
\epsscale{1.2}
\plotone{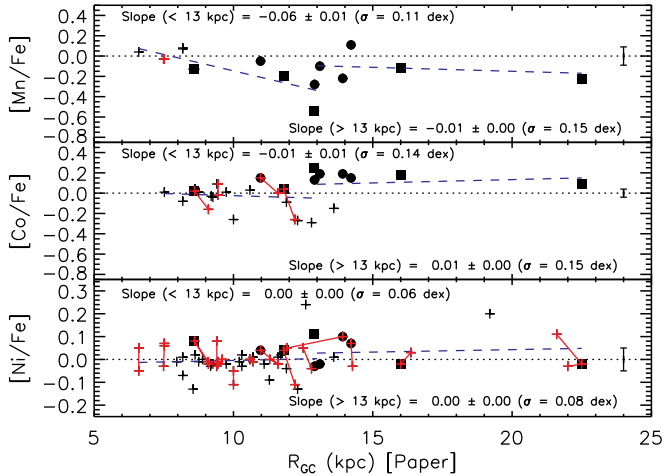}
\caption{Same as Figure \ref{fig:alpha1} but for the Fe-peak 
elements [Mn/Fe], [Co/Fe], and [Ni/Fe]. 
\label{fig:fepeak1}}
\end{figure}

\begin{figure}[t!]
\epsscale{1.2}
\plotone{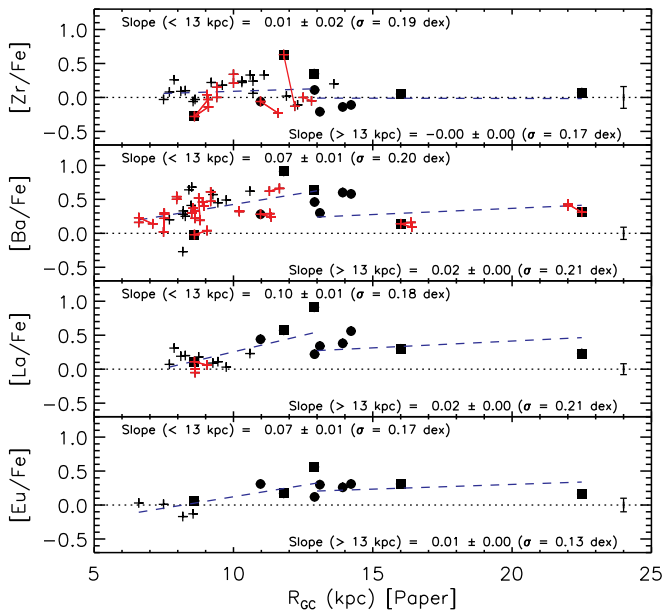}
\caption{Same as Figure \ref{fig:alpha1} but for the neutron-capture 
elements [Zr/Fe], [Ba/Fe], [La/Fe], and [Eu/Fe]. 
\label{fig:neutron1}}
\end{figure}

We plot the light elements 
(Na and Al; Figure \ref{fig:nafe1}), 
the Fe-peak elements 
(Mn, Co, and Ni; Figure \ref{fig:fepeak1}), 
and the neutron-capture elements 
(Zr, Ba, La, and Eu; Figure \ref{fig:neutron1}) 
versus distance. 
For the neutron-capture elements discussed here, we note that 
Zr, Ba, and La are predominantly produced via the $s$-process 
while Eu is produced via the $r$-process \citep{simmerer04}. 
For the neutron-capture elements Zr, Ba, and La, we include the 
abundance data from \citet{dorazi09} and \citet{maiorca11} and 
distances, when not listed in those references or \citet{salaris04}, 
from the \citet{chen03} catalogue. 
We again compute and overplot the linear fit to the local (\rgc\ $<$ 13 kpc) 
and distant (\rgc\ $>$ 13 kpc) samples. 
While most elements show no significant difference in gradient 
between the local and distant samples, within the limited data 
there is a suggestion that Mn, Ba, La, and Eu have different 
gradients between the local and distant samples. 

For all $\alpha$ elements, the gradients are positive (or zero). 
The maximum value, 0.02 $\pm$ 0.01 dex kpc$^{-1}$, occurs for O and Si 
in the range \rgc\ $<$ 13 kpc. Therefore, the $\alpha$ elements show 
only a small trend with increasing \rgc. 
Similarly, Na and Al have positive (or zero) gradients suggesting a 
slight trend with \rgc. The maximum value is 0.03 $\pm$ 0.01 dex kpc$^{-1}$ 
for Al in the range \rgc\ $<$ 13 kpc. 
For the Fe-peak elements Mn, Co, and Ni, the gradients are both 
positive and negative. The largest value is for Mn in the range \rgc\ $<$ 
13 kpc, $-$0.06 $\pm$ 0.01 dex kpc$^{-1}$. All other gradients are 
small. 
The neutron-capture elements Ba, La, and Eu show large positive 
gradients in the range \rgc\ $<$ 13 kpc. In the outer disk, \rgc\ $>$ 13 kpc, 
their gradients are relatively flat with maximum values of 0.02 dex 
kpc$^{-1}$. Zr shows no trend with Galactocentric distance. 

Another way to compare the abundances between the inner and outer disk 
is to consider the mean [X/Fe] ratios in the two regions. 
We again choose a boundary value of \rgc\ = 13 kpc between the 
inner and outer disks. We then measure the mean [X/Fe] ratio 
as well as the standard error of the mean. By adding the two standard 
errors of the mean in quadrature and comparing this value to the 
difference between the two means, we obtain the level of significance. 
In Table \ref{tab:sigma}, we show for each element the level of significance 
that the [X/Fe] ratios differ between the inner and outer disk. 
The elements that differ at the 3-$\sigma$ level or higher are 
O, Na, Al, and Ti. While we note that $\alpha$ is also significant 
at the 3-$\sigma$ level, Mg and Si are less than 1-$\sigma$ 
and Ca is less than 2-$\sigma$. 
In this table, 
we also show results when using 
boundary values of 10 kpc and 15 kpc. 
For some elements, the level of significance depends strongly upon the 
adopted boundary. 

\begin{deluxetable*}{lrrrrrrrrrrrrrrrrr}
\tabletypesize{\scriptsize}
\tablecolumns{17} 
\tablewidth{0pc} 
\tablecaption{Mean [X/Fe] Ratios, Standard Error of the Mean, and 
Level of Significance When Comparing the 
Inner and Outer Disk For Different Boundary Values. 
\label{tab:sigma}}
\tablehead{
\colhead{Species}  & 
\colhead{$\mu$} &
\colhead{s.e.} & 
\colhead{$\mu$} &
\colhead{s.e.} & 
\colhead{Significance} & 
\colhead{} & 
\colhead{$\mu$} &
\colhead{s.e.} & 
\colhead{$\mu$} &
\colhead{s.e.} & 
\colhead{Significance} & 
\colhead{} & 
\colhead{$\mu$} &
\colhead{s.e.} & 
\colhead{$\mu$} &
\colhead{s.e.} & 
\colhead{Significance} \\ 
\colhead{} & 
\multicolumn{2}{c}{Inner} & 
\multicolumn{2}{c}{Outer} & 
\multicolumn{2}{c}{} & 
\multicolumn{2}{c}{Inner} & 
\multicolumn{2}{c}{Outer} & 
\multicolumn{2}{c}{} & 
\multicolumn{2}{c}{Inner} & 
\multicolumn{2}{c}{Outer} \\
\cline{2-6} 
\cline{8-12} 
\cline{14-18} 
\colhead{} & 
\multicolumn{5}{c}{\rgc\ = 10 kpc} &
\colhead{} & 
\multicolumn{5}{c}{\rgc\ = 13 kpc} &
\colhead{} & 
\multicolumn{5}{c}{\rgc\ = 15 kpc}  
}
\startdata
{\rm [O/Fe]}  & $-$0.05 & 0.03 & 0.06 & 0.03 & 2.56 &  & $-$0.02 & 0.02 & 0.17 & 0.05 & 3.52 &  & $-$0.01 & 0.02 & 0.25 & 0.06 & 4.17 \\
{\rm [Na/Fe]}  & 0.14 & 0.03 & 0.20 & 0.02 & 1.62 &  & 0.15 & 0.02 & 0.31 & 0.04 & 3.97 &  & 0.16 & 0.02 & 0.28 & 0.06 & 2.02 \\
{\rm [Mg/Fe]}  & 0.13 & 0.02 & 0.15 & 0.02 & 0.62 &  & 0.14 & 0.02 & 0.17 & 0.04 & 0.71 &  & 0.14 & 0.02 & 0.18 & 0.05 & 0.93 \\
{\rm [Al/Fe]}  & 0.04 & 0.03 & 0.16 & 0.02 & 2.98 &  & 0.09 & 0.02 & 0.20 & 0.03 & 3.10 &  & 0.10 & 0.02 & 0.20 & 0.04 & 2.20 \\
{\rm [Si/Fe]}  & 0.10 & 0.02 & 0.15 & 0.02 & 1.99 &  & 0.13 & 0.01 & 0.13 & 0.04 & 0.02 &  & 0.14 & 0.01 & 0.10 & 0.05 & 0.70 \\
{\rm [Ca/Fe]}  & $-$0.03 & 0.01 & 0.05 & 0.02 & 3.37 &  & 0.00 & 0.01 & 0.06 & 0.03 & 1.66 &  & 0.01 & 0.01 & 0.03 & 0.04 & 0.41 \\
{\rm [Ti/Fe]}  & $-$0.01 & 0.01 & 0.05 & 0.03 & 1.82 &  & $-$0.01 & 0.02 & 0.16 & 0.04 & 4.00 &  & 0.01 & 0.02 & 0.15 & 0.04 & 2.87 \\
{\rm [Mn/Fe]}  & $-$0.02 & 0.03 & $-$0.18 & 0.06 & 2.39 &  & $-$0.11 & 0.05 & $-$0.11 & 0.06 & 0.04 &  & $-$0.10 & 0.05 & $-$0.18 & 0.06 & 1.03 \\
{\rm [Co/Fe]}  & $-$0.02 & 0.03 & 0.00 & 0.05 & 0.55 &  & $-$0.03 & 0.03 & 0.11 & 0.05 & 2.21 &  & $-$0.01 & 0.03 & 0.14 & 0.05 & 2.71 \\
{\rm [Ni/Fe]}  & $-$0.01 & 0.01 & 0.01 & 0.01 & 1.16 &  & 0.00 & 0.01 & 0.04 & 0.02 & 1.67 &  & 0.00 & 0.01 & 0.05 & 0.04 & 1.20 \\
{\rm [Zr/Fe]}  & 0.06 & 0.04 & 0.10 & 0.04 & 0.58 &  & 0.09 & 0.03 & $-$0.02 & 0.06 & 1.62 &  & 0.08 & 0.03 & 0.06 & 0.01 & 0.46 \\
{\rm [Ba/Fe]}  & 0.32 & 0.04 & 0.42 & 0.05 & 1.61 &  & 0.36 & 0.03 & 0.33 & 0.06 & 0.43 &  & 0.37 & 0.03 & 0.26 & 0.06 & 1.76 \\
{\rm [La/Fe]}  & 0.11 & 0.03 & 0.42 & 0.07 & 4.19 &  & 0.22 & 0.06 & 0.36 & 0.06 & 1.83 &  & 0.25 & 0.05 & 0.27 & 0.04 & 0.28 \\
{\rm [Eu/Fe]}  & $-$0.04 & 0.05 & 0.28 & 0.04 & 5.04 &  & 0.11 & 0.07 & 0.27 & 0.03 & 2.02 &  & 0.15 & 0.06 & 0.24 & 0.08 & 0.86 \\
{\rm [$\alpha$/Fe]}  & 0.04 & 0.01 & 0.09 & 0.01 & 3.12 &  & 0.06 & 0.01 & 0.13 & 0.02 & 3.09 &  & 0.06 & 0.01 & 0.12 & 0.03 & 2.07 
\enddata

\end{deluxetable*}

We now re-consider our results when 
we arbitrarily increase our [Fe/H] values by 0.1 dex, 
and therefore decrease our [X/Fe] by 0.1 dex. 
Recall that this shift in [Fe/H] is motivated by 
differences between our [Fe/H] values and 
literature values for clusters in common with the literature. 
We find that while there remain differences 
in the mean abundances between the inner and 
outer disks, the level of significance is lower in some
cases; O, Na, Al, Zr, and $\alpha$ are significant 
at the 2-$\sigma$ level and Ti is significant 
at the 4-$\sigma$ level. 

\subsection{Trends With Age}

\begin{figure}[t!] 
\epsscale{1.2}
\plotone{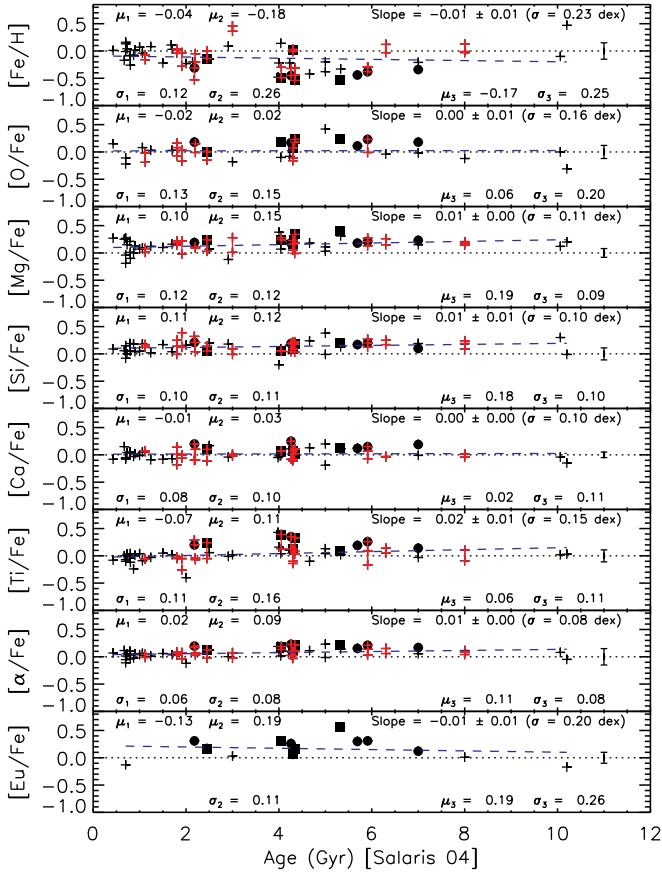}
\caption{Abundance ratios [Fe/H] and [X/Fe] vs.\ age. The symbols are
the same as in Figure \ref{fig:fehab}. 
The dashed blue line shows the linear fit to the data. In each 
panel, we also show the mean abundance ratio and dispersion for 
three age groups: (1) $<$ 2 Gyr, (2) 2 $<$ Age $<$ 5, and (3) $>$ 5 Gyr. 
A representative error bar is included in each panel. 
\label{fig:age1}}
\end{figure}

\begin{figure}[t!] 
\epsscale{1.2}
\plotone{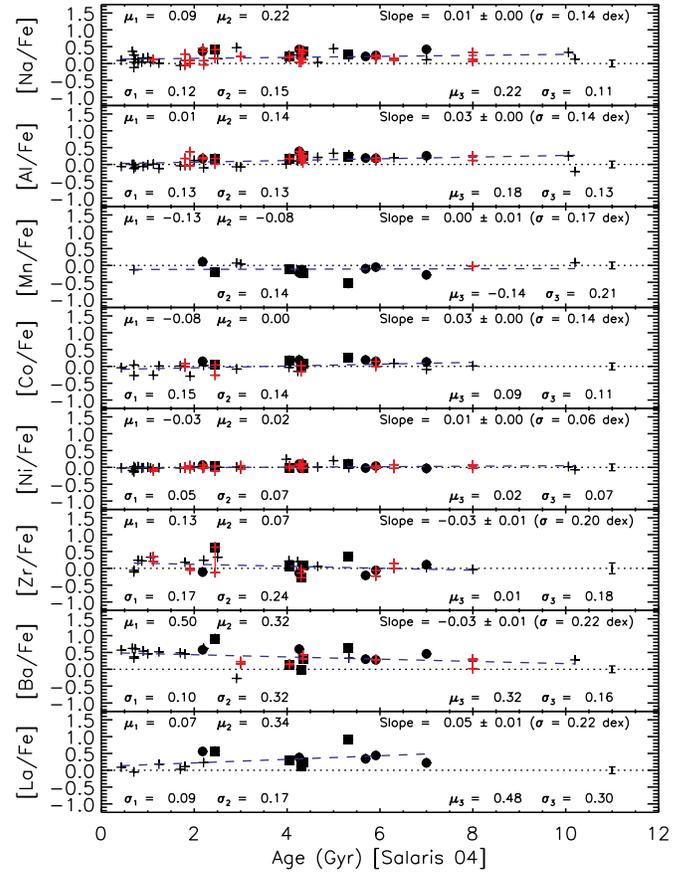}
\caption{Same as Figure \ref{fig:age1} but for different elements. 
\label{fig:age2}}
\end{figure}

Next we explore whether there are any trends between metallicity, 
[Fe/H], and abundance ratios, [X/Fe], versus age 
(Figures \ref{fig:age1} and \ref{fig:age2}). 
In these figures we adopt the ages from \citet{salaris04} and 
individual clusters with multiple measurements 
are connected with red lines. 
For the open clusters not included in \citet{salaris04}, we adopt 
the ages given in the literature sources (see Table \ref{tab:compile} 
for all ages and references). 
In all panels of these figures, 
we compute the linear fit to the data and show the gradient, 
uncertainty, and dispersion about the linear fit. For all 
elements there are no significant trends, at the $>$3-$\sigma$ level, 
between abundance and age. 
The most significant trends we found were for 
Co, +0.03 $\pm$ 0.01 dex/Gyr and Ba, $-$0.03 $\pm$ 0.01 dex/Gyr. 
(Lines from both elements are affected by hyperfine 
and/or isotopic splitting and therefore these elements are 
more likely to be affected by systematic uncertainties than 
other elements.) 

\citet{dorazi09} found that the [Ba/Fe] ratio decreased with increasing age. 
\citet{maiorca11} extended the measurements to additional neutron-capture 
elements and confirmed the trends between decreasing [X/Fe] with 
increasing age. 
We find a similar trend, albeit with a smaller amplitude and significance 
for Zr and Ba. 
A likely explanation for this result is that our new measurements 
of neutron-capture elements in our sample of 
older clusters fall at higher [X/Fe] ratios than the values in  
\citet{dorazi09} and \citet{maiorca11}. 
For La, we find an opposite trend of increasing [La/Fe] 
with increasing age. 

For some elements, primarily Ni, the dispersion about the linear 
fit ($\sigma$ = 0.07 dex) is comparable to the measurement uncertainty 
(0.06 dex). 
We also measure the mean abundance ratio and dispersion for the 
three age groups used earlier: 
$<$ 2 Gyr, 2 Gyr $<$ Age $<$ 5 Gyr, and $>$ 5 Gyr. 
Given the limited data and inhomogeneous analyses, it would be premature 
to conclude that the dispersion differs between the various age subsamples, 
but there are hints for some elements of unusually large, or small, 
dispersions in various age subsamples. The presence, or absence, 
of such trends and dispersions would be revealed from a homogeneous analysis. 

Again, we reconsider the results when we arbitrarily increase our 
metallicities by 0.1 dex, and decreasing our [X/Fe] ratios by 0.1 dex. 
For all elements, the gradients change by $\le$0.01 dex/Gyr and the dispersion 
about the linear trends generally decreases by a very small amount. 

\subsection{Trends With Metallicity}

\begin{figure*}[t!]
\epsscale{1.0}
\plotone{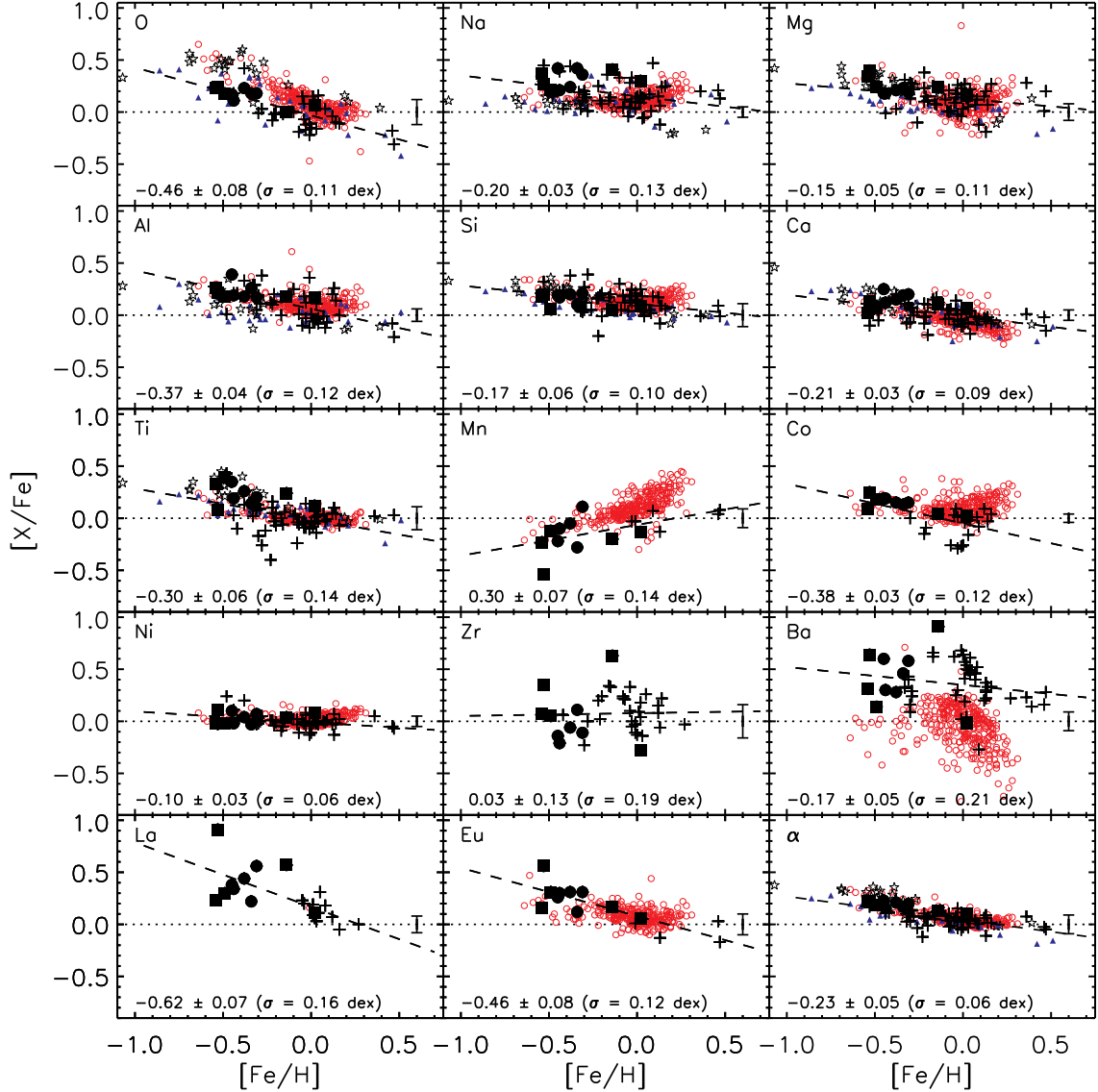}
\caption{[X/Fe] vs.\ [Fe/H]. 
Literature open clusters (plus signs), clusters from this study (closed black 
circles), and clusters from Paper I (closed black squares) are plotted. 
We overplot thin disk (black stars) and 
thick disk (blue triangles) giants from \citet{alvesbrito10} as well as 
(mainly thin disk) giants from \citet{luck07} (red open circles). 
A representative error bar is shown. We plot the linear fit to the
open cluster data 
and write the slope, uncertainty, and dispersion about the slope. 
\label{fig:xfe}}
\end{figure*}

In Figure \ref{fig:xfe} we plot the 
abundance ratios [X/Fe] versus metallicity, [Fe/H]. 
In all panels, 
we plot the linear fit to the data and show the 
slope, uncertainty, and dispersion about the linear fit. 
(As before, the linear fit includes multiple 
measurements of a given cluster such that those clusters in effect 
are given more weight than a cluster with a single measurement.) 
We overplot 
solar neighborhood 
thin and thick disk giants from \citet{alvesbrito10} as well 
as solar neighborhood giants from \citet{luck07} which are primarily thin 
disk objects. 
The open cluster abundance measurements come almost exclusively 
from giant stars. 
Therefore, when comparing the clusters with field stars, we 
chose to include only giants in order to 
minimize any systematic abundance differences. 
For example, comparison of the results for Mg between bulge 
and local thick disk 
stars from 
\citet{fulbright07} and \citet{alvesbrito10} 
illustrates the potential systematic differences 
that may arise when comparing dwarfs with giants. 
Additionally, elements such as Na have important non-LTE 
effects which differ between dwarfs and giants (e.g., \citealt{lind11}). 
Arguably, any dwarf versus giant abundance discrepancies 
may be small relative to the systematic differences arising from 
an inhomogeneous comparison (stellar parameters, atomic data, 
equivalent widths, solar abundances, model atmospheres, 
spectrum synthesis software, etc). 

For almost all elements, the trends between [X/Fe] versus [Fe/H] 
for open clusters and field stars are in fair to good agreement. 
Such agreement would suggest, at face value, that the open 
clusters, which span a range in ages and distances, have experienced 
a similar chemical evolution history as did the local field stars 
which span an unknown (but presumably large) range in age. 
The obvious discrepancies include Mn and Ba (these elements are 
affected by hyperfine and/or isotopic splitting) and perhaps O and Na. 
For Mn, the \citet{luck07} field giants have higher [Mn/Fe] ratios 
than do the open clusters. 
For Ba, the field giants have lower [Ba/Fe] ratios than do the 
open clusters. As noted, \citet{dorazi09} found a trend of 
decreasing [Ba/Fe] with increasing age in the open clusters and 
the lower [Ba/Fe] ratios in field stars could arise if the field 
stars are systematically older than the open clusters. 
In the open clusters, [O/Fe] appears lower than field stars of 
comparable metallicity while [Na/Fe] appears systematically higher 
than in field stars. While 
such abundance patterns resemble the O-Na anticorrelation 
seen in globular clusters \citep{gratton04}, the open clusters 
do not display this abundance pattern \citep{desilva09}. 
The open cluster data are inhomogeneous as are the comparison 
field stars. Until a systematic and homogeneous analysis is performed 
upon a sample of open clusters and field stars, we cannot definitively 
say whether or not 
the open clusters and field stars have the same, or differing, 
chemical abundance trends. 

\begin{figure}[t!]
\epsscale{1.2}
\plotone{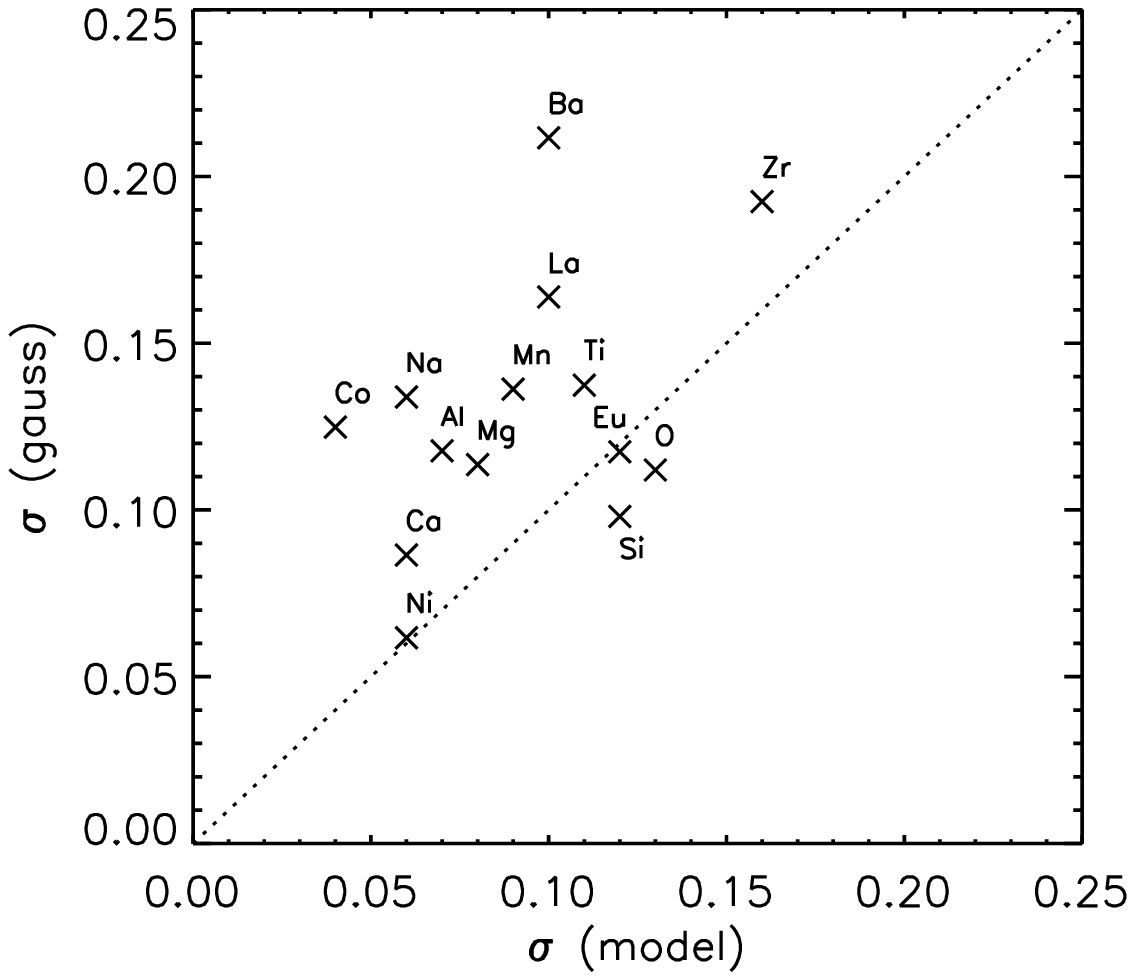}
\caption{Predicted error in [X/Fe] due to uncertainties in the 
model parameters, $\sigma$ (model), versus the dispersions about 
the linear fit to the open cluster data from 
Figure \ref{fig:xfe}, $\sigma$ (gauss). 
The one-to-one relation is shown. 
\label{fig:compare}}
\end{figure}

In Figure \ref{fig:compare}, we compare the predicted error in [X/Fe] 
due to uncertainties in the model parameters (i.e., the values 
in the fourth column of Table \ref{tab:parvar}) with the 
dispersion about the linear fit to open clusters in the [X/Fe] versus [Fe/H] 
plane\footnote{This approach is only meaningful if we expect the 
dependence of [X/Fe] versus [Fe/H] to be linear. Inspection of 
Figure \ref{fig:xfe} would suggest that the distributions are often non-linear. 
In addition, we need to be mindful of potential systematic differences between 
various studies which could serve to increase the observed element abundance 
dispersion.}. 
For O, Si, Ti, Ni, and Eu, these values are in good agreement 
suggesting that the dispersion about the linear fit to [X/Fe] versus 
[Fe/H] can be entirely attributable to the measurement 
uncertainties. 
Given the inhomogeneous nature of the open cluster data, this 
is a slightly surprising result. 
Notably, [Ni/Fe] and [$\alpha$/Fe] show dispersions about the linear fit of 
only 0.07 and 0.06 dex, respectively. 
For other elements, notably Co, Zr, Ba, and La, the predicted 
error in [X/Fe] is considerably smaller than the measured 
dispersion (which reaches values as high as 0.24 dex for Zr and Ba). 
Such a result would indicate that these elements show a 
real dispersion amongst the open clusters or that the 
combined sample is significantly affected by systematic offsets 
between the various individual studies. 
For Ba, the lines are generally quite strong and require consideration 
of hyperfine splitting such that the dispersion will likely be larger than 
the errors arising solely from uncertainties in the stellar parameters. 
Furthermore, the trend between [Ba/Fe] and age found by \citet{dorazi09} 
would introduce additional scatter when considering clusters of all ages 
in the [Ba/Fe] vs.\ [Fe/H] plane. 
Finally, it becomes more difficult to accurately measure the 
strengths of strong lines such as Ba and their strength also means 
that the lines form at shallower optical depths such that the 
LTE assumption is less reliable. 

La may offer a more reliable measure of the $s$-process 
since there are more lines available to measure including a number 
of weaker lines. 
La may show a trend with 
age (a negative trend according to \citep{maiorca11} and a 
positive trend according to this study) 
and therefore inclusion of clusters with a large range of ages 
may serve to increase the dispersion for [La/Fe] vs.\ [Fe/H] beyond 
that expected only from uncertainties in the stellar parameters. Finally, 
the large dispersions for Zr and Co may be due, in part, to the 
small number of lines typically used to derive abundances (although 
some other elements also measured from comparably small numbers 
of lines show agreement between the predicted and observed 
abundance dispersions). 

We re-examine our results when increasing [Fe/H] by 0.1 dex 
and decreasing [X/Fe] by 0.1 dex. For the linear fit to the 
[X/Fe] vs.\ [Fe/H] data, the slopes increase by an average 
of 0.08 $\pm$ 0.01 dex/dex (i.e., the negative slopes 
become flatter). Such a result is readily anticipated given 
that our clusters are amongst the most metal-poor in the combined sample 
and lowering the [X/Fe] ratios will therefore produce flatter 
slopes when fitting [X/Fe] vs.\ [Fe/H]. 
Regarding Ti, our highest [Ti/Fe] ratios are comparable to 
the upper envelope of field star data and some 
open cluster literature 
[Ti/Fe] data lie substantially below the relation defined 
by the field star data. 
When adjusting our [Fe/H] and [X/Fe] ratios by 0.1 dex, 
the dispersion about the linear fit 
is essentially unchanged and the clusters and field stars remain 
in good agreement with the exception of the elements already noted, 
Mn and Ba and perhaps O and Na. 

\subsection{The Formation and Evolution of the Outer Galactic Disk} 

\subsubsection{Insights from Open Clusters} 

We now comment on what we can learn about the origin and evolution 
of the outer disk based on metallicities and chemical abundance ratios 
in open clusters. 
The existing data indicate that the outer Galactic disk open clusters 
(\rgc\ $>$ 13 kpc) 
are uniformly metal-poor, i.e., [Fe/H] $<$ 0.0, albeit with metallicities 
higher than expected based on an extrapolation of the abundance gradient
of local open clusters (\rgc\ $< $13 kpc). 
Interestingly, we note that 
the handful of field red giants in the inner 
Galactic disk \citep{bensby10} are also metal-poor and 
do not lie on a linear extrapolation 
of the metallicity gradient based on the solar neighborhood, although 
these objects have unknown ages and 
may belong to the thick disk rather than the thin disk. 
All recent studies of open clusters 
confirm that the metallicity gradient in the 
outer disk is flatter than the metallicity gradient near 
the solar neighborhood. 
As noted in the introduction, some external galaxies exhibit 
similar behavior in their outer disks. 

We also note that almost all outer disk open clusters have super-solar  
[$\alpha$/Fe] and [Eu/Fe] ratios. Such abundance ratios 
are different from the values found in the more metal-rich 
open clusters in the solar neighborhood. 
As already discussed, 
the [O/Fe], [Ti/Fe], and [$\alpha$/Fe] ratios 
differ between the inner and outer disk at the 3-$\sigma$ level 
or higher (assuming a boundary of 13 kpc). Mg, Si, and Ca do not 
show significant differences for [X/Fe] between the two regions. 
Depending on the choice of the boundary value, [Eu/Fe] differs 
between the inner and outer regions at the 5.0-$\sigma$, 2.0-$\sigma$, 
and 0.9-$\sigma$ level for 10 kpc, 13 kpc, and 15 kpc, respectively. 

As noted, the possibility exists that our [X/Fe] 
measurements are systematically high. 
In order to produce [X/Fe] ratios for O, Ti, and Eu 
that agree within 2-$\sigma$ between the inner and outer disk 
would require decreases of 0.2, 0.4, and 0.1 dex for 
[O/Fe], [Ti/Fe], and [Eu/Fe], respectively.  
Additional data and a homogeneous re-analysis of the existing data 
may be necessary to clarify this intriguing situation of enhanced 
[$\alpha$/Fe] and [Eu/Fe] ratios in the outer disk. 

All $s$-process elements studied show 
a significant abundance 
scatter in the outer disk (and in the solar neighborhood) 
extending to values as high as [X/Fe] $\simeq$ +1.0. 
Such high abundances of the $s$-process elements would suggest that 
asymptotic giant branch stars have played, in some cases, a 
major role in the chemical evolution of open clusters 
relative to field stars. 
Recently, \citet{maiorca12} have suggested that the large 
[$s$-element/Fe] ratios seen in young metal-rich open clusters 
can be explained by low-mass AGB stars with efficient 
$s$-processing. 

When comparing the distant open clusters with solar neighborhood 
field stars at a given metallicity, 
the abundance ratios [$\alpha$/Fe] and [Eu/Fe] 
are in fair agreement. 
Such similarity in abundance ratios would suggest that in the range of 
metallicities spanned by both sets of objects, the relative 
contributions from 
Type II supernovae (SNe II) and Type Ia supernovae (SNe Ia) 
are similar.
The primary 
difference is that the chemical enrichment in the outer disk 
did not yet reach the metallicities of the solar neighborhood. 
Therefore, the conclusion we drew in Paper I that 
``high abundances of $\alpha$-elements indicate 
rapid star formation, such that Type Ia supernovae did not 
have sufficient time to evolve and contribute to the chemical evolution'' 
may not be correct since 
it now appears likely that SNe Ia have made a contribution. 
Two differences between the present work and that of 
Paper I are (1) the number of open clusters with chemical 
abundance measurements has increased 
considerably and (2) the comparison field stars are all giants which 
may remove any systematic abundance differences arising from dwarf 
versus giant comparisons. 
Clearly it would be of great interest to identify and analyze 
more metal-rich distant open clusters. 

The metal-poor open clusters in the outer disk 
have super-solar ratios of the $\alpha$ elements and Eu which 
are primarily produced in massive stars. Since we do not see 
significant contributions from SNe Ia, we may conclude that the 
star formation in the outer disk was not prolonged otherwise we 
would have seen 
[$\alpha$/Fe] substantially lower than in field stars of comparable 
metallicity as found in the more metal-rich stars of 
the nearby dwarf spheroidal galaxies \citep{tolstoy09}. 
As discussed in Paper I, we would naively expect that the lower 
density in the outer disk would result in a slower star formation 
rate relative to the solar neighborhood. In this scenario, 
we would expect the metallicity gradient to be continuous and 
that at a given metallicity, 
the outer disk would have lower [$\alpha$/Fe] ratios than 
in the solar neighborhood. 
As we speculated in Paper I, one possibility for these 
somewhat unexpected characteristics in the outer disk 
is that a merger event 
and/or infall of material triggered a burst of star formation. 
Such a possibility was reinforced by the results from Paper III 
in which we found that the young outer disk Cepheids are more metal-poor 
than the older outer disk open clusters and that some of the Cepheids had 
enhanced [$\alpha$/Fe] ratios. 

Flat abundance gradients could also be produced by radial mixing 
(e.g., \citealt{roskar08,sb09,minchev11,minchev12}) 
and it is suspected that these processes may play a role in 
producing the thick disk 
(e.g., \citealt{sellwood02,haywood08,schoenrich09}). 
\citet{jilkova12} have shown that the Galactic bar and 
spiral arms may be responsible for moving NGC 6791 from 
the inner disk to its current location, although they 
regard the probability to be very low. While it may be 
possible for strongly bound clusters to survive the migration, 
to our knowledge no detailed study has yet been performed. 

Finally, we caution that any interpretation of the abundance 
ratios from compilations 
needs to acknowledge that the open clusters were 
studied by various authors who adopt different 
analysis techniques. 
We also note that for M67, a recent study by 
\citet{onehag10} showed that this cluster has 
a metallicity [Fe/H] = +0.02 and abundance ratios [X/Fe] 
within 0.03 dex of the solar values. Their strictly differential analysis 
of one dwarf star with almost identical stellar parameters 
to the Sun, a so-called solar twin, enabled very high precision. 
For comparison, in Paper I our analysis of M67 giants 
found near solar metallicity, [Fe/H] = 0.02, 
but [X/Fe] ratios that differed from the solar value. 
The elements with [X/Fe] $\ge$ 0.1 dex are Na, Mg, Al, Ti, and La 
and those with [X/Fe] $\le$ $-$0.1 dex are Mn and Zr.  
Table \ref{tab:compile} includes an additional study 
of M67 giants by Friel \& Jacobson 
(e.g., \citealt{friel05,jacobson08,jacobson09,friel10,jacobson11a}) 
that also found [X/Fe] ratios that differ from solar by 0.1 dex or more, 
including Ti for which they find [Ti/Fe] $\le$ $-$0.1. 
Therefore, the \citet{onehag10} 
results serve to highlight any systematic abundance offsets 
between studies that have included this target. 
We also note that our Paper I 
analysis included the bright giant Arcturus 
such that future studies can identify abundance offsets. 
Additionally, 
membership has not been unambiguously determined 
for some key clusters, e.g., Be 31 \citep{friel10}. 

\subsubsection{Insights from Field Red Giant Stars}

Chemical abundance measurements exist for a number of field stars 
beyond \rgc\ = 10 kpc. These objects 
include Cepheids (which we shall discuss in Sec 4.4.3), 
young OB stars \citep{daflon04} as well as field red giants 
\citep{carney05,bensby11} 
which we assume to have ages comparable to the open clusters. 
We focus here on the field red giants and compare the abundances 
obtained in these objects to the open cluster data. 
The main issue we seek to address here is whether the field star population 
is chemically distinct from the open clusters. While we do not necessary 
expect any chemical differences between the two samples, 
it is important to conduct the comparison. 

Before continuing, we caution that outer disk field red giants 
may be affected by selection biases. Stars beyond say \rgc\ = 10 kpc 
are likely chosen to lie well above, or below, the 
Galactic plane to avoid reddening. We speculate that such 
selection criteria may therefore give strong weight towards 
thick disk stars over thin disk stars (although see \citealt{bensby11} for 
a discussion of the scale length and scale height of the 
thick disk). 

\begin{figure}[t!] 
\epsscale{1.2}
\plotone{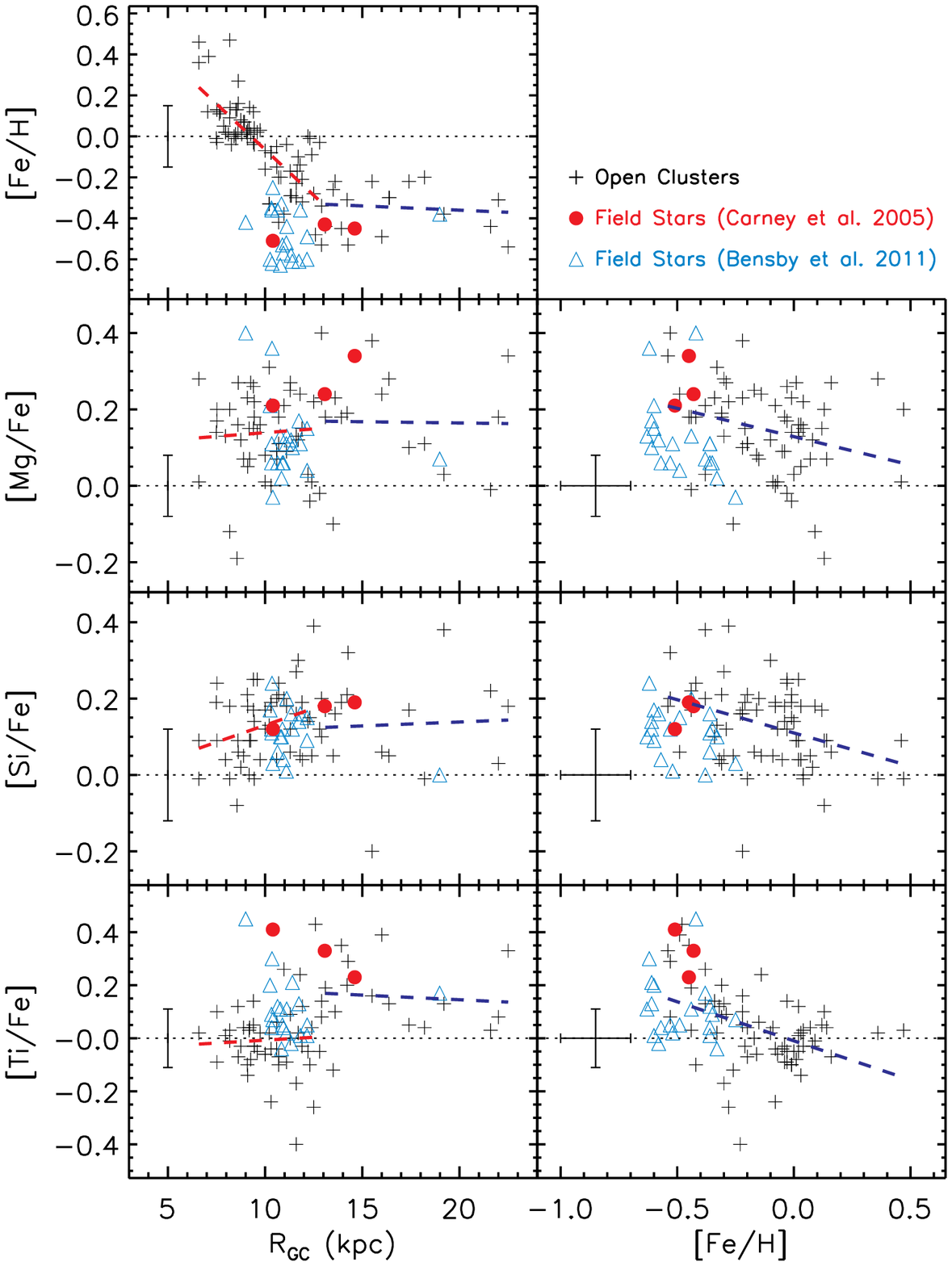}
\caption{[Fe/H], [Mg/Fe], [Si/Fe], and [Ti/Fe] vs. \rgc\ (left panels) and 
[Mg/Fe], [Si/Fe], and [Ti/Fe] vs.\ [Fe/H] (right panels) 
for open clusters (plus signs), field stars from \citet{carney05} (filled red 
circles), and field stars from \citet{bensby11} (blue open triangles). 
In the left panels, we fit the local (\rgc\ $<$ 13 kpc) and distant (\rgc\ $>$ 13 kpc) 
open clusters and in the right panels, we fit the open cluster data. 
Representative error bars are included in each panel. 
\label{fig:field}}
\end{figure}

In Figure \ref{fig:field}, 
we plot [Fe/H], [Mg/Fe], [Si/Fe], and [Ti/Fe] vs.\ \rgc\ (left panels) 
for open clusters and field stars \citep{carney05,bensby11}.  
In these panels, we plot the linear fit to the 
local and distant open cluster samples. The first notable aspect from 
this figure is that the field star samples from \citet{carney05} and 
\citet{bensby11} do not extend to the same large distances as do the open clusters. 
The second point we highlight is 
that the field stars are, on average, more metal-poor than the open clusters 
at the same distances. 
For open clusters with 10 kpc $\le$ \rgc\ $\le$ 12 kpc the 
mean metallicity is [Fe/H] = $-$0.20 and the dispersion is 
$\sigma$[Fe/H] = 0.11 dex. 
For the field stars in the same range of Galactocentric distances, 
we find a mean metallicity of [Fe/H] = $-$0.48 and dispersion of 
$\sigma$[Fe/H] = 0.13 dex. 
It is not obvious why the metallicity difference exists between 
the open clusters and field stars at the same Galactocentric distances. 
Systematic offsets between the various studies at the 
$\sim$0.3 dex level would be necessary to explain these differences 
or the field stars could be thick disk objects. 
Alternatively, it may be that 
the difference in metallicity between the samples are  
due to selection biases in the field stars. 
Raising our open cluster metallicities by 0.1 dex would only 
serve to increase the discrepancy in [Fe/H] 
between open clusters and field stars. 
The third point we note is that 
the dispersion in [Fe/H] for the \citet{bensby11} data (0.13 dex) is comparable 
to the dispersion in [Fe/H] for the open clusters at similar distances (0.11 dex). 

We now compare the distributions of [Mg/Fe], [Si/Fe], and [Ti/Fe] vs.\ \rgc\ for the 
field stars and open clusters. Taking objects 
in the range 10 kpc $\le$ \rgc\ $\le$ 12 kpc, we find mean values and 
dispersions of [Mg/Fe] = 0.15 ($\sigma$ = 0.09) for the open clusters and 
[Mg/Fe] = 0.11 ($\sigma$ = 0.09) for the field stars. For Si, we find 
mean values and dispersions of [Si/Fe] = 0.16 ($\sigma$ = 0.08) and 
[Si/Fe] = 0.12 ($\sigma$ = 0.06) for the open clusters and field 
stars, respectively. For Ti, we find mean values and dispersions of 
[Ti/Fe] = $-$0.03 ($\sigma$ = 0.15) and 
[Ti/Fe] = 0.09 ($\sigma$ = 0.09) for the open clusters and field 
stars, respectively. While 
[X/Fe] is in very good agreement 
between open clusters and field stars in the same range of Galactocentric 
distances, there are very few field stars beyond \rgc\ = 13 kpc with 
chemical abundance measurements. There is a clear need to study more distant 
field stars to understand the chemical properties of the outer Galactic disk. 
Had we decreased our [X/Fe] ratios by 0.1 dex, arguably the agreement 
in [X/Fe] between field red giants and open clusters would remain satisfactory. 

In the right panels of Figure \ref{fig:field}, we compare 
[X/Fe] vs.\ [Fe/H] between the field stars and the open clusters. 
It seems that the field stars have lower [X/Fe] 
ratios, on average, than do the open clusters at the same metallicity. 
Within the limited samples and bearing in mind the measurement uncertainties 
and possible systematic offsets, 
the field stars and open clusters seem to follow the same trends between 
abundance ratios [Fe/H] and [X/Fe] vs.\ \rgc\ and between [X/Fe] vs.\ [Fe/H]. 
Therefore, our tentative conclusion is that 
the interpretation of the outer Galactic disk does not depend 
significantly upon whether we use open clusters or field stars 
although we remind the reader that selection biases for the field stars 
may give much stronger weight to thick disk objects. 

\subsubsection{Insights from Cepheids}

Studies of the chemical abundances of Cepheids provide a complementary view  
on the radial abundance gradients in the disk. With their high masses and 
short lifetimes, the Cepheids likely reflect the present-day 
chemical composition of the disk. A number of studies within the 
last decade have explored the 
radial abundance gradients as traced by Cepheids 
(e.g., 
\citealt{andrievsky02a,andrievsky02b,andrievsky02c,andrievsky04,lemasle08,luck03,luck11,ll11,pedicelli09,y06}). 

While the time variation in radial abundance 
gradients can be obtained by comparing the younger and older 
open clusters, we note that the sample sizes are modest (e.g., 
our compilation has 24 and 43 clusters with ages below 1 Gyr and 
above 2.5 Gyr, respectively, and of the 24 clusters younger than 1 
Gyr, only 3 lie beyond \rgc\ = 12 kpc and none beyond 14 kpc).  
Instead, comparison between Cepheids and old open clusters 
offers a larger sample (e.g., \citealt{ll11} have a homogeneous 
sample of 339 Cepheids including 40 and 15 objects beyond 12 kpc 
and 14 kpc, respectively) and arguably a more robust measure 
of the time variation of radial abundance gradients. 
(We note that planetary nebulae also offer important insights  
into the chemical evolution of the Galactic disk, although 
their distant uncertainties are large relative to 
those of open clusters and Cepheids. We refer 
the interested reader to \citet{maciel03}, 
\citet{costa04}, and \citet{maciel09} and 
references therein.) 

One concern is that there may be large systematic differences between the 
abundances obtained from open cluster red giants and those from Cepheids. 
Indeed, caution must be exercised when directly comparing 
abundances between open clusters and Cepheids. 
However, abundance gradients basically represent 
a differential abundance comparison between similar stars 
but at different Galactocentric distances. If each sample (open cluster 
giants or Cepheids) is analyzed uniformly, then the systematic errors 
in the analysis of a particular type of star 
should largely cancel thereby enabling the radial gradient comparisons 
we seek to conduct. 

\begin{figure}[t!]
\epsscale{1.2}
\plotone{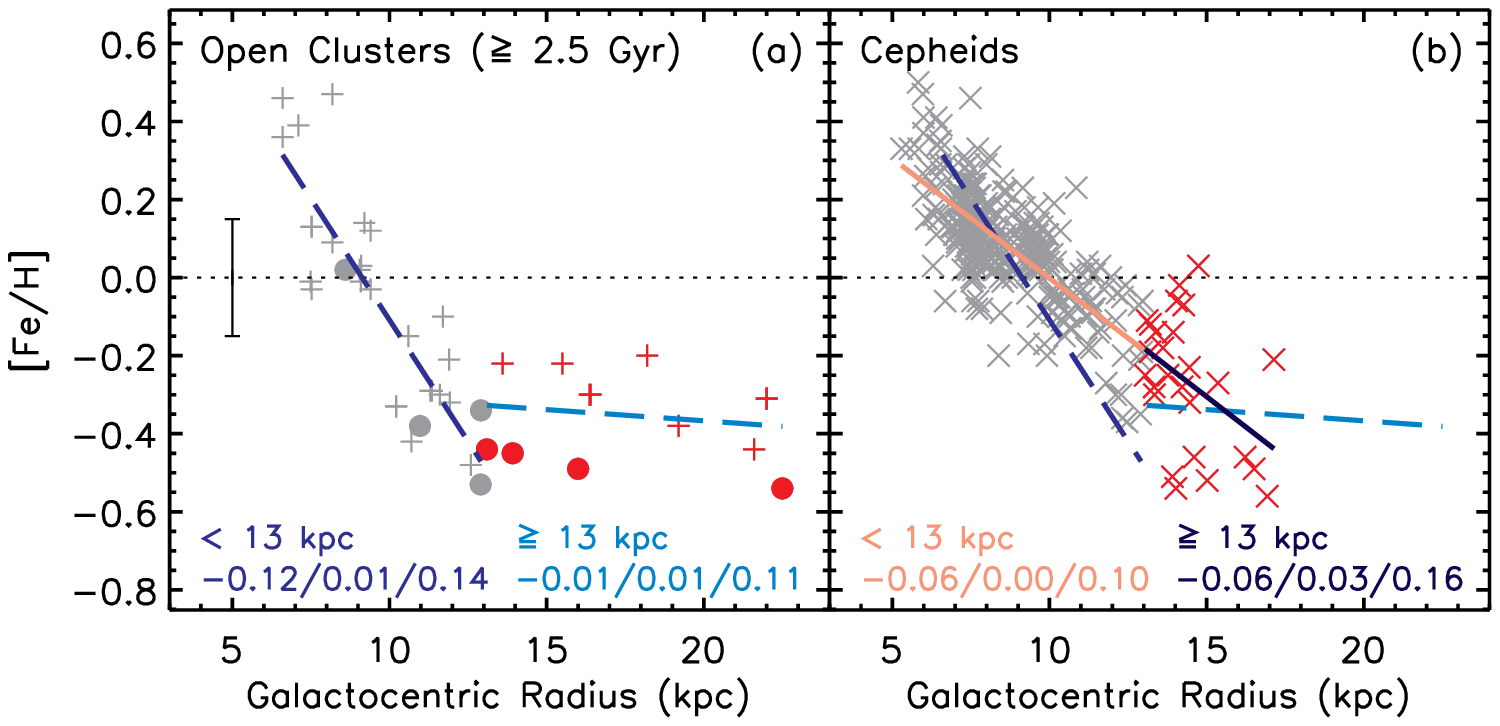}
\caption{[Fe/H] vs.\ \rgc\ for the older 
open clusters (age $\ge$ 2.5 Gyr; left panel) and Cepheids 
(right panel).  Data with \rgc\ $<$ 13 kpc are in grey while data with 
\rgc\ $>$ 13 kpc are in red.  Open clusters from our work are shown as 
filled circles while the Cepheid data are from \citet{ll11}. We determine 
the linear fit, uncertainty, and dispersion about the linear fit to the 
local and distant open clusters and Cepheids. For the Cepheid data, we 
superimpose the linear fits to the open cluster data. 
\label{fig:radialfe}}
\end{figure}

In Figure \ref{fig:radialfe}, we plot [Fe/H] vs.\ \rgc\ for the open clusters 
and for the Cepheids from \citet{ll11}. This figure enables us to address 
whether the older open clusters have a different [Fe/H] radial gradient 
than the young Cepheids. Therefore, we only plot open clusters with ages 
greater than 2.5 Gyr. Additionally, we divide the open clusters and Cepheids 
into a local (\rgc\ $<$ 13 kpc) and distant (\rgc\ $>$ 13 kpc) samples. 
We measure the linear fit to all subsamples and write the slope, 
uncertainty, and dispersion about the linear fit in Figure \ref{fig:radialfe}. 
As noted in previous studies of Cepheids, there is no sudden change in gradient 
between [Fe/H] and \rgc. In contrast to the open clusters, the Cepheids 
decrease smoothly in metallicity with increasing distance. The same result 
would apply had we overplotted our Cepheid data from Paper III. Since 21 of 
those 24 stars were studied by \citet{ll11}, for convenience we rely only upon 
their data to ensure a homogeneous sample. 
Next, we note that for the inner samples, the older open clusters show a much 
steeper abundance gradient ($-$0.12 $\pm$ 0.01 dex/kpc) than the younger Cepheids 
($-$0.06 $\pm$ 0.00 dex/kpc). 

\begin{figure}[t!]
\epsscale{1.2}
\plotone{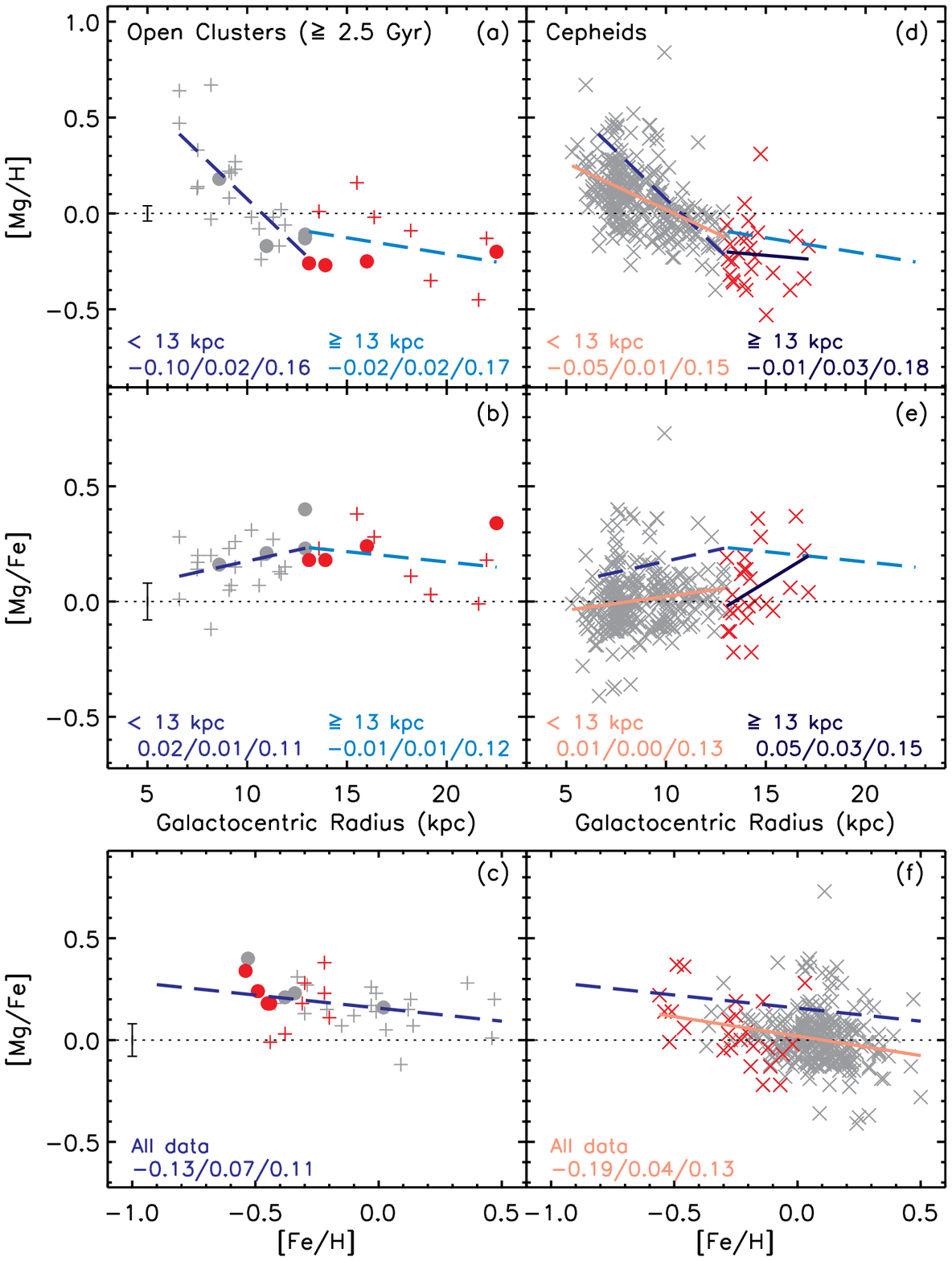}
\caption{[Mg/H] and [Mg/Fe] vs. \rgc\ and [Fe/H] 
for the open clusters (left panel) and Cepheids 
(right panel). Data with \rgc\ $<$ 13 kpc are in grey while data with 
\rgc\ $>$ 13 kpc are in red. Open clusters from our work are shown as 
filled circles while the Cepheid data are from \citet{ll11}. We determine 
the linear fit, uncertainty, and dispersion about the linear fit to the 
local and distant open clusters and Cepheids. For the Cepheid data, we 
superimpose the linear fits to the open cluster data. 
\label{fig:radialmg}}
\end{figure}

\begin{figure}[t!]
\epsscale{1.2}
\plotone{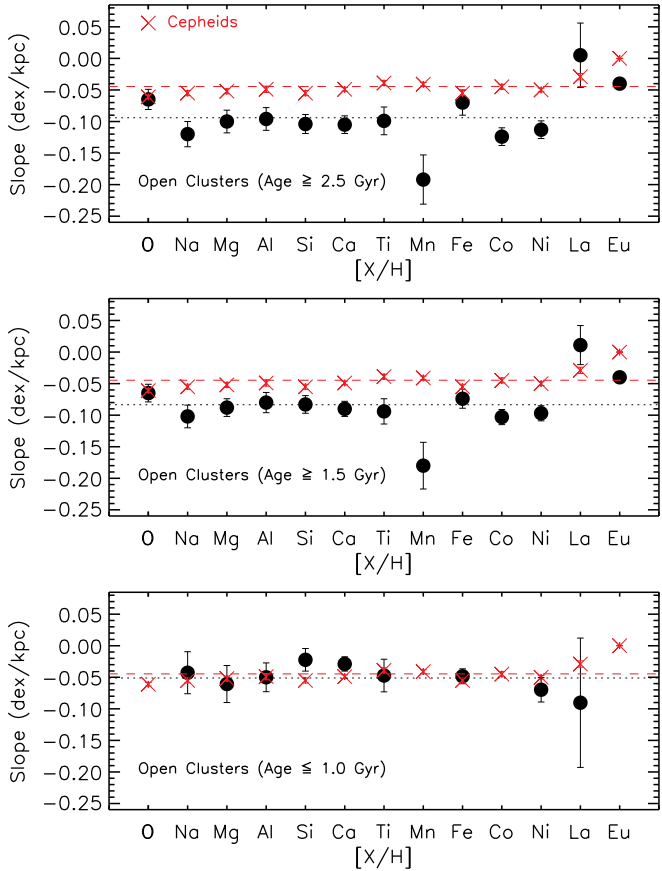}
\caption{
Slope of linear fit (dex/kpc) to [X/H] vs. \rgc\ 
for the open clusters (filled circles) and Cepheids (red crosses) 
for \rgc\ $<$ 13 kpc. 
The upper, middle, and lower panels show values 
for open clusters with ages $\ge$ 2.5 Gyr, $\ge$ 1.5 Gyr, 
and $\le$ 1.0 Gyr, respectively. 
\label{fig:radialcompare}}
\end{figure}

In Figure \ref{fig:radialmg}, 
we plot [Mg/H] and [Mg/Fe] vs.\ Galactocentric distance as well as 
[Mg/Fe] vs.\ [Fe/H] for the open cluster and Cepheid samples. We again 
determine the linear fit to the local and distant samples. 
For the local sample (\rgc\ $<$ 13 kpc), the older open clusters 
show a steeper gradient for [Mg/H] compared to the younger Cepheids, 
a result also seen for [Fe/H]. 
This analysis was repeated for each element and in Figure 
\ref{fig:radialcompare}, we compare the gradients for [X/H] vs.\ \rgc\ between 
open clusters and Cepheids with \rgc\ $<$ 13 kpc. 
For the open clusters, we consider only clusters with 
ages $\ge$ 2.5 Gyr (upper panel) and $\ge$ 1.5 Gyr (middle panel). 
For both cases, we find that the gradient for [X/Fe] vs.\ \rgc\ 
is always shallower for the Cepheids compared to the older open clusters, 
with the exception of La. 
Such a time variation in radial abundance gradient is in the same sense as 
predicted by chemical evolution models 
(e.g., \citealt{hou00,fu09,pilkington12}) and 
is believed to be due to the ``inside-out'' formation of the disk.  

As a further comparison, we measured the radial abundance gradients 
in the youngest open clusters. The 24 open clusters with ages 
$\le$ 1.0 Gyr all lie within \rgc\ = 13 kpc. The radial abundance 
gradient for [Fe/H] is $-$0.05 $\pm$ 0.01 dex/kpc, a value in 
excellent agreement with the local Cepheids ($-$0.06 $\pm$ 0.00 dex/kpc). 
In the lower panel of Figure \ref{fig:radialcompare}, we compare the 
radial abundance gradients, [X/H], for the young Cepheids and youngest 
open clusters (we only show data for those elements for which 
there were four or more clusters). 
The excellent agreement in radial abundance gradients for 
the two sets of young objects validates 
our expectation that radial gradients represent a differential comparison 
within which the systematic errors cancel. 

One issue we can address with these data is whether the outer disk open clusters 
(\rgc\ $>$ 13 kpc) share the same [X/H] abundance 
gradient as the inner disk open clusters 
(\rgc\ $<$ 13 kpc). For open clusters with ages $\ge$ 1.5 Gyr, we find that for all 
elements except La and Eu, the difference in gradients ([X/H] vs.\ \rgc) 
between the local and distant samples is significant at the 3-$\sigma$ level 
or higher. The same result holds when we consider open clusters with 
ages $\ge$ 2.5 Gyr. 
Similarly, we can compare the abundance gradients ([X/H] vs.\ \rgc) 
for the local and distant Cepheids. In contrast to the open clusters, we find 
that for all elements, the abundance gradients agree within 1.8-$\sigma$. 
Furthermore, six of the 12 elements considered show agreement 
below the 0.9-$\sigma$ level between 
gradients for the local and distant Cepheids. 
In summary, the local and distant Cepheids follow the 
same [X/H] vs.\ \rgc\ relation.  
There is, therefore, a fundamental difference in the behavior of 
radial abundance gradients, [X/H] vs.\ \rgc, 
between the older open clusters and the Cepheids. 

Next, we compare the [X/Fe] radial 
abundance gradient between the local and distant 
samples. For the open clusters with ages $\ge$ 1.5 Gyr, the local and distant 
samples show the same [X/Fe] vs.\ \rgc\ gradient with the exceptions of 
Mn (2-$\sigma$), La (3-$\sigma$), and Eu (3-$\sigma$). When considering 
the open clusters with ages $\ge$ 2.5 Gyr, with the exceptions of 
Mn (2-$\sigma$) and Eu (2-$\sigma$), the local and distant samples show 
the same abundance gradients. 
A similar comparison for the Cepheids reveals that with the exception of 
Al (2-$\sigma$), the local and distant samples show identical abundance 
gradients for [X/Fe] vs.\ \rgc\ at the 1.8-$\sigma$ level or lower. 

When we consider the trends between [X/Fe] vs.\ [Fe/H], our qualitative 
impression is that the local and distant open clusters follow the same 
relation. Similarly, the local and distant Cepheids appear to follow 
the same [X/Fe] vs.\ [Fe/H] relation. Given the moderate ranges 
in [Fe/H] covered by the distant open cluster and Cepheid samples, 
these are qualitative rather than quantitative comparisons. 
For the linear fit to [X/Fe] vs.\ [Fe/H] across the full data range, 
we are willing to quantitatively compare the gradients between 
the open clusters and Cepheids. As before, we assume that 
the systematic errors in the analysis of a particular type of star 
will cancel such that we can reliably compare gradients between 
the two different classes of objects (red giants vs.\ Cepheids). 
We find that the open clusters and Cepheids share the same 
gradients ($<$3-$\sigma$) 
from the linear fit to [X/Fe] vs.\ [Fe/H] for all elements 
except O (3-$\sigma$), Mn (4-$\sigma$), and Ni (3-$\sigma$). 

At the risk of over-interpreting these data, the results of this 
comparison ([X/Fe] vs.\ [Fe/H]) 
would suggest that the chemical enrichment histories of the 
older open clusters and Cepheids are surprisingly comparable. 
This similarity applies to the more metal-rich local samples as well 
as to the more metal-poor distant samples. 
Our conclusions would be unchanged had we increased our 
[Fe/H] ratios by 0.1 dex and decreased our [X/Fe] ratios by 0.1 dex 
for the open clusters in Paper I and in this study. 
Similarity in [X/Fe] vs.\ [Fe/H] relations between local and distant stars 
that span a large range in ages cannot be the result of the evolution 
of a closed system. It remains to be seen how such abundance ratios can be 
explained within a self-consistent chemical evolution model. 
One requirement may be the infall of gas onto the outer disk. 
However, that material must have experienced the same ratio of 
SNe II to SNe Ia, at a given metallicity, as the gas from which 
the local disk formed, both at early times (i.e., at the epoch of 
open cluster formation) and later times when the Cepheids formed. 

\section{CONCLUSIONS AND FUTURE WORK}

Previous papers in this series studied the chemical compositions 
of open clusters, field stars, and Cepheids in the outer disk. 
In this paper we present radial velocities and chemical abundances 
for a new sample of outer disk open clusters. 
This paper includes the first analysis of the outer disk clusters 
Be 18 and PWM 4. 
We compiled a set of chemical abundances for a sample of 49 unique 
clusters drawn from our studies and from the literature. 
Using this sample, we studied trends between chemical 
abundances and distance, age, and metallicity. 

We confirm the flattening of the metallicity gradient in the outer disk 
and that the outer disk open clusters are uniformly metal-poor with 
super-solar ratios for [$\alpha$/Fe] and [Eu/Fe]. 
For some elements there are hints that the 
local (\rgc\ $<$ 13 kpc) and distant (\rgc\ $>$ 13 kpc) 
samples may have different radial [X/Fe] abundance gradients. 

We confirm that there are no significant trends between 
metallicity, or abundance ratios [X/Fe], and age (with the likely 
exception of the $s$-process elements already noted by 
\citealt{dorazi09} and \citealt{maiorca11}). 
Compared to a sample of solar neighborhood field giant stars, we 
find that the open clusters share rather similar trends for [X/Fe] versus 
[Fe/H] for almost all elements. 
We quantify the linear trends between [X/Fe] and metallicity and find 
that the scatter about the mean relation, as low as 0.06 dex for Ni, 
is comparable to the measurement 
uncertainties for some elements. 
For other elements including Co, Zr, Ba, and La, 
the scatter about the linear trend is 
significantly higher than the measurement uncertainties 
which may suggest a real dispersion in abundance ratios. 
We note that for lines that are strong and/or affected by 
hyperfine structure (e.g., Ba), the measurement uncertainties may 
underestimate the true errors. Additionally, some elements 
including Co and Zr 
are measured from small numbers of lines such that the 
measured dispersions are likely larger than the errors from 
stellar parameter uncertainties alone. Finally, 
our analysis suggests that this 
inhomogeneous sample includes $\sim$ 0.1 dex systematic offsets for 
some elements. 

The flattening of the metallicity gradient, differences 
in metallicity, and the enhancements in 
[$\alpha$/Fe] (and perhaps [Eu/Fe]) suggest that the outer disk 
formed from gas with a different star formation history than 
the solar neighborhood. 
We reiterate that the individual $\alpha$ elements do not necessarily 
follow identical patterns. In particular, [O/Fe] and [Ti/Fe] 
are strongly enhanced in the outer disk relative to the 
inner disk while [Mg/Fe], [Si/Fe], and [Ca/Fe] show roughly 
constant ratios as a function of Galactocentric distance. 
Outer disk field red giant stars, which cover a more limited range in 
Galactocentric distance, share similar [X/Fe] and [Fe/H] 
vs.\ \rgc\ trends as the open clusters at the same range of distances. 
When compared to Cepheids, the old open clusters (ages $\ge$ 1.5 Gyr) 
show steeper [X/H] vs.\ \rgc\ trends suggesting that the Galactic 
disk grew via an ``inside-out'' process. The Cepheids and open 
clusters share very similar [X/Fe] vs.\ [Fe/H] trends. Such 
similarity between samples of stars with very different 
ages cannot arise from the chemical evolution of a closed 
system. Understanding the chemical abundances of old distant open 
clusters and young distant 
Cepheids represents a challenge for future chemical evolution models. 

Ultimately, a definitive statement about the origin and evolution 
of the outer disk requires a homogeneous analysis of larger samples 
of stars in larger numbers of clusters based on high quality 
spectra. Additional efforts should be undertaken to increase the samples 
of distant field stars and Cepheids with chemical abundance 
analyses for a more complete picture of the outer disk. 
Clearly, the comparison 
field stars also need to be analyzed on the same system as the 
open clusters. 
For example, the work by Przybilla, Nieva, and collaborators 
(e.g., \citealt{przybilla08,nieva12}) 
on unevolved early type stars shows that 
very high precision abundances, 10\%, can be obtained for objects 
spanning a large range in distance. 
Extremely careful analyses may one day provide similar precision 
in abundances in 
open clusters to explore the origin and evolution 
of the outer disk. 

\acknowledgments

D.Y.\ thanks Robert Sharp for assistance with the MPFIT tasks, 
Sergio Ortolani for providing photometry of PWM 4, Heather Jacobson 
for sending electronic data, and 
Giovanni Carraro, 
Ken Freeman, Brad Gibson, Earl Luck, and Ivan Minchev for helpful discussions. 
We thank the anonymous referee for thoughtful comments. 
The authors wish to recognize and acknowledge the very significant 
cultural role and reverence that the summit of Mauna Kea has always 
had within the indigenous Hawaiian community.  We are most fortunate 
to have the opportunity to conduct observations from this mountain. 
We are extremely grateful to the National Science Foundation for
their financial support through grants AST 96-19381, AST 99-88156, 
and AST 03-05431 to the University of North Carolina. 
DY acknowledges travel support through the Access to Major Research 
Facilities Program which is supported by the Commonwealth of Australia
under the International Science Linkages program. 
This research has made use of the WEBDA database, operated at the 
Institute for Astronomy of the University of Vienna.

\appendix
%\section{Appendix}
\subsection{A.1~ SIGNAL-TO-NOISE CONSIDERATIONS}

As noted in Section 4.1, the compilation of open cluster abundances 
includes data that span a range of S/N (and sample sizes per cluster). 
We now re-examine whether the trends between [Fe/H] vs.\ \rgc\ and [X/Fe] vs.\ \rgc\ 
change as we make cuts based on S/N. 

\begin{figure}[th!]
\epsscale{0.5}
\plotone{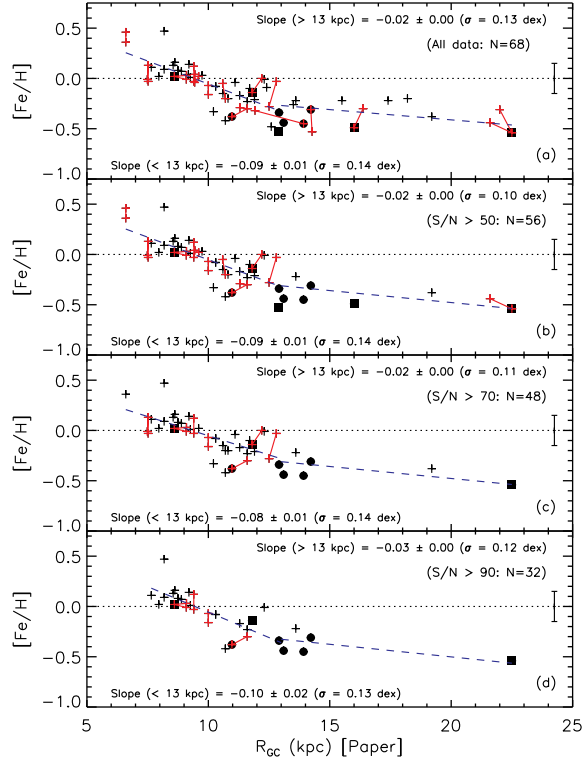}
\caption{[Fe/H] vs.\ \rgc\ for different S/N cuts. 
From top to bottom, the panels include 
(a) all clusters, (b) S/N $>$ 50, (c) S/N $>$ 70, and (d) S/N $>$ 90. 
The symbols are the same as in Figure \ref{fig:fehab}. 
\label{fig:snfe}}
\end{figure}

In Figure \ref{fig:snfe}, we show [Fe/H] vs.\ \rgc\ again applying a linear fit 
to the local (\rgc\ $<$ 13 kpc) and distant (\rgc\ $>$ 13 kpc) samples. 
In this figure, we show (a) all data, (b) S/N $>$ 50, (c) S/N $>$ 70, and 
(d) S/N $>$ 90. We find that the slopes and dispersions about the linear fit do not change 
substantially as we exclude data based on the S/N. One important point we 
highlight is that for the most stringent cut, S/N $>$ 90, there is only 
one cluster beyond \rgc\ = 15 kpc. Panel (d) in this figure underscores the need 
to re-observe the most distant open clusters at higher S/N ratios in order to 
obtain more accurate and precise chemical abundance ratios. We also stress 
the importance of observing multiple stars per cluster, when possible, to 
increase the statistics and to ensure membership. Indeed, a similar figure 
could be generated taking into account the S/N, spectral 
resolution, and number of stars per cluster. 

\begin{figure}[th!]
\epsscale{0.5}
\plotone{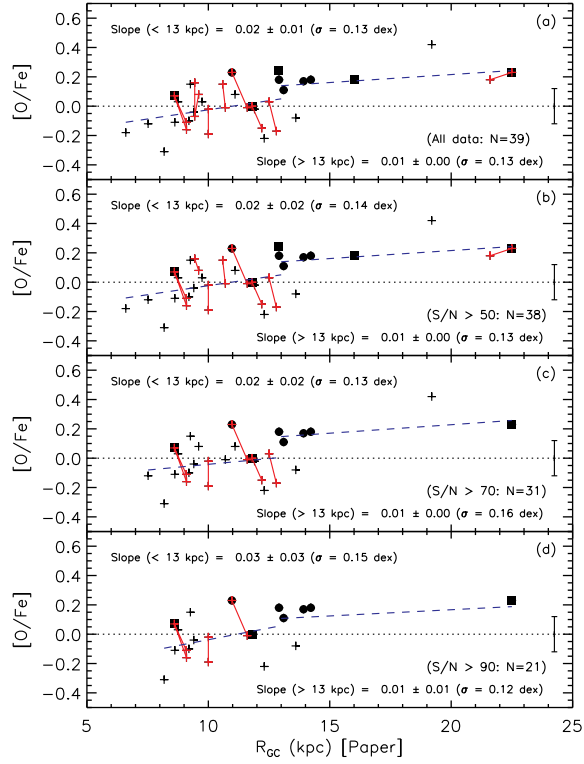}
\caption{Same as Figure \ref{fig:snfe} but for [O/Fe]. 
\label{fig:sno}}
\end{figure}

\begin{figure}[th!]
\epsscale{0.5}
\plotone{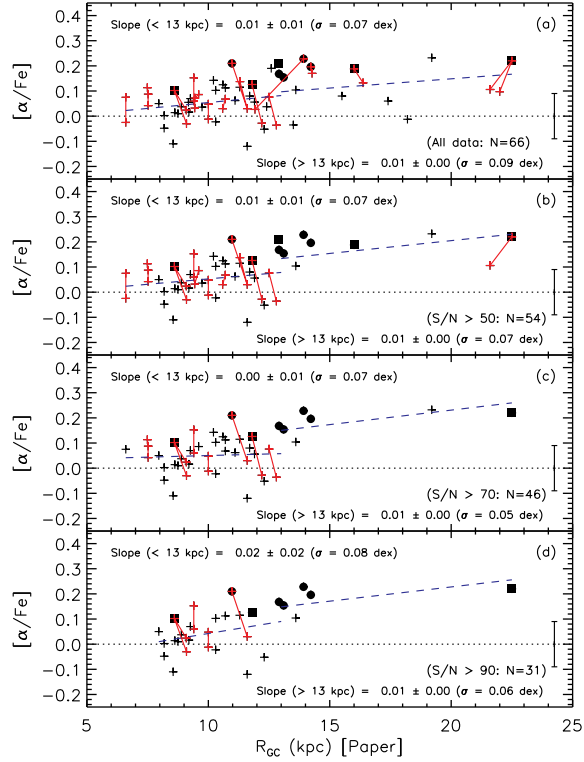}
\caption{Same as Figure \ref{fig:snfe} but for [$\alpha$/Fe]. 
\label{fig:snalpha}}
\end{figure}

\begin{figure}[th!]
\epsscale{0.5}
\plotone{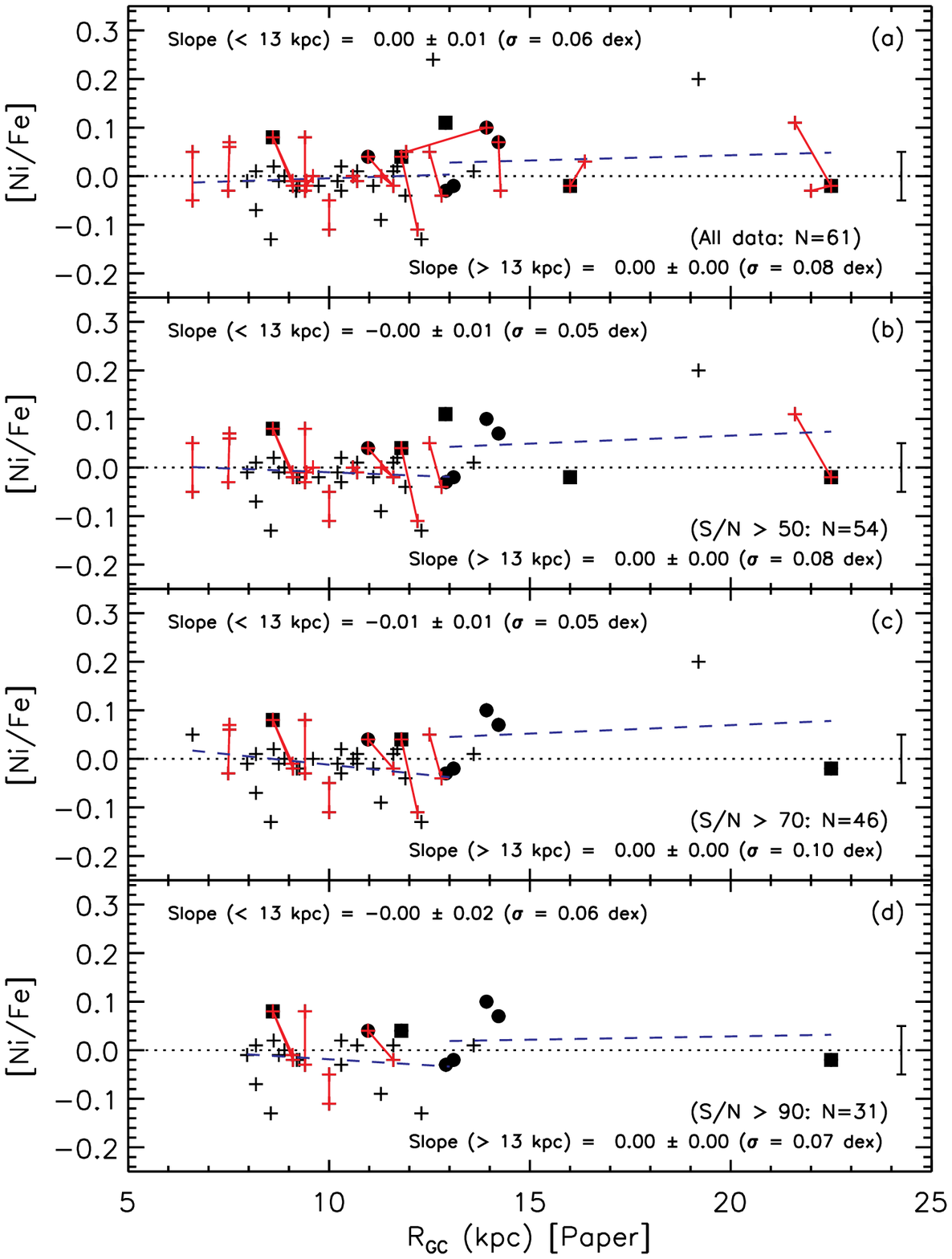}
\caption{Same as Figure \ref{fig:snfe} but for [Ni/Fe]. 
\label{fig:snni}}
\end{figure}

In Figures \ref{fig:sno} to \ref{fig:snni}, we plot [O/Fe], [$\alpha$/Fe], 
and [Ni/Fe] vs.\ \rgc. For these representative elements, 
we again find that the slopes and dispersions 
about the linear fit to the local and distant samples do not significantly 
change as we restrict ourselves to only the highest S/N data. 
Nevertheless, we again argue that 
accurate and precise chemical abundance measurements are obtained from 
higher quality spectra and currently, there is a clear need to obtain 
such data for the most distant open clusters. 

\subsection{A.2~ THE TRANSITION FROM INNER TO OUTER DISK}

\citet{twarog97} were the first to note that the radial abundance 
distribution, as traced by open clusters, shows a sharp transition 
near \rgc\ = 10 kpc. Numerous studies, including this series of papers, 
have sought to further understand the transition from the inner to 
outer disk. The location, and nature, of the transition can potentially 
provide crucial constraints upon the formation and evolution of the 
Galactic disk. 
While most 
studies have performed quantitative comparisons between 
the various (chemical) properties of the inner disk and outer disk 
open clusters, to our knowledge 
the transition region, or transition radius, 
is selected in most, if not all, studies 
through a qualitative visual inspection. 
In this appendix, we seek to quantify the location 
of the transition radius between the inner and outer disk. 

The data we shall use to quantify the transition radius are [Fe/H] vs.\ 
\rgc\ for the complete sample listed in Table \ref{tab:compile}. 
We assume that there are two regions, an inner disk and an 
outer disk. We also assume that each region can be described by a 
linear relation. Thus, there are five parameters to be 
determined: (1 and 2) the slope and intercept of the linear fit 
to the inner disk region, 
(3 and 4) the slope and intercept of the linear fit to 
the outer disk region, and (5) the transition radius. 

\begin{figure}[th!]
\epsscale{0.5}
\plotone{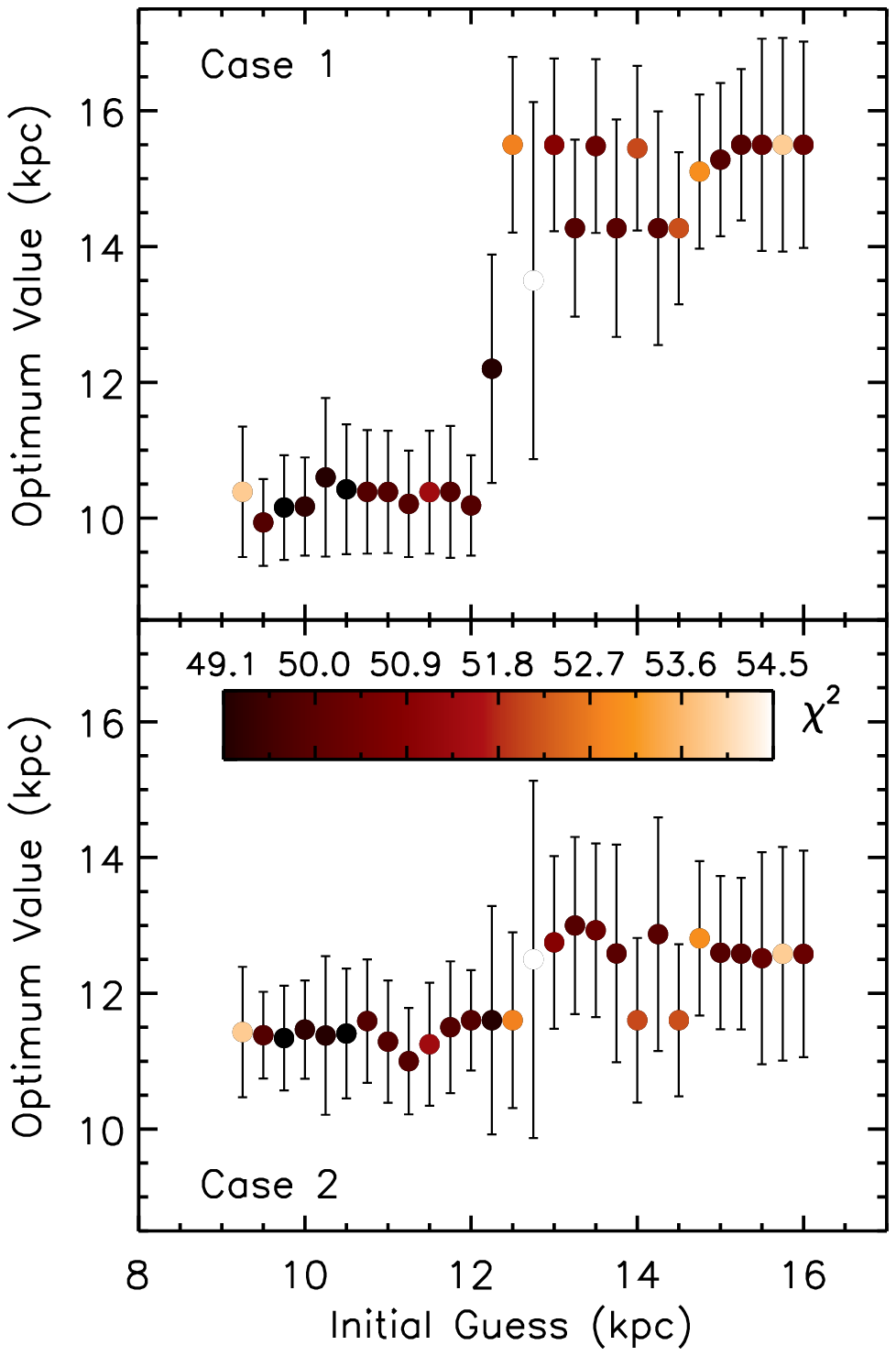}
\caption{Optimum value for the transition radius as a function 
of the initial guess. Case 1 (upper) is when we do not require 
the two linear fits to intersect at the transition radius. 
Case 2 (lower) is when we require the two linear fits to 
intersect at the transition radius. 
The colors represent the $\chi^2$ value for each data point. 
\label{fig:rt}}
\end{figure}

\begin{figure}[th!]
\epsscale{0.5}
\plotone{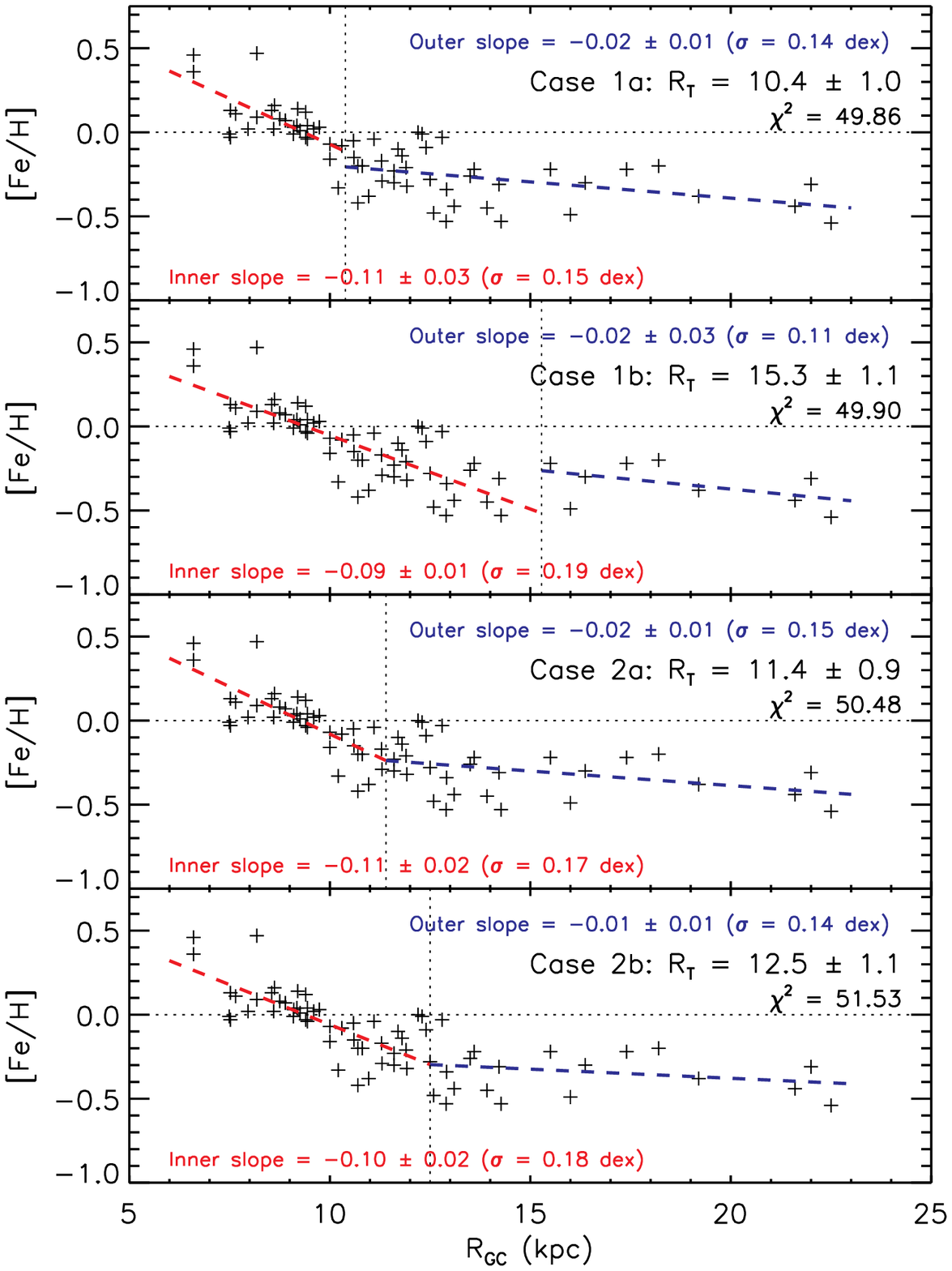}
\caption{[Fe/H] vs.\ Galactocentric distance for the complete 
sample. In each panel, we overplot the optimum solution 
from the MPFIT \citep{markwardt09} routines. Case 1 is when 
we do not require the linear fits to intersect while Case 2 
is when we require the linear fits to intersect. 
For each case, there are two solutions for the transition 
radius, depending on the initial guess (see Figure \ref{fig:rt}). 
We show the two 
solutions for each case and note the slope and error, dispersion 
about the linear fit, the transition radius and error, and 
$\chi^2$. 
\label{fig:rt2}}
\end{figure}

We determine these parameters using the IDL MPFIT routine 
\citep{markwardt09} which uses the Levenberg-Marquardt technique 
for least squares minimization. One consideration is that an
initial guess for the best parameters is required. 
On applying the MPFIT routines, this concern appears to be valid. 
We tested a number of initial guesses for the five parameters. 
The only parameter that showed a dependence on the initial guess was 
the transition radius. In the upper panel of 
Figure \ref{fig:rt} we plot the initial guess 
(which ranges from 9 kpc to 16 kpc in steps of 0.25 kpc) and 
the optimum value. In this figure, there are two best solutions for the 
transition radius, 
10.4 $\pm$ 1.0 kpc and 
15.3 $\pm$ 1.3 kpc,  
and the optimum value depends on the choice of initial 
guess. (The Levenberg-Marquardt technique involves 
gradient descent. If there is a saddle point, it is 
therefore not surprising that the final solution depends 
on our initial guess.) 
We note that when using either of the two best solutions 
as the initial guess, their $\chi^2$ values are almost identical, 
49.86 (10.4 kpc) and 49.90 (15.3 kpc). In the upper two panels of 
Figure \ref{fig:rt2}, we overplot the two solutions to our data. 
The first conclusion we draw from this quantitative analysis is that there 
does not appear to be a single value for the optimum transition radius 
if we do not place any constraints on the linear fits to the inner 
and outer regions (i.e., Case 1). Rather, there are two equally 
good solutions for the transition radius at 10.4 $\pm$ 1.0 kpc or 
15.3 $\pm$ 1.3 kpc. 

\begin{figure}[th!] 
\epsscale{0.5}
\plotone{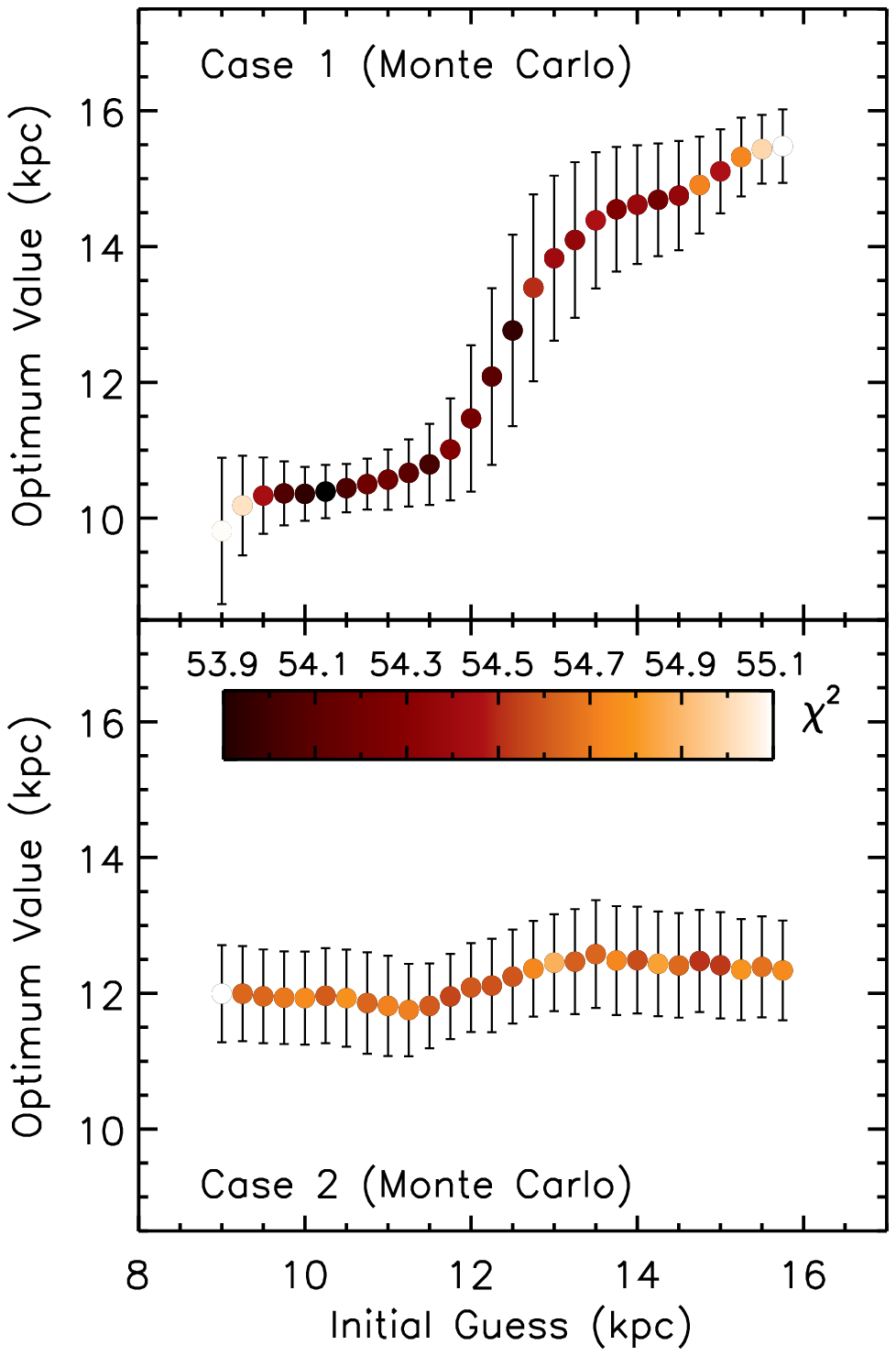}
\caption{Same as Figure \ref{fig:rt} but based on a Monte Carlo 
simulation for N = 1,000 realizations. In each realization, the 
\rgc\ for a given data point was replaced by a random number drawn 
from a normal distribution of width 0.5 kpc centered at the original 
value. The data and error bars represent the average value and 
dispersion from the 1,000 realizations respectively. The colors 
represent the average $\chi^2$ at each data point. 
\label{fig:rtmonte}}
\end{figure}

The MPFIT routines include consideration of the errors 
for the Y axis ([Fe/H], for which we adopt a uniform value of 
0.15 dex), 
but not for the X axis (\rgc). 
In order to explore the effect of uncertainties in the distances, 
we adopt a Monte Carlo approach in which we replaced each distance 
with a random number drawn from a normal distribution of width 0.5 kpc 
centered at the \rgc\ of the given data point. We repeated this 
process for each data point in the sample and determined the optimum 
parameters. We repeated this process for 1,000 new random samples. 
As above, for each sample we trialed a range of initial guesses for 
the transition radius (from 9 kpc to 16 kpc in steps of 0.25 kpc). 
In the upper panel of Figure \ref{fig:rtmonte}, 
we plot the average optimum transition radius as a function of 
the initial guess. In this figure, the error bars represent the standard 
deviation of the distribution of 1,000 values. This figure again 
highlights that there are two preferred values which depend upon 
the initial guess. 

We note that thus far we have not required the two linear fits to intersect 
at the transition radius. 
We now consider how the results change if we 
constrain the two linear functions to intersect at the transition 
radius (i.e., Case 2), and this may be a more appealing way to 
describe the behavior of metallicity with Galactocentric distance. 
In this scenario, there are only four free parameters 
(the slope and intercept of the fit to the inner region, the 
transition radius, and the slope of the fit to the outer region). 
Once again, we test to see whether the optimum value for the 
transition radius shows a dependence on the initial guess. 
In the lower panel of Figure \ref{fig:rt}, 
there is a suggestion that there are again two solutions which 
depend on the initial guess, 11.4 $\pm$ 0.9 kpc and 12.5 $\pm$ 1.1 kpc. 
In the lower panels of Figure \ref{fig:rt2}, we overplot these 
two solutions and note again that the $\chi^2$ values 
(when adopting the 
optimum values as the initial guesses) for the two solutions are 
very similar, 50.48 (11.4 kpc) and 51.53 (12.5 kpc). 

Finally, we apply the Monte Carlo approach described above to Case 2. 
In the lower panel of Figure \ref{fig:rtmonte} 
we plot the optimum transition 
radius as a function of the initial guess where the error bars 
represent the standard deviation 
of the distribution of 1,000 values per initial guess 
and the color-bar represents the average $\chi^2$. 
In contrast to Case 1, the Monte Carlo simulations 
suggest that for Case 2 there is a single best solution for the transition 
radius of 12.1 $\pm$ 0.7 kpc. 
The second conclusion we draw from this analysis is that if we 
require the linear fits to the inner and outer regions to intersect 
at the transition radius, there is a single optimum value 
for the transition radius of 12.1 $\pm$ 0.7 kpc. 

Finally, we note that within the uncertainties, the slopes and linear 
fits we obtain from the MPFIT routines for the inner and outer regions 
agree with the slopes we measure in the main text based on our 
qualitatively selected 13 kpc transition radius.

\end{document}